\documentclass[12pt,a4paper]{article}

\usepackage[a4paper,vmargin=2.3cm,hmargin=2cm]{geometry}
\usepackage{lmodern}
\usepackage[T1]{fontenc}
\usepackage[intlimits]{amsmath}
\usepackage{amssymb}
\usepackage[bookmarks=true,hyperfigures=true]{hyperref}
\usepackage{graphicx} 
\usepackage[font=small,labelfont=bf,width=0.85\textwidth]{caption} 
\usepackage[nosort]{cite} 
\usepackage[bulletsep]{collref} 
\usepackage{fixmath} 
\usepackage{bbm}
\usepackage{tikz} 
\usepackage{dsfont}
\usepackage{mathtools}

\usepackage{microtype}
\usepackage{hep}
\providecommand{\microtypesetup}[1]{}

\newcommand\realization[2]{
\par\medskip
\noindent\ovalbox{\begin{minipage}{.95\textwidth}
\textbf{#1.}
\textit{#2}%
\end{minipage}}
}

\newcommand\theorem[2]{
\par\bigskip
\noindent\underline{\textbf{#1}}
\textit{#2}
}
\usepackage{fancybox}

\graphicspath{{./figures/}}

\makeatletter \let\@keywords\@empty \let\@subject\@empty
\providecommand{\keywords}[1]{\gdef\@keywords{#1}}
\providecommand{\subject}[1]{\gdef\@subject{#1}}
\def\thetitle{\@title}
\def\theauthor{\@author}
\def\thesubject{\@subject}
\def\thedate{\@date}
\def\thekeywords{\@keywords}
\makeatother
\AtBeginDocument{%
\hypersetup{pdftitle={\thetitle}}%
\hypersetup{pdfauthor={\theauthor}}%
\hypersetup{pdfsubject={\thesubject}}%
\hypersetup{pdfkeywords={\thekeywords}}%
}
\providecommand{\hypersetup}[1]{}
\providecommand{\texorpdfstring}[2]{#1}

\hypersetup{plainpages=false}
\hypersetup{pdfpagemode=UseNone}
\hypersetup{bookmarksnumbered=true}
\hypersetup{pdfstartview=FitH} 
\hypersetup{colorlinks=false}
\hypersetup{citebordercolor={.5 1 .5}}
\hypersetup{urlbordercolor={.5 1 1}}
\hypersetup{linkbordercolor={1 .7 .7}}

\numberwithin{equation}{section}

\let\oldbfseries=\bfseries
\let\oldmdseries=\mdseries
\let\oldnormalfont=\normalfont
\renewcommand{\bfseries}{\oldbfseries\boldmath}
\renewcommand{\mdseries}{\oldmdseries\unboldmath}
\renewcommand{\normalfont}{\oldnormalfont\unboldmath}

\newcommand{\sfrac}[2]{{\textstyle\frac{#1}{#2}}}
\newcommand{\half}{\sfrac{1}{2}}
\newcommand{\ihalf}{\sfrac{i}{2}}
\newcommand{\quarter}{\sfrac{1}{4}}

\newdimen\yysquaresize
\newdimen\yyrsquaresize
\newdimen\yythickness
\newdimen\yyskip
\def\yysquare#1{%
\setlength{\yyrsquaresize}{\yysquaresize}%
\addtolength{\yyrsquaresize}{-2\yythickness}%
\vrule width \yythickness%
\vbox to \yysquaresize{%
  \hrule height \yythickness\vss%
  \hbox to \yyrsquaresize{\hss#1\hss}%
  \vss\hrule height \yythickness}%
\vrule width \yythickness}

\def\yyyoung#1{\vtop{\baselineskip0pt\lineskip-\yythickness\halign{\tabskip-\yythickness&\yysquare{##}\cr #1}}}

\newcommand{\young}[1]{\hskip\yyskip\mbox{\yyyoung{#1}}\hskip\yyskip}

\newcommand{\nln}{\nonumber\\}

\newcommand{\shift}{\mathrm{U}}
\newcommand{\transfer}{\mathfrak{t}}
\newcommand{\mon}{\mathrm{T}}
\newcommand{\SR}{{}}
\newcommand{\charge}{\mathcal{Q}}

\newcommand{\perm}[1]{[#1]}

\newcommand{\biloc}[2]{[#1|#2]}

\DeclareMathOperator{\PP}{\mathds{P}}
\newcommand{\permop}{\PP}
\DeclareMathOperator{\idm}{\mathds{1}}
\newcommand{\idop}{\idm}

\newcommand{\defeq}{\mathrel{:=}}

\newcommand{\cyc}{\circ}
\newcommand{\open}{\dashv}
\newcommand{\spc}[1]{\mathbb{#1}}

\newcommand{\cas}{\mathcal{C}}
\newcommand{\cox}{\mathfrak{c}_2}
\newcommand{\killing}{\kappa}
\newcommand{\norm}{\mathcal{N}}
\newcommand{\tensorcas}{\mathcal{C}_\otimes}

\newcommand{\boo}{\mathcal{B}}
\newcommand{\rmat}{\mathcal{R}}
\newcommand{\lax}{\mathrm{L}}
\newcommand{\rop}{\mathrm{R}}
\newcommand{\sop}{\mathrm{S}}

\newcommand{\spin}{X}
\newcommand{\levz}{\mathrm{J}}
\newcommand{\levo}{\mathrm{\widehat J}}
\newcommand{\cev}[1]{\reflectbox{\ensuremath{\vec{\reflectbox{\ensuremath{#1}}}}}}
\newcommand{\boost}[1]{\mathcal{B}[#1]}
\newcommand{\op}[1]{\mathrm{#1}}
\newcommand{\ham}{\mathrm{H}}

\newcommand{\gen}[1]{\mathrm{#1}}

\newcommand{\alg}[1]{\mathfrak{#1}}
\newcommand{\grp}[1]{\mathrm{#1}}
\newcommand{\superN}{\mathcal{N}}
\newcommand{\sym}{$\superN=4$ SYM theory}

\newcommand{\Tr}{\mathop{\mathrm{Tr}}}
\newcommand{\tr}{\mathop{\mathrm{Tr}}}
\newcommand{\order}[1]{\mathcal{O}(#1)}
\newcommand{\sgn}{\mathop{\mathrm{sign}}}

\newcommand{\dd}{\mathrm{d}}

\newcommand{\eps}{\varepsilon}

\newcommand{\tx}{t,x}
\newcommand{\ty}{t,y}

\newcommand{\hel}{\gen{B}}
\newcommand{\superfield}{\Phi}
\newcommand{\eq}{\earel{=}}
\newcommand{\earel}[1]{\mathrel{}&\hspace{-2\arraycolsep}#1\hspace{-2\arraycolsep}&\mathrel{}}
\newcommand{\deltad}[1]{\delta^{#1}}

\newcommand{\tprods}[2]{\langle#1#2\rangle}

\newcommand{\gf}[1]{\mathcal{#1}}
\newcommand{\gi}[1]{\mathfrak{#1}}
\newcommand{\struc}{f}

\newcommand{\field}{X}
\newcommand{\Da}{{\delta\gen{D}}}

\newcommand{\scalar}{\Phi}
\newcommand{\cket}[1]{|#1\rangle}

\ifx\genfrac\sdflkaj\else\fi

\newcommand{\supup}[1]{^{\mathrm{#1}}}

\newcommand{\brk}[1]{(#1)}

\newcommand{\bigbrk}[1]{\bigl(#1\bigr)}

\newcommand{\comm}[2]{[#1,#2]}

\newcommand{\bigcomm}[2]{\bigl[#1,#2\bigr]}

\newcommand{\acomm}[2]{\{#1,#2\}}

\newcommand{\scomm}[2]{[#1,#2\}}

\newcommand{\abs}[1]{|#1|}

\newcommand{\spaa}[1]{{\langle#1\rangle}}

\newcommand{\spbb}[1]{{[#1]}}

\newcommand{\quoting}[2]{
\bigskip
\begin{center}
\begin{minipage}{14cm}\small{\it
``#1''}
#2
\end{minipage}
\end{center}
\bigskip
}

\makeatletter
\def\mr@ignsp#1 {\ifx\:#1\@empty\else #1\expandafter\mr@ignsp\fi}%
\newcommand{\multiref}[1]{\begingroup
\xdef\mr@no@sparg{\expandafter\mr@ignsp#1 \: }%
\def\mr@comma{}%
\@for\mr@refs:=\mr@no@sparg\do{\mr@comma\def\mr@comma{,}\ref{\mr@refs}}%
\endgroup}
\makeatother

\newcommand{\hypref}[2]{\ifx\href\asklfhas #2\else\href{#1}{#2}\fi}
\newcommand{\secref}[1]{Section~\multiref{#1}}
\newcommand{\appref}[1]{Appendix~\multiref{#1}}
\newcommand{\tabref}[1]{Table~\multiref{#1}}
\newcommand{\figref}[1]{Figure~\multiref{#1}}
\renewcommand{\eqref}[1]{(\multiref{#1})}

\makeatletter
\newlength{\apb@width}
\newcommand{\autoparbox}[2][c]{\settowidth{\apb@width}{#2}\parbox[#1]{\apb@width}{#2}}
\newcommand{\includegraphicsbox}[2][]{\autoparbox{\includegraphics[#1]{#2}}}
\makeatother

\title{Lectures on Yangian Symmetry}

\author{%
Florian Loebbert}

\begin{document}

\pdfbookmark[1]{Title Page}{title}

\thispagestyle{empty}

\begin{flushright}\footnotesize
\texttt{HU-EP-16/12}%
\end{flushright}
\vspace{1cm}

\vfill

\vspace{-3cm}

\begin{center}

\begingroup\LARGE\bfseries\thetitle\par\endgroup

\bigskip

\begingroup\scshape\large
Florian Loebbert
\endgroup

\bigskip

\begingroup\itshape

Institut f\"ur Physik and IRIS Adlershof, Humboldt-Universit\"at zu Berlin, \\
Zum Gro{\ss}en Windkanal 6, D-12489 Berlin, Germany

\smallskip
\par\endgroup

\smallskip

{\ttfamily
\href{mailto:loebbert@physik.hu-berlin.de}{loebbert@physik.hu-berlin.de}
}

\vfill

\textbf{Abstract}

\vspace{7mm}

\begin{minipage}{13.4cm}

In these introductory lectures we discuss the topic of Yangian symmetry from various perspectives. Forming the classical counterpart of the Yangian and an extension of ordinary Noether symmetries, first the concept of nonlocal charges in classical, two-dimensional field theory is reviewed. We then define the Yangian algebra following Drinfel'd's original motivation to construct solutions to the quantum Yang--Baxter equation. Different realizations of the Yangian and its mathematical role as a Hopf algebra and quantum group are discussed. We demonstrate how the Yangian algebra is implemented in quantum, two-dimensional field theories and how its generators are renormalized.
Implications of Yangian symmetry on the two-dimensional scattering matrix are investigated. We furthermore consider the important case of discrete Yangian symmetry realized on integrable spin chains.
Finally we give a brief introduction to Yangian symmetry in planar, four-dimensional super Yang--Mills theory and indicate its impact on the dilatation operator and tree-level scattering amplitudes. These lectures are illustrated by several examples, in particular  the two-dimensional chiral Gross--Neveu model, the Heisenberg spin chain and $\superN=4$ superconformal Yang--Mills theory in four dimensions. This review arose from lectures given at the Young Researchers Integrability School at Durham University (UK).

\end{minipage}

\end{center}

\vfill

\setcounter{page}{0}

\newpage
\setcounter{tocdepth}{2}
\hrule height 0.75pt
\pdfbookmark[1]{\contentsname}{contents}
\microtypesetup{protrusion=false}
\tableofcontents
\microtypesetup{protrusion=true}
\vspace{0.8cm}
\hrule height 0.75pt
\vspace{1cm}

\newpage
\section{Introduction}

\quoting{I got really fascinated by these (1+1)-dimensional models that are solved by the Bethe ansatz and how mysteriously they jump out at you and work and you don't know why. I am trying to understand all this better.}{R.\ Feynman 1988 \cite{Feynman,batchelor2007bethe}}

\noindent The possibility to grasp physical models, to efficiently compute observables and to explain mysterious simplifications in a given theory is largely owed to the realization of symmetries. In quantum field theories these range from discrete examples like parity, over spacetime Poincar\'e or super-symmetry to global and local internal symmetries. In the most extreme case, a theory has as many independent symmetries as it has degrees of freedom (possibly infinitely many). Roughly speaking, this is the defintion of an \emph{integrable} model. The concept of integrability has many faces and can be realized or formulated in a variety of different and often equivalent ways. As we will see below, this symmetry appears in certain two- and higher-dimensional field theories or in quantum mechanical models like spin chains. While integrability in classical theories is rather well understood, quantum integrability still asks for a universal definition \cite{2004bmfy.book.....S,Caux:2010by}.
 The nature of what we call a (quantum) integrable system can be identified by unveiling typical mathematical structures which have been subject to active research for many decades.
 \medskip

One realization of integrability is the \emph{Yangian symmetry}, representing a generalization of Lie algebra symmetries in physics. This Hopf algebra was introduced by Vladimir Drinfel'd in order to construct solutions to the famous quantum Yang--Baxter equation\cite{Drinfeld:1985rx,Drinfeld:1986two,Drinfeld:1986in,Drinfeld:1987sy}. Moreover, the Yangian algebra forms part of the familiy of quantum groups introduced by Drinfel'd  and Michio Jimbo \cite{Drinfeld:1985rx,Jimbo:1985zk,Faddeev:1987ih}. These provide the mathematical framework underlying the quantum inverse scattering method and the algebraic Bethe ansatz, which were developed by the Leningrad school around Ludwig Faddeev, see e.g.\ \cite{1993qism.book.....K}. Hence, the Yangian represents a central concept within the framework of physical integrable models and their mathematical underpinnings.
 \medskip

The most common occurrence of Yangian symmetry in physics is the case of two-dimensional quantum field theories or discrete spin chain models. Here a global (internal) Lie algebra symmetry $\alg{g}$ is typically enhanced to a Yangian algebra $Y[\alg{g}]$. This Yangian combined with the Poincar\'e symmetry yields constraints on physical observables. These constraints following from the underlying Hopf algebra structure often allow to bootsrap a quantity of interest, first of all the scattering matrix.
One of the most prominent statements about symmetries of the S-marix is the famous four-dimensional Coleman--Mandula theorem \cite{Coleman:1967ad}. It states that the spacetime and internal symmetries of the S-matrix may only be combined via the trivial direct product. Hence it is by no means obvious that an internal and a spacetime symmetry can be combined in a nontrivial way. In certain 1$+$1 dimensional field theories, however, it was shown that the Lorentz boost of the Poincar\'e algebra develops a nontrivial commutator with the internal Yangian generators. Thus, the internal and spacetime symmetry are coupled to each other \cite{Bernard:1990jw,LeClair:1991cf}. This interconnection  implies stronger constraints on observables than a direct product symmetry, since the boost maps different representations of the Yangian to each other. 
That this nontrivial relation of the Yangian and the spacetime symmetry is possible can be attributed to the fact that the Yangian generators do not act on multi-particle states via a trivial tensor product generalization of their action on single particle states; they have a non-trivial coproduct, which violates the assumptions of the Coleman--Mandula theorem. Interestingly, the internal Yangian and the Poincar\'e algebra are linked in such a way that the Lorentz boost realizes Drinfel'd's automorphism of the Yangian algebra, which was originally designed to switch on the spectral parameter dependence of the quantum R-matrix.
 \medskip

The physical implementation of the abstract mathematical Yangian Hopf algebra can in fact be observed in the case of several interesting examples.
A very intriguing physical system and a two-dimensional prime example in these lectures is the so-called chiral Gross--Neveu model \cite{Gross:1974jv}. This theory of interacting Dirac fermions $\psi$ provides a toy model for quantum chromodynamics and features a plethora of realistic properties whose implementation by a simple Lagrangian is remarkable. 
In particular, the model has a conserved current of the form $j^\mu=\bar \psi \gamma^\mu \psi$. The local axial current given by $j_\text{axial}^\mu=\bar \psi \gamma^\mu\gamma_5 \psi$ is not conserved in this model.
Remarkably, however, it is possible to repair this property by adding nonlocal terms to the axial current, resulting in a conserved nonlocal current. Hence, one finds an additional hidden symmetry that is realized in a more subtle way than the naive local Noether current $j^\mu$. Commuting the corresponding nonlocal conserved charges with each other, one finds an expression which is not proportional to either of the two original charges, but rather generates a new symmetry operator. Importantly, this procedure can be iterated, inducing more and more new generators and thereby an infinite symmetry algebra. As we will see, this algebra furnishes a  realization of the Yangian and a way to formulate the integrability of this quantum field theory.
 \medskip

Another prominent occurence of Yangian symmetry is the case of integrable spin chain models. Here the action of the symmetry generators can be understood as a straightforward generalization of the above field theory operators to the case of a discrete underlying Hilbert space. Spin chains are typically defined by a Hamiltonian whose Yangian symmetry may be tested by commutation with the symmetry generators. Notably, the exact Yangian symmetry strongly depends on the particular boundary conditions of the system under consideration. While Yangian symmetry is exact on infinite spin chains (no boundaries), the symmetry is typically broken by periodic, cyclic or open boundary conditions.%
\footnote{The same applies to two-dimensional field theories which, however, are typically defined on the infinite line.}
Though this breaking implies that the spectrum is not organized into Yangian multiplets, the bulk Hamiltonian is still strongly constrained by requiring a vanishing commutator with the generators modulo boundary terms. Notably, the Lorentz boost of two-dimensional field theories can be generalized to the case of spin chain models, where the Poincar\'e algebra extends to the algebra containing all local conserved charges \cite{Tetelman:1981xx,Thacker:1985gz}. These local charges furthermore allow to define generalized boost operators which in turn generate integrable spin chains with long-range interactions \cite{Bargheer:2008jt,Bargheer:2009xy}.
 \medskip

Interestingly, the above long-range spin chains play an important role in an a priory unexpected context, namely for a four-dimensional quantum field theory which represents another toy model for QCD. The planar maximally supersymmetric Yang--Mills theory in four dimensions%
\footnote{This theory, further discussed in the main text, goes under the name planar $\superN=4$ superconformal Yang--Mills theory. Here $\superN=4$ refers to the number of supercharges. The planar limit corresponds to the limit $N\to \infty$ of an infinite number of colors of the $\grp{SU}(N)$ gauge symmetry.}
is a conformal gauge theory that is believed to be integrable.
The Hamiltonian of this theory in form of the (asymptotic) dilatation operator maps to an integrable long-range spin chain Hamiltonian \cite{Minahan:2002ve,Beisert:2003tq,Beisert:2003yb}. In consequence, the spectrum of local operators, i.e.\ the spectrum of this quantum field theory, can be obtained using the powerful toolbox of integrability in two dimensions. In fact, this Hamiltonian of a $\alg{psu}(2,2|4)$ symmetric (the symmetry of the Lagrangian) spin chain features a bulk Yangian symmetry $Y[\alg{psu}(2,2|4)]$ \cite{Dolan:2003uh,Dolan:2004ps}.
 \medskip

Indications for the Yangian symmetry of $\superN=4$ superconformal Yang--Mills theory were found in the form of Ward identities for various `observables'. In fact, also the \emph{four-dimensional} S-matrix of the Yang--Mills theory features a Yangian symmetry. This can most clearly be seen on color-ordered tree-level scattering amplitudes \cite{Drummond:2009fd} and extends to loop-level when including anomalous contributions into the symmetry equation \cite{Bargheer:2009qu,Beisert:2010gn,CaronHuot:2011kk}. Here the color order of scattering amplitudes plays an important role since it implements two-dimensional characteristics within this four-dimensional Yang--Mills theory. In consequence, the representation of the Yangian generators on the S-matrix resembles the representation on spin chains or the 2d S-matrix.
 \medskip

This review is published in a collection of lecture notes on integrability \cite{Alessandro,Florian,Diego,Fedor,Stijn,Stefano} introduced by \cite{Intro}.
The structure of the present lectures is as follows:
In \secref{sec:classint} we investigate how \emph{classical} integrability makes an appearence in two-dimensional field theories, i.e.\ we discuss the classical analogue of Yangian symmetry. Then, in \secref{sec:yangalg}, we consider the Yangian algebra, its relation to the Yang--Baxter equation and its embedding into mathematical terminology. This section is more formal than the rest of the notes; in particular one may skip \secref{sec:hopfandQG} and \secref{sec:secondandthird} without missing prerequisites for the subsequent sections. We continue by studying how Yangian symmetry is realized in two-dimensional \emph{quantum} field theories, and we discuss some of the implications of the Yangian on the 2d scattering matrix. In \secref{sec:spinchains} we consider the case of discrete spin chain models and point out similarities to the field theory case. Finally we introduce how Yangian symmetry plays a role in four-dimensional superconformal Yang--Mills theory. We finish with a summary and a brief outlook.

\section{Classical Integrability and Non-local Charges in 2d Field Theory}
\label{sec:classint}

In this section we briefly review how ordinary symmetries are related to conserved Noether currents in classical field theories. We will see that assuming the associated local current to be flat, we may construct additional nonlocal currents, which are also conserved. We investigate how these nonlocal currents relate to classical integrability and the Lax formalism. Finally, we consider the example of the Gross--Neveu model and comment on the implementation of nonlocal charges as Noether symmetries. The nonlocal charges considered in this section form the classical version of the Yangian \cite{MacKay:1992he,Bernard:1992ya}.

\subsection{Local and Bilocal Symmetries}

\begin{figure}
\begin{center}
$\biloc{\idop}{\ham}=\displaystyle\int\limits_{y<x}$\quad 
\includegraphicsbox[scale=1]{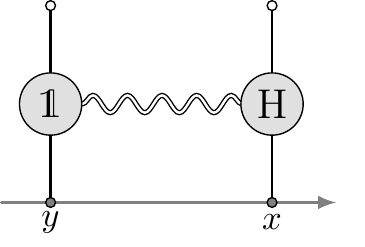}
\qquad
$\biloc{j_a}{ j_b}=\displaystyle\int\limits_{y<x}$\quad
\includegraphicsbox[scale=1]{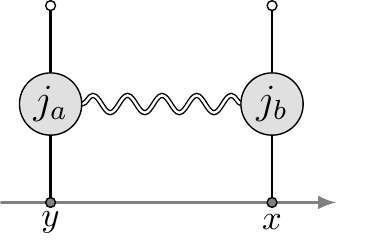}
\end{center}
\caption{Left hand side: Boost generator written as bilocal integral. Right hand side: Bilocal operator composed of two charge densities.}
\label{fig:boostandbi}
\end{figure}

Consider a field theory with a Lagrangian $\mathcal{L}(\phi_A,\partial \phi_A)$. Here $\phi_A$ represents the fields of the theory, which we do not specify for the moment. Suppose the Lagrangian has a continous internal or spacetime symmetry which is infinitesimally realized by a variation $\delta\phi_A$, and for which the Lagrangian changes at most by a total derivative:
\begin{equation}
\delta \mathcal{L}=\partial_\mu f^\mu.
\end{equation}
Via Noether's theorem this symmetry induces a \emph{conserved current} $j_\mu$ which obeys the conservation law
\begin{equation}\label{eq:conslaw}
\partial_\mu j^\mu=0,
\end{equation}
and takes the generic form
\begin{equation}\label{eq:Noethercurrent}
j^\mu=\frac{\partial \mathcal{L}}{\partial (\partial_\mu \phi_A)}\delta\phi_A(x) - f^\mu(\phi_A).
\end{equation}
Depending on the symmetry, it can be convenient to expand the current in terms of the symmetry generators according to $j^\mu=j^\mu_a \,t_a$. Here the symmetry algebra $\alg{g}$ is generated by the operators~$t_a$ which we assume to be anti-hermitian, i.e.\ $t_a=-t_a^\dagger$.%
\footnote{Here we think of an internal symmetry, e.g.\ $\grp{SU}(N)$.}
The generators obey the commutation relations
\begin{equation}
\comm{t_a}{t_b}=f_{abc}\, t_c,
\end{equation}
and for simplicity of the displayed expressions, we refrain from distinguishing upper and lower adjoint indices $a,b,c,\dots$.%
\footnote{In general, these indices are raised and lowered by the Killing form $\kappa_{ab}$, which, in a certain basis, is related to the structure constants and the algebra's dual Coxeter number $\cox$ via%
\begin{align}
\killing_{ad}=f_{abc}f_{bcd}=\cox\,\delta_{ad}.
\end{align}
Alternatively, these algebraic quantities are often expressed in terms of the quadratic Casimir operator $\cas$ in the adjoint representation:
\begin{equation}
-\cas \delta_{ad}=f_{abc}f_{bcd}=(t_b^\text{adj})_{ac}\, (t_b^{\text{adj}})_{cd},
\end{equation}
and we have $\cox=-\cas$.
In these notes we use either the symbol for the quadratic Casimir $\mathcal{C}$ or the dual coxeter number $\cox$ depending on the typical convention in the respective context.}
The above conserved current gives rise to a \emph{conserved charge} defined by the space integral over its time component%
\footnote{Often the (nonlocal) conserved charges are denoted by the letter $\charge$. Since the literature on integrability is full of $\charge$'s anyways, we will use the capital $\levz$ here and save the $\charge$ for later.}
\begin{equation}
\levz(t)=\int \dd^{d-1}x \,j^0(\tx).
\end{equation}
Due to the conservation law \eqref{eq:conslaw} the conserved charge obeys the equation 
\begin{equation}
\frac{\dd \levz(t)}{\dd t}=-\int\limits_V \dd^{d-1}x\, \vec{\nabla} \cdot \vec{j}(\tx)=-\int\limits_Sd\vec{s}\cdot \vec{j}(\tx).
\end{equation}
If we specify the considered situation to $d=2$ spacetime dimensions, we find that the conserved charge obeys
\begin{equation}
\frac{\dd \levz(t)}{\dd t}=j^1(t,S_-)-j^1(t,S_+),
\end{equation}
where $S_\mp$ denotes the boundaries of space.
We can now furthermore assume that the current falls off at the spatial boundaries, i.e.\
\begin{equation}\label{eq:jlocbound}
j^\mu(\tx)\xrightarrow{x\to S_\pm}0,
\end{equation}
and thus the charge $\levz$ is time independent: $\frac{\dd}{\dd t} \levz(t)=0$. In the following the canonical choice will be to consider an infinite volume $V$ with $S_\pm\to\pm \infty$.
\paragraph{Lorentz boost.} 
Consider a Lorentz transformation as an example of a Noether symmetry. Infinitesimally, this transformation can be represented by
\begin{equation}
\Lambda^\mu{}_\nu=\delta^\mu{}_\nu+\lambda^\mu{}_\nu,
\end{equation}
where $\lambda^{\mu\nu}=-\lambda^{\nu\mu}$. For illustration, let us assume that we are dealing with scalar fields $\phi_A$, on which the Lorentz transformation acts as
\begin{equation}
\phi_A(x)\to \phi_A(\Lambda^{-1}x)=\phi_A(x)-\lambda^\mu{}_\nu x^\nu \partial_\mu\phi_A(x).
\end{equation}
Hence we have $\delta\phi_A=-\lambda^\mu{}_\nu x^\nu \partial_\mu \phi_A$. The Lagrangian then transforms according to
\begin{equation}
\delta \mathcal{L}=-\lambda^\mu{}_\nu x^\nu \partial_\mu \mathcal{L}=-\partial_\mu(\lambda^\mu{}_\nu x^\nu \mathcal{L}),
\end{equation}
and the corresponding Noether current takes the form
\begin{equation}
j^\mu =-\lambda^\rho{}_\nu\bigg[\frac{\partial \mathcal{L}}{\partial(\partial_\mu \phi_A)}x^\nu \partial_\rho \phi_A-\delta^\mu{}_\rho x^\nu \mathcal{L}\bigg]=-\lambda^\rho{}_\nu T^\mu{}_\rho x^\nu.
\end{equation}
Here $T^{\mu\nu}$ denotes the energy momentum tensor defined by
\begin{equation}
T^{\mu}{}_\nu=\frac{\partial \mathcal{L}}{\partial(\partial_\mu \phi_A)}\partial_\nu \phi_A-\delta^\mu{}_\nu\mathcal{L}.
\end{equation}
Note that due to the arbitrariness of the infinitesimal transformation $\lambda^\rho{}_\nu$, the above current in $d$ spacetime dimensions in fact contains $d(d-1)/2$ conserved quantities:
\begin{equation}
(j^\mu)^{\rho\sigma}=x^\rho T^{\mu\sigma}-x^\sigma T^{\mu\rho},
\end{equation}
which obey $\partial_\mu(j^\mu)^{\rho\sigma}=0$. For $\rho, \sigma=i,j$ both being spatial indices, the Lorentz transformation corresponds to a rotation, while for $\rho,\sigma=0,i$ being a combination of the time and one spatial component, the transformation represents a Lorentz boost. Since we are particularly interested in two spacetime dimensions, where only one single Lorentz transformation (a boost) exists, we consider the latter case which gives rise to a conserved charge of the form
\begin{equation}
\levz^{0i}=\int \dd^{d-1} x\,(x^0 T^{0i}-x^i T^{00}).
\end{equation}
Note that if the fields have a non-trivial spin as opposed to the considered scalars, i.e.\ the fields transform non-trivially under the Lorentz group, an extra term has to be added to the above boost transformation.
In the case at hand, we may take into account that the Hamiltonian density is defined as the ${00}$-component of the energy-momentum tensor:
\begin{align}
\ham(x)&=\pi^A(x)\dot \phi_A(x)-\mathcal{L}(x),
&
\pi^A(x)&=\frac{\partial \mathcal{L}(x)}{\partial \dot \phi_A}.
\end{align}
Moreover, since the above charge $\levz^{0i}$ is conserved, its value is time-independent and we may simply choose $t=x^0=0$. Then, in $d=1+1$ dimensions,%
\footnote{We use the conventions $(\eta_{\mu\nu})=\text{diag}(1,-1)$ and $\epsilon_{01}=1$.}
we can rewrite the above boost charge as the first moment of the Hamiltonian
\begin{equation}\label{eq:boostmoment}
\boo\equiv \levz_{01}=\int \dd x\,x\, \ham(x).
\end{equation}
Suppose the above integral runs from $S_-$ to $S_+$, such that we can formally write the conserved boost charge in the form of a bilocal integral given by%
\footnote{For brevity we introduce the ordered product $\biloc{A}{B}=\int\limits_{-\infty}^{+\infty} \dd x\, \int\limits_{-\infty}^x\dd y\, A(y)B(x)$.}
\begin{equation}\label{eq:boostbi}
\boo\simeq\int\limits_{S_-}^{S_+} \dd x\int\limits_{S_-}^x \dd y\, 1\cdot \ham(x)
\equiv\biloc{\idop}{\ham},
\end{equation}
modulo a term $S_-\int_{S_-}^{S_+}\dd x\, \ham(x)$ which is proportional to the conserved energy and does hence not modify the property of the boost to be a conserved charge. Here $\idop\equiv 1$ denotes the identity, cf.\ \figref{fig:boostandbi}.%
\footnote{Note that the discarded term $S_-\int_{S_-}^{S_+}\dd x\, \ham(x)$ diverges in the limit $S_{\pm}\to\pm \infty$. For better readability we refrain here from antisymmetrizing the bilocal integral in order to regularize the expression. In \secref{sec:moreboosts} we will see that this formal bilocal expression $\biloc{\idop}{\ham}$ composed of two local densities $\idop$ and $\ham$ takes a natural place in the class of bilocal charges with nontrivial densities on both of the bilocal legs.}

Note that the above example for a Noether charge deals with a spacetime symmetry. Below we will also encounter examples of internal symmetries and associated charges which may be extended to bilocal symmetries. The motivation for recalling the properties of the Lorentz boost here will become clear when we discuss the Yangian.
\paragraph{Bilocal Symmetry.}

After having refreshed our memory about local symmetries, let us continue the survey on conserved currents and charges in 1$+$1 dimensions. Suppose the local current $j^\mu$ is not only conserved but also \emph{flat}. Here flatness means that the current obeys the equation
\begin{equation}
\comm{\partial_\mu+j_\mu}{\partial_\nu+j_\nu}=0,
\end{equation}
i.e.\ it defines a flat connection.
More explicitly, this can be written as
\begin{equation}
\partial_0 j_1-\partial_1 j_0 +\comm{j_0}{j_1}=0,
\end{equation}
which for $j_\mu=j_{\mu a} \, t_a$ and $\comm{t_a}{t_b}=f_{abc}\, t_c$ reads in components
\begin{equation}
\partial_0 j_{1a}-\partial_1 j_{0a} +f_{abc}\,j_{0b} j_{1c}=0.
\end{equation}
Under the above flatness or zero-curvature condition, we may define an additional \emph{bilocal} conserved current of the form
\begin{equation}\label{eq:bilocj}
\widehat j_{a}^\mu(\tx)=\epsilon^{\mu\nu}j_{\nu a}(\tx)-\half f_{abc}\, j_{b}^\mu(\tx)\int_{-\infty}^x \dd y\,j_{c}^0(\ty) ,
\end{equation}
which can be seen to be conserved modulo the conservation of the local current $j^\mu$ and the flatness condition:
\begin{align}
\partial_\mu \widehat j_{a}^\mu(\tx)&=\partial_\mu \epsilon^{\mu\nu}j_{a\nu}(\tx)-\half f_{abc}(\partial_\mu j_b^\mu(\tx))\int_{-\infty}^x \dd y\, j_c^0(t,y)+\half f_{abc}\,\epsilon_{\mu\nu}\,j_b^\mu(\tx)j_c^\nu(\tx)\nonumber\\
&=-\partial_0 j_{1 a}(\tx)+\partial_1 j_{0 a}(\tx)-\comm{j_0(\tx)}{j_1(\tx)}_a=0.
\end{align}
We will refer to $j^\mu$ as the \emph{level-zero} current and to $\widehat j^\mu$ as the \emph{level-one} current.
As for the local level-zero current, we can define a corresponding \emph{level-one charge} by integration over the time component of the current:
\begin{equation}\label{eq:biloccharge}
\levo_a(t) =\int_{-\infty}^\infty \dd x\,\widehat j_a^0(\tx)=\int_{-\infty}^\infty \dd x \, j^{1}_a(\tx) 
- \half f_{abc}\int_{-\infty}^\infty \int_{-\infty}^x \dd x\,\dd y\, j^0_b(\tx) j^0_c(\ty). 
\end{equation}
The ordered one-dimensional integral has a similar form as \eqref{eq:boostbi}, just that here both legs of the bilocal operator are nontrivial. Again we may write the charge in the compact form (cf.\ \figref{fig:boostandbi})
\begin{equation}
\levo_a(t) = \int_{-\infty}^\infty dx\, j_a^1(\tx)-\half f_{abc}\,\biloc{j_b^0(t)}{ j_c^0(t)}. 
\end{equation}
Let us check explicitly under which conditions this charge is time independent. We find
\begin{align}
\frac{\dd}{\dd t} \levo_a(t)
=&-\int_{-\infty}^\infty \dd x\, \partial_1 j_a^0(\tx)-f_{abc}\int_{-\infty}^\infty \dd x\, j_b^0(\tx)j^1_c(\tx)
\nonumber\\
&\qquad\quad-\half f_{abc}\int_{-\infty}^\infty\int_{-\infty}^x \dd x\,\dd y\, \Big[\big(\partial_1 j_{1b}(\tx)\big)j_c^0(\ty)+j_b^0(\tx)\big(\partial_1j_{1c}(\ty)\big)\Big],
\end{align}
where we have used the flatness and conservation of the current. We can partially integrate to obtain
\begin{align}
\frac{\dd}{\dd t} \levo_a(t)
=&j_a^0(t,-\infty)-j_a^0(t,\infty)-\half f_{abc} \big[j_b^0(t,\infty)\levz_c-\levz_b\, j_c^0(t,-\infty)\big].
\end{align}
Hence, as above in the discussion of the local charge conservation, we assume that \eqref{eq:jlocbound}
\begin{equation}
j_a^0(t,x)\xrightarrow{x\to \pm \infty}0,
\end{equation}
such that indeed
\begin{equation}
\frac{\dd}{\dd t} \levo_a(t)=0.
\end{equation}
Since the charges are time independent, we will no longer display their $t$-dependence in what follows. For the sake of compactness, we may also sometimes drop the explicit time dependence in the argument of the currents.

Notably, the above definition of the bilocal current distinguishes two points $S_\pm=\pm\infty$ in the one-dimensional space and thus allows for an order of the integration variables $x$ and $y$. That this is an important input for the definition of the nonlocal charges can be realized by thinking about a possible generalization to the case of a compact periodic space which has no notion of order. It is also not obvious how to generalize the above definition of the nonlocal current to more than one space dimension.

Finally we note that the bilocal charge \eqref{eq:biloccharge} is often written in the alternative and more symmetric forms
\begin{align}
\levo_a&=\int_{-\infty}^\infty \dd x \, j^{1}_a(\tx) 
- \half\int_{-\infty}^\infty \int_{-\infty}^\infty \dd x\,\dd y\, \theta(x-y) \comm{j^0(\tx)}{j^0(\ty)}_a, \\
&=\int_{-\infty}^\infty dx \, j^{1}_a(\tx) 
- \sfrac{1}{4}\int_{-\infty}^\infty \int_{-\infty}^\infty \dd x\,\dd y\, \epsilon(x-y) \comm{j^0(\tx)}{j^0(\ty)}_a, 
\end{align}
where $\theta$ denotes the step function and $\epsilon$ represents the sign function.


\subsection{Nonlocal Charges and Lax Formulation}

In the above section we have seen that two properties of the local current $j^\mu$, namely to be conserved and flat, lead to a conserved bilocal current and an associated charge. Is this the only nonlocal charge we can construct from the above conditions? Let us understand things in a more systematical fashion along the lines of \cite{Brezin:1979am}.%
\footnote{Cf.\ also \cite{Luscher:1977rq,Luscher:1977uq}.}

Given a flat and conserved current $j_\mu$, we can define a covariant derivative $D_\mu=\partial_\mu+j_\mu$. Conservation and flatness become the statements
\begin{align}
\comm{\partial_\mu}{D^\mu}&=0,
&
\comm{D_\mu}{D_\nu}&=0.
\label{eq:consflat}
\end{align}
Now one may try an inductive approach. Suppose we have constructed a conserved current $j_\mu^{(n)}(x)$ of level $n$. The conservation implies that a function (the associated potential) $\chi^{(n)}(x)$ exists, for which
\begin{equation}\label{eq:jpot}
j_\mu^{(n)}=\epsilon_{\mu\nu}\partial^\nu \chi^{(n)},
\qquad
n\geq 0.
\end{equation}
In consequence, an additional current can be defined by
\begin{equation}\label{eq:jD}
j_\mu^{(n+1)}=D_\mu \chi^{(n)},
\qquad
n\geq-1,
\end{equation}
where we set $\chi^{(-1)}=1$.
This current is conserved since we may use \eqref{eq:consflat} to find
\begin{equation}
\partial^\mu j_\mu^{(n+1)}=\partial^\mu D_\mu \chi^{(n)}= D_\mu \partial^\mu\chi^{(n)}=\epsilon^{\mu\nu}D_\mu D_\nu \chi^{(n-1)}=0,
\qquad
n\geq 0.
\end{equation}
Here we have also used that \eqref{eq:jpot} and \eqref{eq:jD} imply $\partial^\mu \chi^{(n)}=\epsilon^{\mu\nu} j_\nu^{(n)}=\epsilon^{\mu\nu}D_\nu \chi^{(n-1)}$ and that $\epsilon^{\mu\nu}\comm{D_\mu}{D_\nu}=-2\comm{D_0}{D_1}=0$.

The start of the induction is $\chi^{(-1)}=1$ with $j_\mu^{(-1)}=0$ and such that $j_\mu^{(0)}=j_\mu$, which is indeed conserved by assumption. Then we can write
\begin{align}\label{eq:potential0}
j_\mu^{(0)}&=\epsilon_{\mu\nu}\partial^\nu \chi^{(0)},
&
\chi^{(0)}=-\int_{-\infty}^x \dd y\, j^0(y).
\end{align}
and thus%
\footnote{%
Note that the conserved charge corresponding to this current equals the previous version up to level-zero charges since we have
\begin{equation}\label{eq:levzsquare}
\int_{-\infty}^\infty \dd x \,j^0(x)\int_{-\infty}^x \dd y\, j^0(y)=\half \int_{-\infty}^\infty\int_{-\infty}^x \dd x \,\dd y\,\comm{j^0(x)}{j^0(y)}+\quarter \acomm{\levz}{\levz}.
\end{equation}
}
\begin{equation} 
\widehat j_{\mu}\equiv j^{(1)}_{\mu}=D_{ \mu} \chi^{(0)}=\epsilon_{\mu\nu}j^\nu(x)-\, j_{\mu}(x) \int_{-\infty}^x \dd y\, j^0(y).
\end{equation}
Hence, having shown the existence of a conserved current $j^{(0)}_\mu=j_\mu$ that obeys \eqref{eq:consflat}, one can construct $j^{(1)}_\mu=\widehat j_\mu$ and an infinite number of conserved nonlocal currents and consequently an infinite number of conserved nonlocal charges
\begin{equation}
\levz^{(n)}=\int_{{-\infty}}^{+\infty} dx\, j_0^{(n)}(x).
\end{equation}

\paragraph{The spectral parameter.}
Now we have obtained a set of conserved charges. Obviously, any linear combination of these charges will also furnish a conserved charge. We might thus wonder whether one can construct a conserved generating function $ \mon(u)$ whose expansion in $u$ yields the conserved charges constructed above:%
\footnote{Here we follow the usual convention and consider the expansion in $1/u$ instead of $u$ and we set $\levz^{(-1)}=1$.}
\begin{equation}
\label{eq:genfunc}
\mon(u)\simeq\sum_{k=-1}^\infty u^{-k-1} \levz^{(k)}.
\end{equation}
For the below discussion it may be useful to be familiar with some of the standard notions of classical integrability. These are for instance introduced in the review \cite{Alessandro} or in the textbook \cite{2003icis.book.....B}.
Let us try to stay within the geometric picture that is suggested by the appearence of the covariant derivative.
In fact we may define a new covariant derivative $\mathcal{D}_\mu(u)=\partial_\mu-\lax_\mu(u)$, where%
\footnote{Note that in the language of differential forms, this is a linear combination of $j$ and $\star j$, where $\star$ denotes the Hodge star. In this language the form of the Lax connection $\lax$ might appear more natural.}
\begin{equation}\label{eq:Lax}
\lax_\mu(\tx,u)=\frac{1}{u^2-1}\big[j_\mu(\tx)+u\, \epsilon_{\mu\nu}j^\nu(\tx)\big],
\end{equation}
defines the \emph{Lax connection} depending on the \emph{spectral parameter} $u$.
We may then collect both conditions in \eqref{eq:consflat} by requiring that the following equation holds for all $u$:
\begin{equation}\label{eq:flatD}
\comm{\mathcal{D}_\mu(u)}{\mathcal{D}_\nu(u)}=0.
\end{equation}
This furnishes a very compact way of writing the conservation and flatness conditions for the current $j_\mu$.
Note that we can understand the components of $\lax_\mu(u)$ as a one-parameter family of Lax pairs, cf.\ \cite{Alessandro}.

The above equation \eqref{eq:flatD} can be understood as a compatibility condition for the following so-called \emph{auxiliary linear problem} 
\begin{equation}\label{eq:auxlin}
\mathcal{D}_\mu(u)\, \Phi(t,x)=0,
\end{equation}
which represents a system of two differential equations for the function $\Phi(t,x)$.
In fact, applying another covariant derivative $\mathcal{D}_\nu(u)$ to this equation shows that the solution $\Phi$ is only well-defined, if \eqref{eq:flatD} holds. Equation \eqref{eq:auxlin} relates an infinitesimal translation generated by $\partial_\mu$ to the flat connection $\lax_\mu(u)$.

Next we determine the transport matrix $\mon(t,x_0,x;u)$, which transports the solution $\Phi(t,x_0,u)$ along the interval $[x_0,x]$:
\begin{equation}
\Phi(t,x,u)=\mon(t,x_0,x;u)\Phi(t,x_0,u).
\end{equation}
Note that this transport matrix may be defined by the equations (cf.\ e.g.\ \cite{Luscher:1977rq,deVega:1979zy,Eichenherr:1979ci,Zakharov:1978wc,deVega:1983gn}):
\begin{align}\label{eq:deftrans}
\mathcal{D}_1(u)\, \mon(t,x_0,x;u)&=0,
&
\mon(t,x_0,x_0;u)&=1.
\end{align}
We may integrate \eqref{eq:deftrans} along the $x$-coordinate and obtain the explicit path-ordered solution:
\begin{equation}\label{eq:defmon}
\mon(t,x_0,x;u)=\mathcal{P}\, \exp\Bigg[\,\int_{x_0}^{x}\dd x'\, \lax_1(t,x',u) \Bigg].
\end{equation}
Here $\mathcal{P}$ denotes path-ordering with greater $x$ to the left.
Based on this expression, we define the \emph{monodromy matrix}%
\footnote{In ancient greek we have: $\mu \'o \nu o \varsigma $  [\emph{``monos''}]: single and $\delta \rho \'o \mu o \varsigma$ [\emph{``dromos''}]: course, path, racetrack.}
$\mon(t;u)$ as the transport matrix along the whole $x$-axis:
\begin{equation}
\mon(t;u)\equiv\mon(t,-\infty,\infty;u).
\end{equation}
In order to evaluate the expansion of $\mon(t;u)$ in powers of $1/u$, we note that for $v=1/u$ we have
\begin{align}
\lax_\mu(\tx,v)\big|_{v=0}&=0,
&
\sfrac{\dd }{\dd v} \lax_\mu(\tx,v)\big|_{v=0}&=\epsilon_{\mu\nu}j^\nu(\tx),
&
\sfrac{\dd^2}{\dd v^2} \lax_\mu(\tx,v)\big|_{v=0}&=2j_\mu(\tx),
\end{align}
and thus we obtain
\begin{align}\label{eq:monexpcont}
\mon(t;u)&=1-\frac{1}{u}\int_{-\infty}^\infty \dd x \,j_0(t,x) +\frac{1}{u^2}\Bigg[\int_{-\infty}^\infty \dd x\, j_1(t,x)+\int_{-\infty}^\infty \dd x\int_{-\infty}^x \dd y \,j_0(t,x) j_0(t,y) \Bigg]+\mathcal{O}\big(\sfrac{1}{u^3}\big).
\end{align}
Hence, we find indeed the level-zero and level-one charges as the first coeffcients of the expansion \eqref{eq:genfunc}, cf.\ also \eqref{eq:levzsquare}.
Assuming that $j_\mu(x)\xrightarrow{x\to \pm \infty}0$, one can also show that in general
\begin{equation}
\frac{\dd }{\dd t}\mon(t;u)=\lax_0(t,+\infty,u) \mon(t;u)-\mon(t;u)\lax_0(t,-\infty,u)\to0.
\end{equation}
That is the monodromy $\mon(u)\equiv \mon(t;u)$ really furnishes a conserved generating function for infinitely many conserved charges $\levz^{(n)}$. See \cite{Alessandro} for more details on the Lax formalism and the classical monodromy.

For certain models, the above nonlocal charges can be understood as the classical analogues of the Yangian algebra introduced below \cite{MacKay:1992he,Bernard:1992ya}. Whether the charges really form a classical Yangian or another algebra depends on the Poisson algebra of the currents which in turn depends on the model. A classical Yangian can for instance be found in the chiral Gross--Neveu model or the principal chiral model, cf.\ \cite{MacKay:1992he}. In these models it was also shown that the above boost charge \eqref{eq:boostmoment} Poisson-commutes with the charges $\levz_a$ and $\levo_a$:
\begin{align}\label{eq:classicalboost}
\{\boo,\levz_a\}&=0,
&
\{\boo,\levo_a\}&=0.
\end{align}
In \secref{sec:boostauto} we will see that these commutation relations become nontrivial in the quantum theory.

\subsection{Chiral Gross--Neveu Model}

Let us consider some of the above concepts for the case of the 1$+$1 dimensional chiral Gross--Neveu model. This theory introduced in 1974 by Gross and Neveu \cite{Gross:1974jv} represents the two-dimensional version of the four-dimensional Nambu--Jona--Lasinio model \cite{Nambu:1961tp,Nambu:1961fr}. It furnishes a toy model for QCD with a surprisingly rich catalog of features. While conformal at the classical level, masses are generated by quantum corrections. Furthermore the theory is asymptotically free and can be solved in the large-$N$ limit, where $N$ is the parameter of the global symmetry $\alg{u}(N)$.
Remarkably, the theory is also integrable which can be seen as follows.

\paragraph{Local and nonlocal currents.}

We consider the Lagrangian of the $\alg{u}(N)$ symmetric \emph{chiral} Gross--Neveu model%
\footnote{Note that there is also the $\alg{o}(2N)$ symmetric \emph{Gross--Neveu model} (without \emph{chiral}) on the market, whose Lagrangian is given by dropping the $\gamma_5$-term.}
\begin{equation}\label{eq:cGNLagrangian}
\mathcal{L}
=\sum_{\alpha=1}^{N} \bar \psi^\alpha(i \slashed{\partial})\psi_\alpha+ \frac{g^2}{2} \bigg[\Big(\sum_{\alpha=1}^{N} \bar \psi^\alpha \psi_\alpha\Big)^2-\Big(\sum_{\alpha=1}^{N} \bar \psi^\alpha\gamma_5 \psi_\alpha\Big)^2\bigg],
\end{equation}
with $\slashed{\partial}=\gamma^\mu\partial_\mu$. The Dirac fermions are denoted by $\psi_{\alpha j}$ and $\bar\psi^{\alpha}_j=\psi^{\dagger\alpha}_{ i}(\gamma^0)_{ij}$  with $i,j=1,2$ and with fundamental or anti-fundamental $\alg{u}(N)$ indices $\alpha$, respectively. The two-dimensional gamma matrices in the Weyl representation take the form
\begin{align}
\gamma_0=&\sigma_1
=
\begin{pmatrix}
0&1\\
1&0
\end{pmatrix},
&
\gamma_1
&=i\sigma_2
=
\begin{pmatrix}
0&1\\
-1&0
\end{pmatrix},
&
\gamma_5
&=\gamma^0\gamma^1
=
\begin{pmatrix}
-1&0\\
0&1
\end{pmatrix},
\end{align} 
and obey the Clifford algebra $\acomm{\gamma_\mu}{\gamma_\nu}=2\eta_{\mu\nu}$.
The Lagrangian also has a chiral $\alg{u}(1)$ symmetry 
\begin{align}
\psi_\alpha\to e^{i \theta \gamma_5}\psi_\alpha,
\end{align}
which is not  broken at the quantum level since the massive particles generated by spontaneous symmetry breaking are not charged under this symmetry, and the particles carrying a chiral charge decouple.%
\footnote{Therefore this mass generation mechanism is not in contradiction with Coleman's theorem forbidding Goldstone bosons in two dimensions \cite{Coleman:1973ci}.}

Alternatively, the above Lagrangian can be written in the form
\begin{equation}\label{eq:GNLagrangian}
\mathcal{L}
= \bar \psi(i \slashed{\partial})\psi+g^2\Big[\big( \bar \psi\gamma_\mu t_a \psi\big)\big( \bar \psi \gamma^\mu t_a \psi\big)\Big],
\end{equation}
where we do not display the sum over double indices $a,b,\dots=1,\dots,N^2$ and $\alpha,\beta,\dots=1,\dots,N$ from now on. Here $t_a=-t_a^\dagger$ represent the $N^2$ generators of $\alg{u}(N)$. In the following we will refer to \eqref{eq:GNLagrangian} as the chiral Gross--Neveu Lagrangian. For practical reasons one sometimes considers the case of generators $t_a$ of $\alg{su}(N)$  instead of $\alg{u}(N)$.

The equivalence of the above Lagrangians can be shown by using the Fierz identity 
\begin{equation}
(\gamma_\mu)_{ij}(\gamma^\mu)_{kl}=\delta_{il}\delta_{kj}-(\gamma_5)_{il}(\gamma_5)_{kj},
\end{equation}
as well as the following identity for the $\alg{u}(N)$ generators:%
\footnote{For $\alg{su}(N)$ symmetry the Lagrangian \eqref{eq:cGNLagrangian} gets an extra $1/N$ term coming from the $\alg{su}(N)$ identity $(t_a)_\alpha{}^\beta(t_a)_\gamma{}^\delta=-\half\delta_\gamma^\beta\delta_\alpha^\delta+\sfrac{1}{2N}\delta_\alpha^\beta\delta_\delta^\gamma$. For a more transparent illustration of the equivalence of the two Lagrangians we have considered the $\alg{u}(N)$ symmetric Lagrangian here.}
\begin{equation}\label{eq:unident}
(t_a)_\alpha{}^\beta(t_a)_\gamma{}^\delta=-\half\delta_\gamma^\beta\delta_\alpha^\delta.
\end{equation}
The (Euler--Lagrange) equations of motion read
\begin{align}
0&=i\partial_\mu \bar \psi^\alpha \gamma^\mu-2g^2 (\bar \psi \gamma^\mu t_a \psi) (\bar \psi \gamma_\mu t_a)^\alpha ,
&
0&=i\gamma^\mu \partial_\mu \psi_\alpha+2g^2(\gamma_\mu t_a \psi)_\alpha(\bar \psi \gamma^\mu t_a \psi).
\end{align}
Now we multiply these equations by $\psi$ and $\bar\psi$, respectively, and use again the identity \eqref{eq:unident}. Combining the two equations of motion then yields
\begin{equation}
i(\partial_\mu \bar \psi^\alpha)\gamma^\mu \psi_\beta+i\bar \psi^\alpha\gamma^\mu \partial_\mu \psi_\beta=0,
\end{equation}
which directly implies that the following current is conserved \cite{deVega:1984wk}:
\begin{equation}\label{eq:currGN}
j_{a}^\mu=-2g^2i(\bar \psi^\alpha \gamma^\mu (t_a)_{\alpha}{}^\beta \psi_\beta).
\end{equation}
Here the normalization is chosen for later convenience.
In order to see the \emph{flatness} of this current, we note that the equations of motion imply
\begin{equation}
 \epsilon^{\mu\nu} i\partial_\mu(\bar \psi^\alpha\gamma_\nu\psi_\beta)=2g^2 \epsilon^{\mu\nu}(\bar \psi^\alpha \gamma_\mu \psi_\gamma)(\bar \psi^\gamma \gamma_\nu \psi_\beta),
\end{equation}
where we used that $\acomm{\gamma_5}{\gamma_\mu}=0$ and $\gamma_\mu\gamma_5=-\epsilon_{\mu\nu}\gamma^\nu$ as well as the identity \eqref{eq:unident}.
In terms of the current and contracting with a generator $t_a$, this takes the form 
\begin{equation}
\epsilon^{\mu\nu}\partial_\mu(j_\nu)^{\alpha}{}_\beta \,(t_a)_{\alpha}{}^\beta=\epsilon^{\mu\nu}(j_\mu)^{\alpha}{}_\gamma (j_\nu)^{\gamma}{}_{\beta}\,(t_a)_{\alpha}{}^\beta,
\end{equation}
and thus yields the flatness condition
\begin{equation}\label{eq:flatcGN}
\partial_0j_{1a}-\partial_1j_{0a}+\comm{j_0}{j_1}_a=0.
\end{equation}
In consequence, we can construct a bilocal current $\widehat j$ according to the procedure described above.

\paragraph{Axial current.}
Note that as a starting point to obtain a bilocal current we might also have considered the axial current
\begin{equation}
(j_\text{axial})^\mu_{a}=-2g^2i\bar\psi\gamma^5\gamma^\mu t_a\psi=\epsilon^{\mu\nu}j_{\nu a}
\end{equation}
which is familiar from our quantum field theory course, but which is not conserved in this model since (cf.\ \eqref{eq:flatcGN})
\begin{equation}
\partial_\mu (j_\text{axial})^\mu_{a}=\partial_\mu \epsilon^{\mu\nu}j_{\nu,a}=-\partial_0 j_{1a}+\partial_1 j_{0a} \neq 0.
\end{equation}
However, the bilocal current constructed from the conserved current $j^\mu_{a}$ can be understood as a nonlocal completion of this axial current which is then conserved as seen above, cf.\ \cite{Curtright:1981ww}:
\begin{equation} 
\widehat j_{\mu a}(x)=(j_\text{axial})_{\mu  a}-\half \int_{-\infty}^x \dd y\, \comm{ j_{\mu}(x)}{j_0(y)}_{a}.
\end{equation}

\paragraph{Poisson algebra and Lax formalism.}
In order to study the symmetry algebra that is generated by the above currents, we have to define
a Poisson bracket for the Dirac fermions \cite{deVega:1983gn}:
\begin{equation}
\{F,G\}=i\int \dd x \mathop{\sum_{\alpha=1}}_{j=1,2}^{N}F
\bigg(
 \frac{\cev\delta }{\delta \psi^{\dagger\alpha}_{j}(x)}\frac{\vec\delta }{\delta \psi_{\alpha,j}(x)}
 +
 \frac{\cev\delta }{\delta \psi_{\alpha,j}(x)}\frac{\vec\delta }{\delta \psi^{\dagger\alpha}_j(x)}
 \bigg)G.
\end{equation}
Here the arrows are introduced to take care of the Gra{\ss}mann statistics of the fields and they indicate whether the variation acts on the function $F$ or $G$.
Using this definition of the Poisson bracket one can show that the current \eqref{eq:currGN} obeys the algebra relations
\begin{equation}
\acomm{j^\mu_{a}(x)}{j^\nu_{b}(y)}=2g^2\delta(x-y)f_{abc}\,j_{c}^{\abs{\mu-\nu}},
\end{equation}
with the $\alg{su}(N)$ structure constants $f_{abc}$.
The Lax connection and monodromy matrix can be defined as in \eqref{eq:Lax} and \eqref{eq:defmon}, respectively. Their commutators with the classical R-matrix of the chiral Gross--Neveu model (see e.g.\ \cite{Alessandro,2003icis.book.....B} for these notions of classical integrability)
\begin{equation}
r(u,v)=\frac{\tensorcas}{u-v}\,,
\end{equation}
may then be considered as the fundamental integrability equations of this physical system, cf.\ \cite{deVega:1983gn}. For $\alg{g}=\alg{u}(N)$ and generators $t_a$ in the fundamental representation, the tensor Casimir is given by $\tensorcas=\permop$, with $\permop$ representing the permutation operator that acts on a state $a\otimes b$ according to
\begin{equation}
\permop a\otimes b=b\otimes a,
\end{equation}
and on an operator $A\otimes B$ by conjugation:
\begin{equation}
\permop A\otimes B\permop=B\otimes A.
\end{equation}
We will encounter the permutation operator in its role as the tensor Casimir several times in this review.

\subsection{Nonlocal Symmetries as Noether Charges}
\label{sec:nonlocandnoether}

A very valid question is whether also nonlocal symmetries can be understood as Noether symmetries. At least for particular cases this question has been answered with a `yes', cf.\ e.g.\ \cite{Dolan:1980kz,Hou:1981hn}. 
For illustration let us briefly review some results of \cite{Dolan:1980kz} and consider the so-called \emph{principal chiral model} in two dimensions with Lagrangian
\begin{equation}
\mathcal{L}=\frac{1}{16}\tr \partial_\mu g(x) \partial^\mu g^{-1}(x).
\end{equation}
Here the field $g(x)$ is group-valued, i.e.\ an element of a group $\grp{G}$.
The equations of motion take the form of a conservation equation
\begin{equation}
\partial_\mu j^\mu=0,
\end{equation}
for the current 
\begin{equation}
j_\mu\equiv g^{-1}\partial_\mu g=-(\partial_\mu g^{-1})g.
\end{equation}
This current is also flat. 
As discussed in \cite{Dolan:1980kz}, one may define the following nonlocal field variation
\begin{align}
\delta ^{(1)}_\rho g&=-4g\comm{\chi^{(0)}}{\rho},
&
\chi^{(0)}(x)&=\half\int_{-\infty}^\infty \dd y \,\epsilon(x-y)j_0(y),
\end{align}
where $\rho=t^a\rho^a$, with $t^a$ denoting the generators of the group $\grp{G}$ and $\rho^a$ being some constants. Here $\chi^{(0)}$ represents again the potential associated to the level-zero current of \eqref{eq:potential0}. The Lagrangian is invariant under this transformation up to a total derivative:
\begin{equation}
\delta^{(1)}_\rho\mathcal{L}=\sfrac{1}{2}\tr j_\mu \partial^\mu \comm{\chi^{(0)}}{\rho}=\partial^\mu \half \tr\big[\big(\half \epsilon_{\mu\nu}\comm
{\partial^\nu \chi^{(0)}}{\chi^{(0)}}+\epsilon_{\mu\nu} j^\nu \big)\, \rho\big].
\end{equation}
Importantly, the equations of motion have not been used to arrive at this form.
This level-one symmetry (cf.\ \eqref{eq:Noethercurrent}) yields the conserved level-one Noether current
\begin{equation}\label{eq:lev1PCM}
j_\mu^{(1)}=-\epsilon_{\mu\nu}j^\nu+\comm{j_\mu}{\chi^{(0)}}-\half \epsilon_{\mu\nu}\comm{\partial^\nu \chi^{(0)}}{\chi^{(0)}}.
\end{equation}
The conservation of this level-one current implies the flatness of the level-zero current, which is very much in agreement with our intuition gained in the previous subsections:
\begin{equation}
\partial^\mu j_\mu^{(1)}\simeq -\partial_0j_1+\partial_1j_0-\comm{j_0}{j_1}.
\end{equation}
Interestingly, the current \eqref{eq:lev1PCM} does not have the standard form of \eqref{eq:bilocj}. 
In fact, the current is conserved without making use of the equations of motion. It is thus conserved on the set of all fields, i.e.\ off-shell. Using the equations of motion such that
\begin{equation}
\partial_\mu\chi^{(1)}=-\epsilon_{\mu\nu}j^\nu,
\end{equation}
\eqref{eq:lev1PCM} reduces to the standard form \eqref{eq:bilocj} of the level-one current. Note that one might also have started with an ansatz of the form \eqref{eq:lev1PCM} in order to determine $\chi^{(0)}$ such that $j^{(1)}$ is conserved,
cf.\ \cite{Polyakov:1980ca}.
Notably, the above symmetries may be extended to a one-parameter family of nonlocal Noether symmetries  \cite{Hou:1981hn}. As the monodromy considered above, this family furnishes a generating function for the parameter independent symmetries.
Before we discuss the physical realization of the quantum version of the classical nonlocal symmetries considered in the previous subsections, we will now introduce the Yangian.


\section{The Yangian Algebra}
\label{sec:yangalg}

This section follows the line of the beautiful original papers by Drinfel'd who introduced the notion of Yangians in the context of quantum groups. In 1990 he was awarded the Fields Medal for his work on quantum groups and for his work in number theory. We will discuss three different realizations of the Yangian, which means three different mathematical definitions of the same algebraic structure that are related by isomorphisms. As opposed to the rest of these notes, in this section we sometimes distinguish between abstract algebra elements, e.g.\ a generator $\levz$, and their representation, e.g.\ $\rho(\levz)$.

\subsection{Yang's R-matrix and the First Realization}
\label{sec:firstreal}

One of the most important concepts underlying integrable models in general is the famous quantum Yang--Baxter equation. This equation was found to emerge in the context of a one-dimensional scattering problem by Yang in 1967 as well as for the eight-vertex model by Baxter in 1972 \cite{Yang:1967bm,Baxter:1972hz} (see also \cite{1964JMP.....5..622M}). In fact also the Yangian was defined in order to determine solutions to this equation. Let us see how this happened.
\paragraph{Yang's solution to the Yang--Baxter equation.}
In the paper  \cite{Yang:1967bm} (see also \cite{1967PhLA...24...55G,Sutherland:1968zz,2006JPhA...39.1073O,2013RvMP...85.1633G}) Yang considered the following one-dimensional Hamiltonian for $n$ interacting particles in a delta-function potential:
\begin{align}\label{eq:YangHam}
H&=-\sum_{k=1}^n\frac{\partial^2}{\partial x_k^2}+2 c \hspace{-.2cm}\sum_{1\leq j<k\leq n} \delta(x_j-x_k),
&
c&>0.
\end{align}
He made a (coordinate) Bethe ansatz%
\footnote{The Bethe ansatz is named after Hans Bethe's solution to the Schrödinger equation for a spin chain \cite{Bethe:1931hc}.}
(cf.\ \cite{Fedor})
for the wavefunction of this quantum mechanical problem, which in the domain $0<x_{k_1}<\dots<x_{k_n}<L$  takes the form
\begin{equation}
\Psi(x_{k_1}<\dots<x_{k_n})=\quad\sum_{\mathclap{\{j_1,\dots,j_n\}\in \text{Perm}\{1,\dots,n\}}} \quad M_{k_1,\dots,k_n,j_1,\dots,j_n} \exp i[p_{j_1}x_{k_1}+\dots+p_{j_n}x_{k_n}],
\end{equation}
with the sum running over all $n!$ permutations of $1,\dots,n$. Here $M$ can be organized as an $n!\times n!$ matrix spanned by the $n!$ column vectors $\xi$:
\begin{equation}
M=
\begin{pmatrix}
\xi_{I_1},\xi_{I_2},\dots,\xi_{I_{n!}}
\end{pmatrix}.
\end{equation}
These vectors have indices $I_1=\{{1,2,3\dots,n}\}$, $I_2=\{{2,1,3,\dots,n}\}$, \dots, $I_{n!}=\{n,n-1,\dots,1\}$.
Notably, with this general ansatz Yang made no assumption on the symmetries of the wavefunction or the exchange statistics of the particles, respectively. It is however assumed that the scattering is purely elastic, i.e.\ that the values of momenta form a fixed set and are conserved individually. Often, in addition a particular exchange symmetry is assumed which allows to reduce the matrix $M$ in the above ansatz to one row.%
\footnote{%
For identical fermions one would have $\Psi(x_{k_1},\dots,x_{k_i},x_{k_j},\dots,x_{k_n})=-\Psi(x_{k_1},\dots,x_{k_j},x_{k_i},\dots,x_{k_n})$. 
For identical bosons the physical system with the Hamiltonian \eqref{eq:YangHam} is called the Lieb--Liniger model \cite{Lieb:1963rt} and we would have $\Psi(x_{k_1},\dots,x_{k_i},x_{k_j},\dots,x_{k_n})=\Psi(x_{k_1},\dots,x_{k_j},x_{k_i},\dots,x_{k_n})$. }

From the form of the Hamiltonian \eqref{eq:YangHam}, one can deduce by integrating the Schr\"odinger equation 
in center of mass coordinates
that the wavefunction $\Psi$ has to be continuous at $x_{j}=x_{k}$, while its first derivative should have a discontinuity at these points. Yang found that these conditions are satisfied at for instance $x_{k_3}=x_{k_4}$ if the permutation of the momentum labels ${j_3}$ and ${j_4}$ is compensated by a factor of the so-called R-matrix:
\begin{equation}\label{eq:Yangconsistent}
\xi_{j_1,j_2,j_3,j_4, j_5\dots j_{n}}=\permop_{34} \rop_{34}(u_{j_4j_3}) \,\xi_{j_1,j_2,j_4,j_3,j_5,\dots,j_{n}}.
\end{equation}
Here we make the exchange operator $\permop_{34}$ for the particles with coordinates $x_{k_3}$ and $x_{k_4}$ explicit, while it is sometimes included into an alternative definition of the R-operator.%
\footnote{The operator $\permop_{ij}$ represents the permutation operator on the vector $\xi_I$ permuting the entries $k_i$ and $k_j$. An alternative definition of the R-matrix found in the literature is $\check \rop_{ij}=\permop_{ij}\rop_{ij}$ (note that $\permop^2=\idop$). Acting on $\xi$, we have for identical bosons $\permop_{ij}=1$ while for identical fermions $\permop_{ij}=-1$. For a model of identical bosons for instance, whose wavefunction is symmetric under exchange of particles at $x_{k_3}$ and $x_{k_4}$, the permutation operator on the right hand side of \eqref{eq:Yangconsistent} acts as the identity and $ \rop_{34}(u_{j_4j_3})$ represents the scattering matrix for the two bosonic particles $3$ and $4$ with momentum difference $u_{j_4j_3}=p_{j_4}-p_{j_3}$.}
The above R-matrix accounts for the scattering of two particles. 

\begin{figure}
\begin{center}
\includegraphicsbox[scale=1.1]{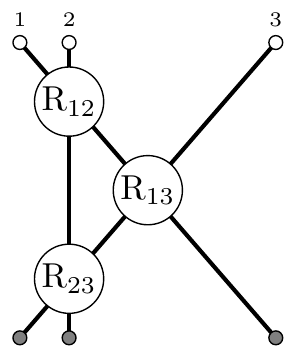}
\quad$=$\quad
\includegraphicsbox[scale=1.1]{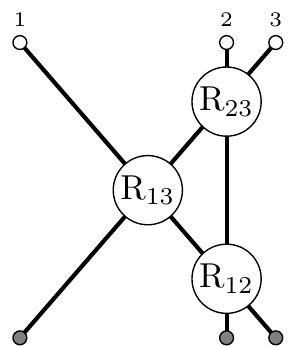}
\end{center}
\caption{Illustration of the Yang--Baxter equation.}
\label{fig:YBE}
\end{figure}

As discussed by Yang, the $n!(n-1)$ equations of the above form \eqref{eq:Yangconsistent} are mutually consistent, if the R-matrix is unitary, i.e.\ if we have $\rop_{\ell m}(u)\rop_{m\ell}(-u)=1$
and if the following \emph{quantum Yang--Baxter equation} is obeyed, cf.\ \figref{fig:YBE}
(see e.g.\ \cite{Jimbo:1989qm} for a nice introduction to the Yang--Baxter equation by Jimbo):%
\begin{equation}
\rop_{12}(u_{12})\rop_{13}(u_{13})\rop_{23}(u_{23})
=
\rop_{23}(u_{23})\rop_{13}(u_{13})\rop_{12}(u_{12}).
\label{eq:YBE}
\end{equation}
For three identical bosons for instance, $\permop$ acts on $\xi$ as the identity, and the Yang--Baxter equation can be understood by noting that via \eqref{eq:Yangconsistent} the expression $\xi_{321}$ can be obtained from $\xi_{123}$ in two different ways, which have to be consistent:
\begin{equation}
\rop_{12}(u_{12})\rop_{13}(u_{13}) \rop_{23}(u_{23})\xi_{123}=\xi_{321}= \rop_{23}(u_{23}) \rop_{13}(u_{13}) \rop_{12}(u_{12})\xi_{123}.
\end{equation}
The quantum Yang--Baxter equation is of central importance for integrable models and appears in many different contexts. In general, it represents an operator equation acting on three spaces $\spc{V}_1\otimes\spc{V}_2\otimes \spc{V}_3$ labeled $1,2$ and $3$.
Each R-matrix (e.g.\ $\rop_{12}$) acts on two spaces (e.g.\ $1$ and $2$), and is a four-index object more explicitly written as%
\footnote{Alternatively, one can write the Yang--Baxter equation as 
\begin{equation}
\rop_{j_1j_3}^{k_2k_1}(u_{12})\rop_{j_2i_3}^{k_3j_3}(u_{13})\rop_{i_1i_2}^{j_2j_1}(u_{23})
=
\rop_{j_3j_2}^{k_3k_2}(u_{23})\rop_{i_1j_1}^{j_3k_1}(u_{13})\rop_{i_2i_3}^{j_2j_1}(u_{12}).
\label{eq:YBE2}
\end{equation}
}
\begin{equation}
\rop_{i_1i_2}^{k_2k_1}=[\rop_{12}]_{i_1i_2}^{k_2k_1}
=\includegraphicsbox{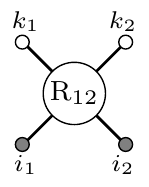}.
\end{equation}
That is, when acting on $n$-dimensional vector spaces $\spc{V}$ with basis vectors $v_1,\dots, v_n$ 
we have%
\footnote{Using a tensor product notation, the R-matrices entering the Yang--Baxter equation can also be written as
\begin{align}
\rop_{12}&=\rop\otimes\idop,
&
\rop_{23}&=\idop\otimes \,\rop,
&
\rop_{13}&=(\permop\otimes \idop) \rop_{23} (\permop\otimes \idop).
\end{align}}
\begin{equation}
\rop(u)[\,v_i\otimes v_j]=\sum_{k,l}\rop^{kl}_{ij}(u) \,v_k\otimes v_l.
\end{equation}

Coming back to the above specific model with delta-function potential, the solution to the quantum Yang--Baxter equation given by Yang takes the form%
\begin{equation}\label{eq:Yangsol}
\rop_{\ell m}(u_{ij})
=\frac{u_{ij}}{u_{ij}+i c}\Big(\idop_{\ell m}-\frac{i c}{u_{ij}}\permop_{\ell m}\Big),
\end{equation}
where the parameter $u_{ij}=p_i-p_j$ is given by the difference of the particle momenta. Note again that for instance for the symmetry algebra $\alg{u}(N)$ with generators $\levz_a$ in the fundamental representation, the permutation operator can be written as the tensor Casimir operator $\permop=\tensorcas=\levz_a\otimes \levz_a$.
For this reason, the solution 
\begin{equation}\label{eq:Yangsform}
\rop(u)=\idop+\frac{c}{u}\,\tensorcas,
\end{equation}
to the quantum Yang--Baxter equation is called \emph{Yang's R-matrix}. Here $c$ denotes some constant.
\paragraph{The Yangian.}
Almost twenty years after Yang, in 1985, Drinfel'd studied the quantum Yang--Baxter equation in order to develop an efficient method for the construction of its solutions  \cite{Drinfeld:1985rx}. Drinfel'd was one of the pioneers in introducing the related concept of quantum groups which he motivates as follows:

\bigskip
\begin{center}
\begin{minipage}{14cm}\small{\it
``Recall that both in classical and in quantum mechanics there are two basic concepts: state and observable. In classical mechanics states are points of a manifold $M$ and observables are functions on $M$. In the quantum case states are $1$-dimensional subspaces of a Hilbert space $H$ and observables are operators in $H$ (we forget the self-adjointness condition). The relation between classical and quantum mechanics is easier to understand in terms of observables. Both in classical and in quantum mechanics observables form an associative algebra which is commutative in the classical case and non-commutative in the quantum case. So quantization is something like replacing commutative algebras by noncommutative ones.''}
V.\ Drinfel'd 1986 \cite{Drinfeld:1986in}.
\end{minipage}
\end{center}
\bigskip
For Drinfel'd the starting point to understand the quantum R-matrix $\rop(u,\hbar)$ was its classical counterpart $r(u)$ obtained in the limit $\hbar\to 0$ from
\begin{equation}\label{eq:classlim}
\rop(u,\hbar)\simeq \idop+\hbar \, r(u) +\order{\hbar^2}.
\end{equation}
Subject to the quantum Yang--Baxter equation \eqref{eq:YBE}, the classical R-matrix $r(u)$ satisfies the classical Yang--Baxter equation:
\begin{equation}
\comm{r_{12}(u_{12})}{r_{13}(u_{13})}+\comm{r_{12}(u_{12})}{r_{23}(u_{23})}+\comm{r_{13}(u_{13})}{r_{23}(u_{23})}=0.
\end{equation}
Drinfel'd considered Yang's solution
\begin{equation}
r(u)=\frac{1}{u}\tensorcas=\frac{1}{u}\levz_a\otimes \levz_a
\end{equation}
of the classical Yang--Baxter equation.  Here $\tensorcas=\levz_a\otimes \levz_a$ again represents the tensor Casimir operator of the underlying finite dimensional simple Lie algebra $\alg{g}$ with generators  $\levz_a$.
Given a representation $\rho: \alg{g}\to \text{End}(\spc{V})$ of the Lie algebra, Drinfel'd's intention was to show that solutions to the quantum Yang--Baxter equation exist, which have the form of quantum deformations around the classical R-matrix $r(u)$.
Assuming that $\hbar \sim \frac{1}{u}$, equation \eqref{eq:classlim} can be translated into an $\hbar$-independent form. 
The precise question then becomes whether rational solutions to the quantum Yang--Baxter equation exist which have the form
\begin{equation}\label{eq:RYangform}
(\rho\otimes\rho)(\rmat(u))=\idop+\frac{1}{u}\,\rho(\levz_a)\otimes \rho(\levz_a)+\sum_{k=2}^\infty \frac{\rop_k(u)}{u^k}.
\end{equation}
Here we now distinguish between an abstract, more universal algebra element $\mathcal{R}$ and its representation $(\rho\otimes\rho)(\mathcal{R})$. 

Before coming to the actual definition of the Yangian algebra, we have to introduce another piece of notation. Since the above operators act on tensor product states, one important question is how to generally promote representations from one site or vector space, to two or more sites. In physics language this might for instance be the question of how to go from one-particle to multi-particle representations in the context of scattering processes. The mathematical answer to this question is given by the so-called \emph{coproduct} $\Delta$ acting for example on the elements of a Lie algebra $\alg{g}$ according to
$
\Delta: \alg{g}\to \alg{g}\otimes \alg{g}.
$
In the particular case of a \emph{Lie algebra} with generators $\levz_a$, the (primitive) coproduct is simply given by the tensor product action:
\begin{equation}
\Delta(\levz_a)=\levz_a\otimes \idop+\idop\otimes \levz_a=\levz_{a,1}+\levz_{a,2}.
\end{equation}
In scattering processes relating asymptotic in- to out-states, on which the coproduct acts differently  (see also \secref{sec:smatrix}), it is useful to also define an \emph{opposite coproduct} $\Delta^\text{op}$ via%
\footnote{The permutation or transposition of factors is sometimes alternatively denoted by $\sigma$ acting as $\sigma\circ (a\otimes b)=b\otimes a$. That is we can alternatively write $\Delta^\text{op}\equiv\sigma\circ \Delta$.}
\begin{equation}\label{eq:oppcop}
\Delta^\text{op}\equiv\permop\Delta\,\permop.
\end{equation}
Here $\permop$ again denotes the permutation operator that acts on the coproduct by conjugation. 

Looking at \eqref{eq:RYangform}, we see that at least the first order of the expansion of the rational R-matrix is completely specified by Lie algebra generators $\levz_a$. In order to define an abstract object $\mathcal{R}(u)$ that obeys the Yang--Baxter equation and has a rational form \eqref{eq:RYangform}, one may thus wonder whether also the higher orders of the expansion can be defined in terms of some (possibly generalized) algebra. This is indeed the case.
Inspired by Yang's first rational solution \eqref{eq:Yangsol} to the quantum Yang--Baxter equation \eqref{eq:YBE}, Drinfel'd 
introduced the following Hopf algebra as the \emph{Yangian} \cite{Drinfeld:1985rx}.
\bigskip

\cornersize{.1} 
\realization{First Realization}{
Given a finite-dimensional simple Lie algebra $\alg{g}$ with generators $\levz_a$, the Yangian $Y[\alg{g}]$ is defined as the algebra generated by $\levz_a$ and $\levo_a$ with the relations
\begin{align}\label{eq:Yangcomms}
\comm{\levz_a}{\levz_b}&=f_{abc}\levz_c,
&
\comm{\levz_a}{\levo_b}&=f_{abc}\levo_c,
\end{align}
and the following Serre relations constrain the commutator of two level-one generators%
\addtocounter{footnote}{1}\footnotemark[\thefootnote]
\begin{align}
\comm{\levo_a}{\comm{\levo_b}{\levz_c}}-\comm{\levz_a}{\comm{\levo_b}{\levo_c}}
&=\hbar^2 g_{abcdef}\{\levz_d,\levz_e,\levz_f\},\label{eq:Yang3}
\\
\comm{\comm{\levo_a}{\levo_b}}{\comm{\levz_r}{\levo_s}}
+\comm{\comm{\levo_r}{\levo_s}}{\comm{\levz_a}{\levo_b}}
&=\hbar^2(g_{abcdef}f_{rsc}+g_{rscdef}f_{abc})\{\levz_d,\levz_e,\levz_f\}.\label{eq:Yang4}
\end{align}
Here the $f_{abc}$ denote the structure constants of the algebra $\alg{g}$ and we have
\begin{align}
g_{abcdef}&=\frac{1}{24}f_{ad i}f_{be j}f_{cf k}f_{ijk},
&
\{x_1,x_2,x_3\}=\sum_{i\neq j\neq k}x_ix_jx_k.
\end{align}
For completeness we already note that the Yangian defined by the above relations is a Hopf algebra (discussed in more detail below) with the coproduct%
\addtocounter{footnote}{1}\footnotemark[\thefootnote]
\begin{align}\label{eq:Yangcoprod}
\Delta(\levz_a)&=\levz_a\otimes \idop+\idop\otimes\, \levz_a,
&
\Delta(\levo_a)&=\levo_a\otimes \idop+\idop\otimes \,\levo_a-\half \hbar f_{abc}\,\levz_b\otimes \levz_c.
\end{align}
}
\addtocounter{footnote}{-1}\footnotetext[\thefootnote]{These Serre relations are sometimes called \emph{Drinfel'd's terrific relations} since Drinfel'd referred to the ``terrific right-hand sides'' of \eqref{eq:Yang3} and \eqref{eq:Yang4} in the proceedings \cite{Drinfeld:1986two}. Note that in a later version of those proceedings, the word ``terrific'' was exchanged for ``horrible'' \cite{Drinfeld:1986in}. The left hand side of \eqref{eq:Yang3} may also be written as a three-term expression of the form of the Jacobi identity, cf.\ \cite{Chari:1994pz}.}\addtocounter{footnote}{1}
\footnotetext[\thefootnote]{Here we could alternatively write $-\hbar f_{abc}\levz_b\otimes \levz_c=\hbar\comm{\levz_a\otimes 1}{\tensorcas}$, where $\tensorcas=\levz_a\otimes \levz_a$ denotes the tensor Casimir operator of the underlying Lie algebra $\alg{g}$. }
\bigskip

\noindent Strictly speaking the Yangian was defined as the above algebra with $\hbar=1$. This can usually be achieved by a rescaling of the level-one generators. Still it is elucidating to sometimes make the quantum deformation parameter $\hbar$ of this quantum group explicit. 

Note that for $\alg{g}=\alg{sl}(2)$ the relations \eqref{eq:Yangcomms} imply \eqref{eq:Yang3}. For $\alg{g}\neq\alg{sl}(2)$ \eqref{eq:Yang4} follows from \eqref{eq:Yangcomms} and \eqref{eq:Yang3} as already noted by Drinfel'd. Hence, we can neglect \eqref{eq:Yang4} for most cases and we will refer to \eqref{eq:Yang3} as the \emph{Serre relations} in what follows.

As opposed to more familiar commutation relations of Lie algebras, the above definition does not specify the commutators of all generators. Rather we obtain a new generator from evaluating $\levz^{(2)}_a\simeq f_{abc}\comm{\levo_b}{\levo_c}$ in addition to $\levz^{(0)}_a\equiv \levz_a$ and $\levz^{(1)}_a\equiv \levo_a$. In this way one may iteratively obtain an infinite set of generators that defines the infinite dimensional Yangian algebra. The Serre relations furnish consistency conditions on this procedure as is discussed in some more detail below.

\paragraph{Yang--Baxter equation and Boost automorphism.}
Let us come back to Drinfel'd's original motivation for introducing the Yangian, namely the construction of rational solutions to the quantum Yang--Baxter equation. In order to do so, he defined the automorphism $\boo_u$ of the Yangian algebra $Y[\alg{g}]$ with the property \cite{Drinfeld:1985rx}
\begin{align}\label{eq:boostauto}
\boo_u(\levz_a)&=\levz_a,
&
\boo_u(\levo_a)&=\levo_a+u \,\levz_a,
\end{align}
for all $u\in\mathbb{C}$. 
The mathematical importance of this operator is due to its role for the below construction of solutions to the Yang--Baxter equation from the Yangian. Physically, this automorphism $\boo_u$ is realized in 1$+$1 dimensional models by the Lorentz boost of rapidity $u$.
In these theories the above nontrivial action of $\boo_u$ thus couples the internal Yangian symmetry with the spacetime symmetry. Due to this physical role we will refer to $\boo_u$ as the \emph{boost automorphism} in what follows.%
\footnote{In the literature one also finds the names evaluation-, translation- or shift-automorphism for  $\boo_u$.}
 Subject to the properties of this operator, the following theorem due to Drinfel'd holds.

\theorem{Theorem 1}{
There is a unique formal series 
\begin{align}\label{eq:uniqueformal}
\mathcal{R}(u)&=\idop+\sum_{k=1}^\infty \mathcal{R}_k\frac{1 }{u^k},
&
\mathcal{R}_k\in Y[\alg{g}]\otimes Y[\alg{g}].
\end{align}
such that 
\begin{align}\label{eq:theo1}
(\Delta\otimes \idop)\mathcal{R}(u)&=\mathcal{R}_{13}(u)\mathcal{R}_{23}(u),
&
(\idop\otimes \Delta)\mathcal{R}(u)&=\mathcal{R}_{13}(u)\mathcal{R}_{12}(u),
\end{align}
and with $\Delta^\text{op}(a)=\permop \Delta(a)\permop$ we have
\begin{align}
( \boo_u\otimes \idop)\Delta^\text{op}(a) &=\mathcal{R}(u)(\boo_u\otimes \idop)\Delta(a)\mathcal{R}^{-1}(u),
\label{eq:linetheo3}
\end{align}
for $a\in Y[\alg{g}]$. The operator $\mathcal{R}(u)$ satisfies the quantum Yang--Baxter equation.
In addition, the so-called pseudo-universal R-matrix $\mathcal{R}(u)$ satisfies a unitarity condition of the form $\mathcal{R}_{12}(u)\mathcal{R}_{21}(-u)=\idop$ and can be expanded around infinity in the rational form
\begin{equation}\label{eq:expR}
\log \mathcal{R}(u)=\frac{1}{u}\levz_a\otimes\levz_a+\frac{1}{u^2}(\levo_a\otimes\levz_a-\levz_a\otimes \levo_a)+\mathcal{O}\Big(\frac{1}{u^3}\Big).
\end{equation}
Lastly, the R-matrix transforms under the boost automorphism as
\begin{align}
(\boo_v\otimes \idop)\mathcal{R}(u)&=\mathcal{R}(u+v),
&
(\idop\otimes \boo_v)\mathcal{R}(u)&=\mathcal{R}(u-v).
\end{align}
}
\bigskip

\noindent Thus for a given irreducible representation $\rho:Y[\alg{g}]\to \text{Mat}(n,\mathbb{C})$, the operator 
\begin{equation}
\rop^\rho(u)=(\rho\otimes \rho)(\mathcal{R}(u))
\end{equation} 
is a solution to the quantum Yang--Baxter equation in the form of \eqref{eq:RYangform}.%
\footnote{See \eqref{eq:expR} for the explicit application of the single-site representation of $\levz, \levo$ to the R-operator.}

Notably, the above theorem maps the search for rational solutions to the Yang--Baxter equation to the search for representations of the Yangian algebra. However, since in general we do not know the pseudo-universal R-matrix $\mathcal{R}$ (and cannot obtain it easily), this does not allow to straightforwardly construct representations of solutions to the Yang--Baxter equation. For this purpose, another theorem is very interesting \cite{Drinfeld:1985rx,Chari:1994pz}.

\theorem{Theorem 2}{
Given a finite dimensional irreducible representation $\rho:Y[\alg{g}]\to\mathrm{End}(\spc{V})$, the pseudo-universal R-matrix evaluated on this representation $\rop_\rho(z)=(\rho\otimes\rho)(\rmat(z))$  is the Laurent expansion about $z=\infty$ of a rational function in $z$. The operator  
\begin{equation}
\rop(u-v):\rho(\boo_u(\spc{V}))\otimes \rho(\boo_v(\spc{V}))\to \rho( \boo_u(\spc{V}))\otimes\rho( \boo_v(\spc{V})),
\end{equation}
defined by the below constraints, is up to a scalar factor (and up to finitely many $u-v$) the same solution to the quantum Yang--Baxter equation as $\rop_\rho$ obtained from the pseudo-universal R-matrix.
The constraints on $\rop(u,v)=\rop(u-v)$ take the following form:
\begin{align}
&\text{Level zero:}&
&(\rho\otimes\rho)\big[\levz_a\otimes\idop+\idop\otimes \levz_a\big]\rop(u,v)=\rop(u,v)(\rho\otimes \rho)\big[\levz_a\otimes\idop+\idop\otimes \levz_a\big],
\label{eq:theo21}
\\
&\text{Level one:}&
&(\rho\otimes \rho)\big[(\levo_a+u\levz_a)\otimes \idop+\idop\otimes(\levo_a+v\levz_a)
+\half f_{abc}\levz_b\otimes\levz_c\big]
\rop(u,v)
=\nonumber\\
&&
&\hspace{1.2cm} \rop(u,v)(\rho\otimes\rho)\big[(\levo_a+u \levz_a)
\otimes \idop+\idop\otimes(\levo_a+v \levz_a) 
-\half f_{abc}\levz_b\otimes\levz_c\big].
\label{eq:theo22}
\end{align}
These constraints can be evaluated as a finite system of linear equations.
}
\bigskip

\noindent Notably, all rational solutions to the quantum Yang--Baxter equation can be generated from Yangian representations in this way.

\paragraph{Example: $Y[\alg{su}(2)]$.}
For illustration let us consider the rank-one example of $\alg{g}=\alg{su}(2)$ with representation $\rho:Y[\alg{g}]\to\text{Mat}(2,\mathbb{C})$ defined on one site as $\rho(\levz_a)=\levz_a=\frac{\sigma_a}{2i}$ and $\rho(\levo_a)=0$. Here $\sigma_{a=1,2,3}$ denotes the Pauli matrices such that $\comm{\levz_a}{\levz_b}=\epsilon_{abc}\levz_c$. The above constraints at level zero, i.e.\
\begin{equation}
\comm{\levz_a\otimes\idop+\idop\otimes \levz_a}{\rop(u,v)}=0,
\end{equation}
correspond to the ordinary $\alg{su}(2)$ Lie algebra symmetry. For $\rop(u,v): \mathbb{C}^2\otimes  \mathbb{C}^2\to  \mathbb{C}^2\otimes  \mathbb{C}^2$ we have only two independent irreducible representations which are mapped onto themselves by the $\alg{su}(2)$ symmetry, i.e.\ in terms of Young tableaux:
\begin{equation}
\autoparbox{\young{\cr}}\otimes\autoparbox{\young{\cr}}=\autoparbox{\young{\cr\cr}}\oplus \autoparbox{\young{&\cr}}\;.
\end{equation}
In consequence there are also two $\alg{su}(2)$-invariant operators of range two, e.g.\ the projectors onto the two irreducible representations.
We already know that for $\alg{u}(2)$ the tensor Casimir $\tensorcas$ is proportional to the permutation operator $\permop$. Obviously, this operator also commutes with the $\alg{su}(2)$ symmetry. A second invariant operator is the identity $\idop$ (the second Casimir of $\alg{u}(2)$) and hence the level-zero symmetry constrains the R-matrix to be of the form
\begin{equation}\label{eq:Ransatz}
\rop(u,v)=a(u,v)\,\idop +\;b(u,v)\,\permop,
\end{equation}
with arbitrary coefficients $a(u,v)$ and $b(u,v)$.
After multiplication with the permutation operator~$\permop$, the level-one constraint is given by%
\footnote{Sometimes one introduces $\check{\rop}=\permop \rop$ and rephrases the above statements in terms of this operator.}
\begin{equation}
\big(v \levz_a\otimes \idop+u\, \idop\otimes \levz_a-\half\epsilon_{abc} \levz_b\otimes \levz_c\big)\permop \rop(u,v)
=
\permop \rop(u,v)\big(u \levz_a\otimes \idop+v\, \idop\otimes \levz_a-\half \epsilon_{abc} \levz_b\otimes \levz_c\big),
\end{equation}
which implies
\begin{equation}
\half\comm{\epsilon_{abc} \levz_b\otimes \levz_c}{\permop \rop(u,v)}=\big(v \levz_a\otimes \idop+u\,\idop\otimes \levz_a\big)\permop \rop(u,v)-\permop \rop(u,v)\big(u \levz_a\otimes \idop+v\idop\otimes \levz_a\big).
\end{equation}
Furthermore noting that we have
\begin{equation}\label{eq:levoneP}
\comm{ \epsilon_{abc} \levz_b\otimes \levz_c}{\permop}=\levz_a\otimes \idop-\idop\otimes \levz_a,
\end{equation}
and using \eqref{eq:Ransatz}, we thus find
\begin{equation}
 -\half a(u,v)\big(\levz_a\otimes \idop-\idop\otimes \levz_a\big)=(u-v)b(u,v)\big(\levz_a\otimes \idop-\idop\otimes \levz_a\big).
\end{equation}
Hence, we have $-\half a(u,v)= (u-v)b(u,v)$ such that Yangian symmetry fixes the R-matrix up to an overall factor to be of Yang's form \eqref{eq:Yangsform}:
\begin{equation}\label{eq:su2rmat}
\rop(u,v)=a(u,v)\Big(\idop - \frac{1}{2(u-v)}\,\permop\Big).
\end{equation}
Note that for the above normalization of a basis of $\alg{u}(2)$ we have $\tensorcas=-\half\permop$. This example for $\alg{g}=\alg{su}(2)$ illustrates the basic principle of how to fix the matrix structure of an R- or S-matrix from Yangian symmetry and can be generalized to more complicated algebras $\alg{g}$.


\paragraph{Representations and Serre relations.}
If you encounter a symmetry in a physical model that has generators $\levz_a$ and $\levo_a$ and follows the coproduct structure \eqref{eq:Yangcoprod}, this is a promising sign that you are dealing with a Yangian algebra. However, you will have to verify that your generators obey the Serre relations, which is typically hard work. It may thus be useful to understand the nature of these Serre relations a bit better.

The above coproduct is an algebra homomorphism, that is the following relation should hold for $a,b\in Y[\alg{g}]$:
\begin{equation}
\Delta(\comm{a}{b})=\comm{\Delta(a)}{\Delta(b)}.
\end{equation}
This homomorphism property is trivially obeyed  for some commutators of generators with the coproduct structure \eqref{eq:Yangcoprod}, i.e.\ one easily verifies that
\begin{align}
\Delta(\comm{\levz_a}{\levz_b})&=\comm{\Delta(\levz_a)}{\Delta(\levz_b)},
&
\Delta(\comm{\levz_a}{\widehat \levz_b})&=\comm{\Delta(\levz_a)}{\Delta(\widehat \levz_b)}.
\end{align}
Consider for instance the second case, whose left- and right hand sides explicitly evaluate to
\begin{align}
\Delta(\comm{\levz_a}{\widehat \levz_b})&=\Delta(f_{abc}\levo_c)=f_{abc}\idop\otimes\levo_c+f_{abc}\levo_c\otimes \idop-\half f_{abc}f_{cde}\levz_d\otimes \levz_e,
\\
\comm{\Delta(\levz_a)}{\Delta(\widehat \levz_b)}
&=\comm{\idop\otimes \levz_a+\levz_a\otimes\idop}{\idop\otimes \levo_b+\levo_b\otimes\idop-\half f_{bcd}\levz_c\otimes\levz_d}\nonumber\\
&=f_{abc}\idop\otimes\levo_c+f_{abc}\levo_c\otimes\idop-\half (f_{bdc}f_{ace}+f_{bce}f_{acd})\levz_d\otimes\levz_e.
\end{align}
Both sides are equal upon using the Jacobi identity $f_{abc}f_{cde}+f_{dac}f_{cbe}+f_{bdc}f_{cae}=0$.
On the other hand, the relation
\begin{equation}\label{eq:coSerre}
\Delta(\comm{\widehat \levz_a}{\widehat \levz_b})=\comm{\Delta(\widehat \levz_a)}{\Delta(\widehat \levz_b)}
\end{equation}
or when making the representation $\rho$ explicit
\begin{equation}\label{eq:coSerre2}
\rho(\Delta(\comm{\widehat \levz_a}{\widehat \levz_b)})=\comm{\rho(\Delta(\widehat \levz_a))}{\rho(\Delta(\widehat \levz_b))},
\end{equation}
does not trivially follow from the definition of the coproduct, but it implies non-trivial constraints on the representation $\rho$ of the Yangian generators. 
This can be seen by noting that the left hand side of \eqref{eq:coSerre2} forms part of the antisymmetrized tensor product of the adjoint representation with itself
\begin{equation}
(\text{adj}\otimes \text{adj})_\text{asym}=\text{adj}\oplus \mathbb{X}.
\end{equation}
This relation defines the representation $\mathbb{X}$ that does typically not contain the adjoint representation. The adjoint part defines the coproduct for the level-two Yangian generators while the Serre relations furnish a sufficient criterion for the vanishing of the $\mathbb{X}$-component, cf.\ e.g.\ \cite{Bargheer:2010hn} for more details. 
In fact, if the Serre relations are satisfied for the one-site representation, they will also hold for the $n$-site representation since the coproduct preserves the Serre relations. 

\paragraph{Construction of representations.}
As pointed out by Drinfel'd, given a Lie algebra represenation $\rho$ one may choose the following one-site representation $\rho_0$ of the Yangian generators
\begin{align}\label{eq:zerorep}
\rho_0(\levz_a)&=\rho(\levz_a),
&
\rho_0(\levo_a)&=0.
\end{align}
The left hand side of  \eqref{eq:Yang3}  vanishes in this case. In order to show that our representation $\rho_0$ obeys the Serre relations, we have to show that the $\mathbb{X}$-projection of the right hand side of \eqref{eq:Yang3} vanishes for the one-site representation:
\begin{equation}\label{eq:onesitezero}
\rho_0(\{\levz_a,\levz_b,\levz_c\})|_\mathbb{X}=0.
\end{equation}
In \cite{Drinfeld:1985rx} Drinfel'd indicated the existence of such representations for all types of algebras $\alg{g}$ except for $\alg{e}_8$.%
\footnote{%
These representations also play an important role in the AdS/CFT correspondence. The Serre relations were shown for representations of $\alg{psu}(2,2|4)$ \cite{Dolan:2004ps} and $\alg{osp}(4|6)$ \cite{Bargheer:2010hn} that realize the Yangian symmetry in $\superN=4$ super Yang--Mills and $\superN=6$ superconformal Chern--Simons theory.
}
Once \eqref{eq:onesitezero} is shown for the one-site representation, one promotes the representation to multiple sites via the coproduct which preserves the Serre relations.

\paragraph{Evaluation representation.}
For some representations $\rho$ of the Lie algebra $\alg{g}$ there exists a so-called \emph{evaluation representation} $\rho_u$ of the Yangian algebra given by
\begin{align}\label{eq:evalrep}
\rho_u(\levz_a)&=\rho_0(\levz_a),
&
\rho_u(\levo_a)&=u\,\rho_0(\levz_a).
\end{align}
As discussed above for $u=0$, this choice puts the constraint on the representation that the right hand side of the Serre relations vanishes for the one-site representation since the left hand side is trivially zero. 
The evaluation representation can also be defined using the boost automorphism and the above representation $\rho_0$ as follows:
\begin{align}
\rho_u(\levz_a)&=\rho_0(\levz_a),
&
\rho_u(\levo_a)&=\rho_0\big( \boo_u(\levo_a)\big).
\end{align}
This representation is important for evaluating the Yangian symmetry of the two-particle S-matrix $\sop(u,v)$ where $u$ and $v$ represent particle rapidities, cf.\ \secref{sec:smatrix}. For this purpose it will be useful to explicitly evaluate the two-site representation using the coproduct and the two-site boost automorphism 
\begin{align}
\boo_u\otimes \boo_v (\Delta(\levo_a))
=
u\, \levz_a\otimes \idop+v\, \idop\otimes \levz_a+\levo_a\otimes \idop+\idop\otimes \levo_a-\half f_{abc} \levz_b\otimes \levz_c,
\end{align}
which yields%
\footnote{%
The evaluation representation may be used to define Yangian-invariant deformations of scattering amplitudes in $\superN=4$ super Yang--Mills \cite{Ferro:2012xw} and $\superN=6$ superconformal Chern--Simons theory \cite{Bargheer:2014mxa}.
}
\begin{align}
\rho_u\otimes\rho_v\big(\Delta(\levo_a)\big)
=
\rho_0\otimes\rho_0\big(\boo_u\otimes \boo_v (\Delta(\levo_a))\big)
=
\rho\otimes\rho\big(u\, \levz_a\otimes \idop+v\, \idop\otimes \levz_a-\half  f_{abc} \levz_b\otimes \levz_c\big).
\end{align}

\subsection{The Yangian as a Hopf Algebra and Quantum Group.}
\label{sec:hopfandQG}

We continue our study of the mathematical structure behind the Yangian together with Drinfel'd:

\begin{center}
\begin{minipage}{14cm}\small{\it
``Now let us consider the elements of a group $\grp{G}$ as states and functions on $\grp{G}$ as observables.
The notion of group is usually defined in terms of states. To quantize it one has to translate it first into the language of observables. This translation is well known, but let us recall it nevertheless.''} V.\ Drinfel'd 1986 \cite{Drinfeld:1986in}.
\end{minipage}
\end{center}
\begin{table}[t]
\begin{center}
\begin{tabular}{| l || c | c |}\hline\hline
&\textbf{Classical Mechanics}&\textbf{Quantum Mechanics} \\\hline
States&Points on manifold $M$& 1d subspaces of Hilbert space $H$\\\hline
Observables& Algebra of functions on $M$&Algebra of operators on $H$\\\hline\hline
&\textbf{Classical Group}&\textbf{Quantum Group} \\\hline
States&Elements of group $\grp{G}$&?\\\hline
Observables&Algebra of functions on $\grp{G}$&?\\\hline
\end{tabular}
\phantom{aaaaaaaaaaa}\begin{tikzpicture}[>=latex,scale=1.2,cap=round]
\draw[thick,->](-2,.5)--(-2,0)--(-2,0)--(4,0)--(4,.5);
\node at (1,.2) {Quantization};
\node at (1,-.2) {Commutative $\to$ Non-commutative};
\end{tikzpicture}
\caption{What is a quantum group?}
\label{tab:quantumgroup}
\end{center}
\end{table}

\paragraph{Hopf algebras.}
We are now interested in generalizing the quantization of classical mechanics to the case of groups or algebras, cf.\ \tabref{tab:quantumgroup}.
 Following \cite{Drinfeld:1986two,Drinfeld:1986in}, we thus want to understand how the properties of a group considered as the space of states, translate into the language of observables.
 For this purpose we remember that a group is defined as a pair $(\grp{G},f)$ of a set $\grp{G}$ and a group operation $f$ such that 
\begin{equation}
f:\grp{G}\times \grp{G}\to \grp{G}.
\end{equation}
Remember also that a group is defined to be associative which can be conveniently displayed using the following diagram: 
\begin{align}
f(f\times \idop)(x,y,z)&=f(\idop\times f)(x,y,z):
&
&\includegraphicsbox{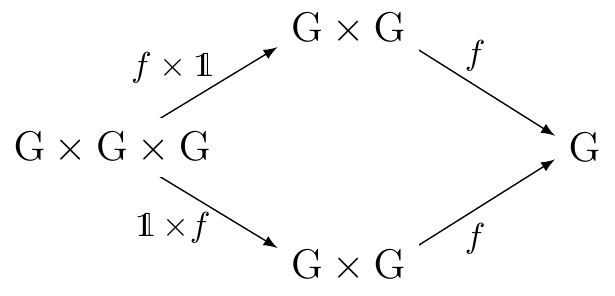},
\end{align}
for $x,y,z\in \grp{G}$.
We consider the algebra $\alg{a}=\text{Fun}(\grp{G})$ (the observables) consisting of functions on the group $\grp{G}$. The group map $f$ induces an algebra homomorphism%
\footnote{
An algebra homomorphism between two algebras $\alg{a}$ and $\alg{b}$ over the field $\mathbb{C}$ is a map $\Delta:\alg{a}\to \alg{b}$ such that for all $k\in \mathbb{C}$ and $a,b\in \alg{a}$:
\begin{align}
\Delta(k a)&=k \Delta(a),
&
\Delta(a+b)&=\Delta(a)+\Delta(b),
&
\Delta(a b)=\Delta(a)\Delta(b).
\end{align}
}
which is dubbed \emph{coproduct} or \emph{comultiplication}%
\footnote{The coproduct is induced via $(\Delta(a))(x y):=a(x y)$ for $a\in \alg{a}$ and $x,y\in \grp{G}$, cf.\ e.g.\ \cite{Fuchs:1992nq} for more details.}
\begin{equation}
\Delta:\alg{a} \to \alg{a}\otimes \alg{a},
\end{equation}
where $\alg{a}\otimes \alg{a}=\text{Fun}(\grp{G}\times \grp{G})$. This is the coproduct we already encounterd above.
As mentioned, in physics the coproduct furnishes a prescription for how to extend the symmetry from one- to multi-particle states in a fashion that is compatible with the underlying algebraic structure.

Translating the associativity of the group map to the coproduct $\Delta$ we find the property of \emph{coassociativity}, i.e.\
\begin{align}
&(\idop\otimes\Delta)\Delta(a)=(\Delta\otimes \idop)\Delta(a):
&
\includegraphicsbox{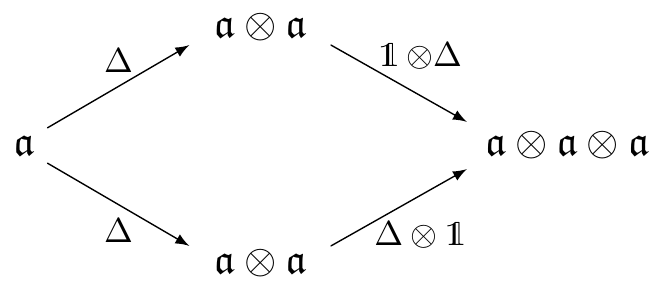}.
\end{align}
Finally we need to find the analogue of the group inversion $x\mapsto x^{-1}$, which is denoted the \emph{antipode}
\begin{equation}
s:\alg{a}\to \alg{a},
\end{equation}
and the analogue of the unit element $e$ of the group, which is denoted the \emph{counit}:
\begin{equation}
\epsilon: \alg{a}\to \mathbb{C}.
\end{equation}
We also have an ordinary multiplication 
$
m:\alg{a}\otimes \alg{a} \to \alg{a},
$
and the unit map $\eta:\mathbb{C}\to \alg{a}$ defined as $\eta:c\mapsto c\cdot \idop$, for $c\in \mathbb{C}$ and $\idop\in \alg{a}$.
These maps should obey the following commutative diagrams which correspond to $e\cdot x=x$ and $x\cdot e=x$ for $x\in \grp{G}$:
\begin{align}
&a=(\idop\otimes\epsilon)\Delta(a):
&
&\includegraphicsbox{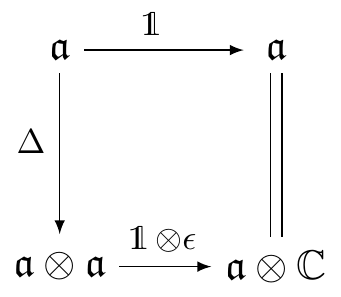},\qquad
&
&a=(\epsilon\otimes\idop)\Delta(a):
&
&\includegraphicsbox{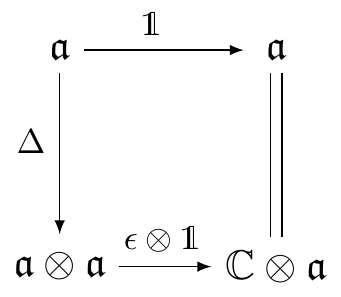}.\nonumber
\end{align}
Lastly, all of the introduced maps should be compatible with each other and obey the relations
\begin{align}
&m(s\otimes\idop)\Delta(a)=\eta(\epsilon(a))=m(\idop\otimes s)\Delta(a):
&
&\includegraphicsbox{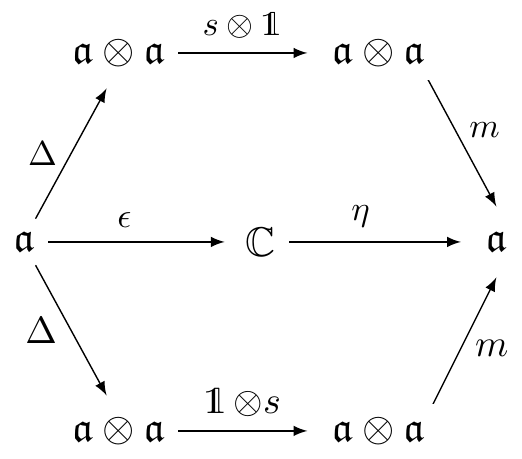}.
\end{align}
The above properties of $(\alg{a},\Delta,s,\epsilon,m,\eta)$ define the commutative \emph{Hopf algebra} $\alg{a}$. This Hopf algebra furnishes a notion of the class of observables in the context of groups. 

Importantly, a Hopf algebra is called \emph{cocommutative} if the opposite coproduct obeys
\begin{equation}
\Delta^\text{op}(a)=\Delta(a),
\end{equation}
for $a\in \alg{a}$. Remember that the opposite coproduct was defined as
$\Delta^\text{op}=\permop\Delta\,\permop$.
The analogy between the group product $f$ and the coproduct $\Delta$ is underlined by the fact that the above Hopf algebra $\alg{a}$ is cocommutative iff the group $\grp{G}$ is abelian. 
In the above spirit, we can thus understand the quantization of a Hopf algebra as the replacement of a cocommutative coproduct by a non-cocommutative coproduct. This is in close analogy to the transition from classical to quantum mechanics, where a commutative product is replaced by a non-commutative one, e.g.\ $\comm{x}{p}=0$ goes to $\comm{\hat x}{\hat p}=i\hbar$.

\paragraph{Example: The universal enveloping algebra  $U[\alg{g}]$.}
Given a Lie algebra $\alg{g}$, one can define the \emph{universal enveloping algebra (UEA) } $\alg{a}=U[\alg{g}]$ as the quotient of the tensor algebra 
\begin{equation}
T(\alg{g})=\bigoplus_{n=0}^\infty \alg{g}^{\otimes^n}=\mathbb{K}\oplus\,\alg{g}\oplus (\alg{g}\otimes\alg{g})\oplus (\alg{g}\otimes\alg{g}\otimes\alg{g})\oplus\dots
\end{equation}
by the elements $a\otimes b-b\otimes a-\comm{a}{b}$ for $a,b\in \alg{g}$. Here $\mathbb{K}$ is the field associated to the Lie algebra~$\alg{g}$. Hence, the UEA may be considered as the space of polynomials of elements of $\alg{g}$ modulo the commutator, i.e.\ for our Lie algebra generators the combination $\levz_a\otimes \levz_b-\levz_b\otimes \levz_a$ is identified with $f_{abc}\levz_c$.
The UEA can be equipped with a Hopf algebra structure by defining
\begin{align}\label{eq:UEAHopf}
\Delta(a)&=a\otimes 1+1\otimes a,
&
s(a)&=-a,
&
\epsilon(a)&=0,
\end{align}
for $a\in\alg{g}$ which naturally extends to $U[\alg{g}]$. Note that this coproduct is cocommutative, i.e.\ invariant under exchanging the factors on the left and right hand side of $\otimes$, which is consistent with our idea of an un-quantized, i.e.\ classical algebra.

\paragraph{Quantum groups.} 

The term \emph{quantum group} introduced by Drinfel'd is generically employed to refer to a deformed algebraic structure. This deformation is typically parametrized by a deformation parameter which we call $\hbar$ to remind of the physical quantum deformation of classical mechanics: $\comm{x}{p}=0 \to \comm{\hat x}{\hat p}=i\hbar$. In particular, the name quantum group does often not refer to a group in the ordinary mathematical sense.
Here we will understand quantum groups as special examples of Hopf algebras,%
\footnote{Strictly speaking quantum groups are rather dual (but equivalent) to Hopf algebras, but this distinction is often not made. See for instance \cite{Aschieri:2007ep} for some explicit discussions.}
namely quantizations $U_\hbar[\alg{g}]$ of the universal enveloping algebra $U[\alg{g}]$ of an underlying algebra $\alg{g}$. In accordance with the relation between classical and quantum mechanics, this quantization goes along with replacing a cocommutative coproduct by a non-cocommutative one.

\paragraph{Example: $U_\hbar[\alg{sl}(2)]$.}
Consider the example of the Lie algebra $\alg{g}=\alg{sl}(2)$ (cf.\ e.g.\ \cite{Chari:1994pz}) with generators $X^+,X^-$ and $H$
obeying 
\begin{align}\label{eq:undeformed}
\comm{X^+}{X^-}&=H,
&
\comm{H}{X^\pm}&=\pm 2 X^\pm.
\end{align}
Based on this algebra, we may define the universal enveloping algebra $U[\alg{sl}(2)]$ as introduced above.
Then the primitive coproduct
\begin{align}\label{eq:primco}
\Delta(X^+)&=X^+\otimes\idop +\idop\otimes X^+,
&
\Delta(X^-)&=X^-\otimes \idop+\idop\otimes X^-,
&
\Delta(H)&=H\otimes \idop+\idop \otimes H,
\end{align}
may be understood as an algebra homomorphism on the universal enveloping algebra $U[\alg{sl}(2)]$.
A deformation $U_\hbar[\alg{sl}(2)]$ of the universal enveloping algebra is induced by deforming the above commutation relations or structure constants, respectively, to
\begin{align}
\comm{X^+}{X^-}&=\frac{e^{\hbar H}-e^{-\hbar H}}{e^\hbar-e^{-\hbar}},
&
\comm{H}{X^\pm}&=\pm 2 X^\pm.
\end{align}
Note that in the classical limit $\hbar \to 0$ we obtain the undeformed algebra \eqref{eq:undeformed}:
\begin{equation}
\lim_{\hbar \to 0}\frac{e^{\hbar H}-e^{-\hbar H}}{e^\hbar-e^{-\hbar}}=H.
\end{equation}
The non-cocommutative coproduct for the three types of generators takes the form
\begin{align}
\Delta(X^+)&=X^+\otimes e^{\hbar H}+\idop\otimes X^+,
&
\Delta(X^-)&=X^-\otimes \idop+e^{-\hbar H}\otimes X^-,
\end{align}
and
\begin{equation}
\Delta(H)=H\otimes \idop+\idop \otimes H.
\end{equation}
For $\hbar\to 0$ the coproduct is the primitive one \eqref{eq:primco}, which is cocommutative.

\paragraph{Quasitriangular Hopf algebras and universal R-matrix.}
Let us briefly introduce some further important concepts related to integrable models and the Yangian.
A Hopf algebra $\alg{a}$ is called \emph{almost cocommutative}, if an element $\mathcal{R}\in \alg{a} \otimes \alg{a}$ exists such that
\begin{equation}\label{eq:triangular}
\Delta^\text{op}(a)=\mathcal{R}\Delta(a)\mathcal{R}^{-1},
\end{equation}
for all $a\in \alg{a}$, where $\Delta^\text{op}=\permop\Delta\,\permop$. That is if the opposite coproduct $\Delta^\text{op}$ and the coproduct $\Delta$ are similar.
Comparing \eqref{eq:triangular} to \eqref{eq:linetheo3} we see that this is not the case for the Yangian (see below paragraph).
An almost cocommutative Hopf algebra $(\alg{a},\mathcal{R})$ is called \emph{quasitriangular} if
\begin{align}
(\Delta\otimes \idop)(\mathcal{R})&=\rmat_{13}\rmat_{23},
&
(\idop\otimes\Delta)(\mathcal{R})&=\rmat_{13}\rmat_{12}.
\end{align}
If $\alg{a}$ is quasitriangular, the element $\rmat$ is called the \emph{universal R-matrix} of $(\alg{a},\rmat)$. The universal R-matrix of a quasi-triangular Hopf algebra satisfies the quantum Yang--Baxter equation as well as the relation
\begin{equation}\label{eq:crossingpode}
(s\otimes \idop)(\rmat)=\rmat^{-1}=(\idop\otimes s^{-1})(\rmat),
\end{equation}
where $s$ denotes the antipode. The property \eqref{eq:crossingpode} is important for physical applications since it represents the \emph{crossing} relation when $\rmat$ is given by a scattering matrix with Hopf algebra symmetry, cf.\ e.g.\ \cite{Diego}.
For completeness let us mention that a quasi-triangular Hopf algebra is called \emph{triangular} if $\rmat_{12}\rmat_{21}=\idop$.

\paragraph{The Yangian as a Hopf algebra and quantum group.}
The Yangian defined above is a Hopf algebra with the coproduct (we may set $\hbar=1$)
\begin{align}\label{eq:Yangcoprod2}
\Delta(\levz_a)&=\levz_a\otimes \idop+\idop\otimes \levz_a,
&
\Delta(\levo_a)&=\levo_a\otimes \idop+\idop\otimes \levo_a-\half \hbar f_{abc}\levz_b\otimes \levz_c.
\end{align}
The antipode acts on the generators according to
\begin{align}
s(\levz_a)&=-\levz_a,
&
s(\levo_a)=-\levo_a+\half \hbar f_{abc}\levz_b\levz_c,
\end{align}
and the counit acts trivially as%
\footnote{For simple Lie algebras $\alg{g}$ one can rewrite $f_{abc}\levz_b\levz_c=\half \cas \levz_a$ with $\cas$ being the quadratic Casimir of $\alg{g}$ in the adjoint representation.
}
\begin{align}
\epsilon(\levz_a)&=0,
&
\epsilon(\levo_a)&=0.
\end{align}

The Yangian is \emph{not} quasitriangular since the \emph{pseudo-universal} operator $\rmat$ of the Yangian is not an element of $Y[\alg{g}]\otimes Y[\alg{g}]$. This requires the introduction of the above boost automorphism \eqref{eq:boostauto}, and \eqref{eq:linetheo3} represents the pseudo-triangularity condition analogous to \eqref{eq:triangular} for the quasitriangular case. 
Alternatively one could consider the so-called \emph{Yangian double} which possesses a universal R-matrix, see e.g.\ \cite{Chari:1994pz}.

On the level of the abstract algebra, the evaluation representation \eqref{eq:evalrep} discussed above may be induced by an \emph{evaluation homomorphism} from the Yangian to the universal enveloping algebra 
\begin{align}
\text{ev}_u: Y[\alg{g}] &\to U[\alg{g}],
&
\text{ev}_u(\levz_a)&=\levz_a,
&
\text{ev}_u(\levo_a)&=u\, \levz_a.
\end{align}
This homomorphism, however, turns out to exist only for $\alg{g}=\alg{sl}(2)$, while it takes a more complicated form for $\alg{sl}(N)$ with $N>2$ and does not exist for symmetry algebras of type different from $\alg{a}_N$ (in the Dynkin classification of simple Lie groups)  \cite{Chari:1994pz}.

\paragraph{The Yangian is a quantum deformation of what?}
The Yangian is a deformation of the UEA of the so-called  \emph{polynomial algebra} $\alg{g}[u]$.
Given a Lie algebra $\alg{g}$, the polynomial algebra $\alg{g}[u]$ is defined as the space of polynomials in $u$ with values in $\alg{g}$. This means that $\alg{g}[u]$ is spanned by monomials of the form $\levz_n^a=u^n \levz^a$ with $n=0,\dots,\infty$. The simplest way to construct representations of the polynomial algebra is via the evaluation homomorphism
\begin{equation}
\text{ev}_u:\alg{g}[u]\to\alg{g},
\end{equation}
which evaluates a polynomial at a fixed point $u\in \mathbb{C}$. The evaluation homomorphism of the Yangian algebra discussed above represents the quantum generalization of this map.
Taking $\hbar\to0$ in the defining relations of the Yangian, one obtains the UEA of $\alg{g}[u]$ with the correct Hopf structure \eqref{eq:UEAHopf}.

\quoting{By the way, we are lucky that $Y[\alg{g}]$ is pseudo-triangular and not triangular:
 otherwise $Y[\alg{g}]$ would be isomorphic (as an algebra) to a universal enveloping algebra and life would be dull.}{V.\ Drinfel'd 1986 \cite{Drinfeld:1986in}}

\subsection{Second and Third Realization}
\label{sec:secondandthird}
While these lectures put more weight on the original, first realization and its connection to physical systems, it should be emphasized, that further notable realizations of the Yangian algebra exist and were discussed by Drinfel'd. In the context of physical systems, in particular the third, so-called  RTT realization establishes a connection to earlier work on integrability and the quantum inverse scattering method, see also \secref{sec:spinchains}.
\subsubsection{Second Realization}
In 1988 Drinfel'd introduced a new realization of the Yangian that will be briefly discussed in this subsection. Drinfel'd's motivation for studying this new realization was some shortcomings of the first realization:

\bigskip
\begin{center}
\begin{minipage}{14cm}\small{\it
``Unfortunately, the realization given in \cite{Jimbo:1985zk}
 and \cite{Drinfeld:1986in} of Yangians and quantized affine algebras%
\addtocounter{footnote}{1}\footnotemark[\thefootnote]
  is not suitable for the study of finite-dimensional representations of these algebras.''} V.\ Drinfel'd 1988 \cite{Drinfeld:1987sy}
\end{minipage}
\footnotetext{Here Drinfel'd refers to the quantum algebras that take a similar role for trigonometric solutions to the Yang--Baxter equation as the Yangian for rational solutions.}
\end{center}
\bigskip
The new realization of the Yangian given below can be used to demonstrate a one-to-one correspondence between irreducible finite-dimensional representations of the Yangian and sets of polynomials \cite{Drinfeld:1987sy}. While this correspondence proves useful for studying Yangian representations, it is beyond the scope of these lectures. 

The second realization is particularly interesting since it specifies the defining relations for all generators as opposed to the first realization. This is important for the construction of a universal R-matrix of the so-called Yangian double.

In order to understand the approach towards this new realization of the Yangian, let us first get some inspiration from ordinary Lie algebras.

\paragraph{Semisimple Lie algebras.}
Due to Serre, every finite dimensional semisimple Lie algebra can be represented in terms of a Chevalley basis of generators. More explicitly, an $n\times n$ Cartan matrix $A=(a_{ij})$ and a set of $3n$ generators $\{X^\pm_i,H_i\}_{i=1}^n$ which satisfy the Serre relations, uniquely define a semisimple Lie algebra $\alg{g}$ of rank $n$.
The generators obey the commutation relations (here $\comm{\cdot}{\cdot}$ denotes the Lie bracket)
\begin{align}
\comm{H_i}{H_j}&=0,
&
\comm{H_i}{X^{\pm}_j}&=\pm a_{ij} X^{\pm}_j,
&
\comm{X^+_i}{X^-_j}=\delta_{ij}H_j,
\end{align}
as well as the Serre relations
\begin{align}
i\neq j:\qquad \text{ad}(X^\pm_i)^{1-a_{ij}}(X^\pm_j)=\comm{X^\pm_i}{\comm{X^\pm_i}{\dots\comm{X^\pm_i}{X^\pm_j}}}=0.
\end{align}
Note that the $H_i$ generate a Cartan subalgebra of $\alg{g}$. The simplest example with $n=1$ is a one-dimensional Cartan matrix (element) $A=a_{11}=2$ such that the above relations yield the well known commutation relations of $\alg{sl}(2)$:
\begin{align}
\comm{H}{X^\pm}&=\pm 2X^\pm,
&
\comm{X^+}{X^-}=H.
\end{align}
One example for infinite dimensional generalizations of semisimple Lie algebras generated in this way are the Kac--Moody algebras. Another example can be defined as follows.

\paragraph{Chevalley--Serre realization of the Yangian.}
In \cite{Drinfeld:1987sy} Drinfeld introduced a second realization of the Yangian that follows the above Chevalley--Serre pattern.

\realization{Second Realization}{
The algebra $\alg{c}$ defined in the following way is isomorphic to $Y[\alg{g}]$.
Given a simple Lie algebra $\alg{g}$ with inner product $(\cdot,\cdot)$,
the associative algebra $\alg{c}$ with generators $x^\pm_{ik}$ and $h_{ik}$ is defined by the relations
(here $\comm{\cdot}{\cdot}$ denotes the commutator in $\alg{c}$)
\begin{align}
\comm{h_{ik}}{h_{jl}}&=0,
&
\comm{h_{i0}}{x^\pm_{jl}}&=\pm a_{ij}x^\pm_{jl},
&
\comm{x^+_{ik}}{x^-_{jl}}&=\delta_{ij} h_{i,k+l}
\end{align}
and 
\begin{align} 
\comm{h_{i,k+1}}{x_{jl}^\pm}-\comm{h_{ik}}{x^\pm_{j,l+1}}&=\pm \half a_{ij}(h_{ik}x^\pm_{jl}+x_{jl}^\pm h_{ik}),\label{eq:Beq1}\\
\comm{x_{i,k+1}^\pm}{x^\pm_{jl}}-\comm{x_{ik}^\pm}{x_{j,l+1}^\pm}&=\pm \half a_{ij}(x_{ik}^\pm x_{jl}^\pm+x_{jl}^\pm x_{ik}^\pm)\label{eq:Beq2},
\end{align}
as well as
\begin{equation}
i\neq j,\,\,m=1-a_{ij}\quad\Rightarrow\quad \mathrm{Sym}_{\{k\}}[x_{ik_1}^\pm,[x_{ik_2}^\pm,\dots [x_{ik_{m}}^\pm,x_{jl}^\pm]]\dots]=0.
\end{equation}
Here $A=(a_{ij})$ denotes the Cartan matrix of $\alg{g}$.
 The indices $i,j$ run over $1,\dots,\mathrm{rank}(\alg{g})$ and $k,l=1,2,\dots$. Furthermore $\mathrm{Sym}_{\{k\}}$ denotes symmetrization in $k_1,\dots, k_m$ with weight $1$.
 }
\medskip

\noindent Let $\{H_i,X_i^\pm\}$ denote a Chevalley--Serre basis of the Lie algebra $\alg{g}$ with $\widehat H_i$ and $\widehat X_i^{\pm}$ representing the level-one generators introduced in the context of the first realization.
Then Drinfel'd's isomorphism $\varphi$ between the Yangian and the algebra $\alg{c}$ takes the form
\begin{align}
\varphi(H_i)&=h_{i0},
&
\varphi(X^+_i)&=x_{i0}^+,
&
\varphi(X^-_i)&=x_{i0}^-,\\
\varphi(\widehat H_i)&=h_{i1}+\varphi(v_i),
&
\varphi(\widehat X^+_i)&=x_{i1}^++\varphi(w_i),
&
\varphi(\widehat X^-_i)&=x_{i1}^-+\varphi(z_i),
\end{align}
with
\begin{align}
v_i&=+\sfrac{1}{4}\sum_\alpha (\alpha,\alpha_i)(e_\alpha e_{-\alpha}+e_{-\alpha}e_\alpha)-\half H_i^2,
\\
w_i&=+\sfrac{1}{4}\sum_\alpha (\comm{X^+_i}{e_\alpha}e_{-\alpha}+e_{-\alpha}\comm{X^+_i}{e_\alpha})-\sfrac{1}{4}(X_i^+H_i+H_iX_i^+),
\\
z_i&=-\sfrac{1}{4}\sum_\alpha (\comm{X^-_i}{e_{-\alpha}}e_\alpha+e_\alpha \comm{X^-_i}{e_{-\alpha}})-\sfrac{1}{4}(X_i^-H_i+H_iX_i^-).
\end{align}
Here $\alpha$ runs over all positive roots and the $e$'s denote the generators of the Cartan-Weyl basis.
 
Drinfel'd furthermore noted that if the right hand side of \eqref{eq:Beq1} and \eqref{eq:Beq2} is set to zero (which corresponds to the classical limit), then the algebra $\alg{c}$ is isomorphic to the universal enveloping algebra $U[\alg{g}[u]]$ with an isomorphism of the structure
\begin{align}
h_{ik}&\mapsto H_i u^k,
&
x_{ik}^+ &\mapsto X^+_i u^k,
&
x_{ik}^-&\mapsto X^-_i u^k.
\end{align}

An explicit expression for the coproduct of the second realization is not known. The boost autmorphism in this realization is given by \cite{Chari:1994pz}
\begin{align}
\boo_u(h_{i,r})&=\sum_{s=0}^r
{r \choose s}
 u^{r-s}\,h_{i,s},
&
\boo_u(x^\pm_{i,r})&=
\sum_{s=0}^r{r \choose s}
u^{r-s}\,x_{i,s}^\pm.
\end{align}

Notably, one may modify the above definition of the Yangian employing only a finite number of the generators of the second realization, a result due to Levendorski$\mathrm{\check \imath}$ \cite{1993JGP....12....1L}.
\subsubsection{Third Realization}
\label{sec:RTT}

We will now consider a third realization of the Yangian that was implicitly studied by the Leningrad school \cite{Takhtajan:1979iv} before Drinfel'd's seminal papers. Drinfel'd himself made the connection to the earlier work explicit: 
\bigskip
\begin{center}
\begin{minipage}{14cm}\small{\it
``Finally, I am going to mention a realization of $Y[\alg{g}]$ which is often useful and which appeared much earlier than the general definition of $Y[\alg{g}]$.''}
V.\ Drinfel'd 1986 \cite{Drinfeld:1986in}
\end{minipage}
\end{center}
\bigskip
In particular, this realization establishes the connection to the fundamental equations underlying the algebraic Bethe Ansatz \cite{Fedor}.

\realization{RTT Realization}{
Fix a nontrivial irreducible representation of the Yangian $\rho:Y[\alg{g}]\to \mathrm{Mat}(n,\mathbb{C})$. Let furthermore $\rop(u)=(\rho\otimes\rho)(\mathcal{R}(u))$, where $\mathcal{R}(u)$ denotes the rational solution to the Yang--Baxter equation given in \eqref{eq:uniqueformal}. Define a Hopf algebra $\alg{a}_\rho$ by the RTT-relations%
\begin{equation}\label{eq:RTT}
\rop_{12}(u-v) \mon_1(u)\mon_2(v)=\mon_2(v)\mon_1(u) \rop_{12}(u-v).
\end{equation}
Here, denoting the $n\times n$ identity matrix by $\idop$, we have
\begin{align}
\mon_1(u)&=\mon(u)\otimes\idop,
&
\mon_2(v)&=\idop\otimes\mon(v),
\end{align}
and with $\mon(u)=\sum_{\alpha\beta}t^\alpha{}_\beta(u) E_\alpha{}^\beta$ and $1\leq \alpha,\beta\leq N$
 the Laurent expansion of the matrix elements $t^\alpha{}_\beta(u)$ takes the form
\begin{equation}\label{eq:tseries}
t^\alpha{}_\beta(u)=\delta^\alpha_\beta+\sum_{k=1}^\infty \frac{(t^{(k)})^\alpha{}_\beta}{u^k}.
\end{equation}
The Hopf algebra $\alg{a}_\rho$ is generated by the operators $(t^{(k)})^\alpha{}_\beta$ with $k=1,2,\dots$ and the coproduct
\begin{equation}
\Delta(t^\alpha{}_\beta(u))=\sum_\gamma t^\alpha{}_ \gamma(u)\otimes t^\gamma{}_\beta(u).
\end{equation}
One has an epimorphism (surjection) $\alg{a}_\rho\to Y[\alg{g}]$ defined by $T(u)\mapsto (\idop\otimes \rho)(\mathcal{R}(u))$, where the expansion of $\mathcal{R}$ in terms of the generators in the first realization was given by \eqref{eq:expR}:
\begin{equation}
\log \mathcal{R}(u)=\frac{1}{u}\levz_a\otimes\levz_a+\frac{1}{u^2}(\levo_a\otimes\levz_a-\levz_a\otimes \levo_a)+\mathcal{O}\Big(\frac{1}{u^3}\Big).
\end{equation}
To obtain $Y[\alg{g}]$ from $\alg{a}_\rho$ (i.e.\ to define a bijection), one generically has to add an auxiliary relation of the form 
\begin{equation}\label{eq:cmap}
c(u)=1,
\end{equation}
where
 $\Delta(c(u))=c(u)\otimes c(u)$ and such that $\comm{a}{c(u)}=0$,
for all $a\in \alg{a}_\rho$.
}
\medskip

\begin{figure}
\begin{center}
\includegraphicsbox{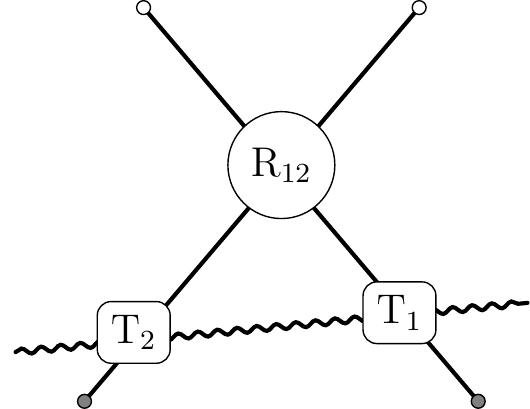}
$=$
\includegraphicsbox{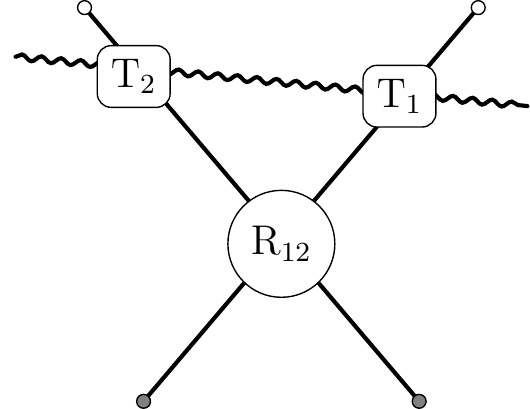}
\end{center}
\caption{Illustration of the RTT-relations.}
\label{fig:RTT}
\end{figure}

%
\noindent Note that this realization makes explicit reference to a representation $\rho$ from the start as opposed to the previous two realizations.
The above RTT-relations \eqref{eq:RTT}, see \figref{fig:RTT}, follow from the Yang--Baxter equation \eqref{eq:YBE} if we identify $\rop_{i3}(u)$ with $\mon_i(u)\equiv \mon_{i3}(u)$ and set $u\equiv u_1$, $v\equiv u_2$ and $u_3\equiv 0$. The RTT-relations relate the products $\mon_1(u)\mon_2(v)$ and $\mon_2(v)\mon_1(u)$ to each other and can thus be considered as generalized commutation relations defining the operator $\mon$ based on a solution $\rop$ of the Yang--Baxter equation.
\paragraph{Example 1.}
Let us consider the most studied example of this realization, namely the case of $\alg{g}=\alg{gl}(N)$ with fundamental generators $\rho(\levz^\alpha{}_\beta)=E^\alpha{}_\beta$ and Yang's R-matrix:
\begin{equation}
\rop(u-v)=\idop+\frac{\permop}{u-v},
\end{equation}
where $\permop=\sum_{\alpha,\beta=1}^N E^\alpha{}_\beta\otimes E_\alpha{}^\beta$ again denotes the permutation operator alias the quadratic tensor Casimir operator. 
We closely follow the lines of \cite{Molev:1994rs} which contains an extensive discussion of the RTT-realization for $\alg{gl}(N)$.
We expand $\mon(u)=\sum_{\alpha\beta} t^\alpha{}_\beta(u) E_\alpha{}^\beta$ as well as \eqref{eq:RTT} using that
\begin{align}
(\mon(u)\otimes \idop) (\idop\otimes \mon(v))&=\sum_{\alpha,\beta,\gamma,\delta} t^\alpha{}_\beta(u)t^\gamma{}_\delta(v)E_\alpha{}^\beta\otimes E_\gamma{}^\delta,
 \\
(\idop\otimes \mon(v))(\mon(u)\otimes \idop) &=\sum_{\alpha,\beta,\gamma,\delta} t^\gamma{}_\delta(v)t^\alpha{}_\beta(u) E_\alpha{}^\beta\otimes E_\gamma{}^\delta.
\end{align}
Applying both sides of the RTT-relations to a basis vector $e_\beta\otimes e_\delta\in\mathbb{C}^N\otimes \mathbb{C}^N$ one finds on the left hand side 
\begin{equation}
\sum_{\alpha,\gamma}t^\alpha{}_\beta(u)t^\gamma{}_\delta(v)e_\alpha\otimes e_\gamma-\frac{1}{u-v}\sum_{\alpha,\gamma}t^\alpha{}_\beta(u)t^\gamma{}_\delta(v)e_\gamma\otimes e_\alpha,
\end{equation}
and on the right hand side
\begin{equation}
\sum_{\alpha,\gamma}t^\gamma{}_\delta(v)t^\alpha{}_\beta(u) e_\alpha\otimes e_\gamma-\frac{1}{u-v}\sum_{\alpha,\gamma}t^\gamma{}_\beta(v)t^\alpha{}_\delta(u) e_\alpha\otimes e_\gamma.
\end{equation}
Multiplication by $u-v$ and equating the coefficients of independent basis elements $e_\alpha\otimes e_\gamma$ yields
\begin{equation}
(u-v)\comm{t^\alpha{}_\beta(u)}{t^\gamma{}_\delta(v)}=t^\gamma{}_\beta(u)t^\alpha{}_\delta(v)-t^\gamma{}_\beta(v)t^\alpha{}_\delta(u).
\end{equation}
Expanding as in \eqref{eq:tseries} then gives
\begin{equation}\label{eq:seconddefining}
\comm{(t^{(r+1)})^\alpha{}_\beta}{(t^{(s)})^\gamma{}_\delta}-\comm{(t^{(r)})^\alpha{}_\beta}{(t^{(s+1)})^\gamma{}_\delta}=(t^{(r)})^\gamma{}_\beta (t^{(s)})^\alpha{}_\delta-(t^{(s)})^\gamma{}_\beta(t^{(r)})^\alpha{}_\delta,
\end{equation}
for $r,s=0,1,\dots$ and $(t^{(0)})^\alpha{}_\beta=\delta^\alpha_\beta$. The relations \eqref{eq:seconddefining} may be taken as an alternative definition of the Yangian algebra spanned by $(t^{(r)})^\alpha{}_\beta$. In particular, one typically has the following relation to the Yangian generators in the first realization:%
\footnote{See \secref{sec:lax} for an expansion of the monodromy $\mon(u)$ in the context of spin chains.}
\begin{align}
(t^{(1)})^\alpha{}_\beta&\simeq \levz^a\, (t^a)^\alpha{}_\beta
&
(t^{(2)})^\alpha{}_\beta&\simeq\levo^a \,(t^a)^\alpha{}_\beta+\dots,
\end{align}
where the dots stand for lower-level generators or the identity, cf.\ \cite{Dolan:2004ys}.


\paragraph{Example 2.}
Notably, in the above example we did not require the auxiliary map $c(u)$ of \eqref{eq:cmap}. Let us thus also 
consider another example with $\alg{g}=\alg{sl}(N)$ where this map is required, c.f.\ \cite{Drinfeld:1987sy}. Then $Y[\alg{g}]$ is isomorphic to the algebra $\alg{a}_\rho$ defined by the relations
\begin{equation}
\Big(\idop+\frac{\permop}{u-v}\Big)\mon_1(u)\mon_2(v)=\mon_2(v)\mon_1(u) \Big(\idop+\frac{\permop}{u-v}\Big),
\end{equation}
and the auxiliary constraint equation
\begin{equation}\label{eq:auxrel}
c(u)=\mathrm{det}_\text{q}\, \mon(u)=1.
\end{equation}
Here $\permop$ denotes again the permutation operator and the so-called quantum determinant is defined as \cite{Kulish:1981bi}
\begin{equation}
\mathrm{det}_\text{q}\,\mon(u)=\,\,\sum_{\mathclap{{\mathrm{Perm}(\alpha_1,\dots,\alpha_n)}}}\,\, \sgn(\alpha_1,\dots,\alpha_n) t^1{}_{\alpha_1}\Big(u+\frac{n-1}{2}\Big)t^2{}_{\alpha_2}\Big(u+\frac{n-3}{2}\Big)\dots t^n{}_{\alpha_n}\Big(u+\frac{1-n}{2}\Big),
\end{equation}
where the sum runs over permutations of $(\alpha_1,\dots,\alpha_n)$ with values in $1,\dots,n$. Note that here $\rop(u)$ takes again the form of Yang's R-matrix. In the above sense, the Yangian algebra for $\alg{sl}(N)$ (Example 2) is given by the Yangian of $\alg{gl}(N)$ (Example 1) modulo the quantum determinant relation \eqref{eq:auxrel}.

\section{Quantum Nonlocal Charges and Yangian Symmetry in 2d Field Theory}

In this section we will rediscover some of the concepts learned about Yangian symmetry in the last chapter. In particular, we will discuss Drinfel'd's first realization.

\subsection{Quantum Nonlocal Charges alias Yangian Symmetry}
\label{sec:QuantumCharges}


Let us start by noting a crucial difference to the case of classical charges considered in \secref{sec:classint}. 
In a quantum field theory, the product of two operators $\mathcal{O}_1(x)\mathcal{O}_2(y)$ is typically divergent in the limit $x\to y$.
That this is a priori a problem in the context of nonlocal symmetries becomes immediately clear when looking at the classical bilocal current:%
\footnote{We assume that the classical conservation of the current is not broken by quantum anomalies. The breaking of symmetries at the quantum level may occur if the symmetry is a symmetry of the action but not of the measure of the path integral. }
\begin{equation}
 (\widehat j_\text{classical})_{a}^\mu(x)=\epsilon^{\mu\nu}j_{a \nu}(x)-\half f_{abc}\, \int_{-\infty}^x \dd y\,j_{b}^\mu(x)j_{c}^0(y).
\end{equation}
The current contains the product $j_{b}^\mu(x)j_{c}^0(y)$ and since $y$ is integrated up to $x$, the problem is apparent. In order to get control over the divergencies, it is useful to employ the so-called \emph{point-splitting} regularization, i.e.\ to split the point $x$ into two points $x$ and $x-\delta$. Then the short-distance singularities of the product $\mathcal{O}_1(x)\mathcal{O}_2(x-\delta)$ can be extracted as the coefficients in the expansion around $\delta=0$. Below we will use this point-splitting regularization in order to define a quantum bilocal current, but first we have to understand the singularities of the current product a bit better.

In general, the question for the behavior of the product of currents in the limit $x\to y$ is addressed by the \emph{operator product expansion} (OPE) which takes the form
\begin{equation}\label{eq:OPEgeneral}
f_{abc}j_\mu^b(x)j_\nu^c(0)=\sum_i c_{\mu\nu}^{(i)}(x) \mathcal{O}_a^{(i)}(0).
\end{equation}
Here the $f_{abc}$ denote again the structure constants of an internal semi-simple Lie algebra symmetry $\alg{g}$ with level-zero charges $\levz_a$ induced by the local currents which obey
\begin{align}\label{eq:chargealg}
\comm{\levz_a}{\levz_b}&=\norm f_{abc}\levz_c.
&
f_{abc}f_{bcd}&=-\cas \delta_{ad}.
\end{align}
We have introduced a (possibly coupling-dependent) normalization $\norm$ and the adjoint Casimir~$\cas$.
Note that understanding the OPE also furnishes a quantum analogue of the classical flatness condition with a proper normal ordering prescription:%
    \footnote{Cf.\ the appendix of \cite{Hauer:1997ig}.}
\begin{equation}
\partial_0 j_1^a-\partial_1 j_0^a +f_{abc} :j_0^b j_1^c:=0.
\end{equation}
\paragraph{L\"uscher's theorem.}
In general, the OPE can be studied by exploiting the fact that both sides of equation \eqref{eq:OPEgeneral} have to obey the same symmetries and carry the same quantum numbers. With the aim to quantize the definition of the bilocal current $\widehat j_\mu$, we follow \cite{Luscher:1977uq,Bernard:1990jw,Hauer:1997ig} and try to understand what can be said about the operator product \eqref{eq:OPEgeneral}  if one makes the following set of assumptions for the two-dimensional quantum field theory under consideration:
\begin{itemize}
\item The theory  is renormalizable (to have a well-defined OPE) and asymptotically free (which determines the scaling behavior of $c_{\mu\nu}^{(i)}(x)$). 
\item The theory has a local conserved 
current $j_\mu^a$.
\item There is only one operator of dimension smaller than $2$ that transforms under the adjoint representation of $\alg{g}$, namely the conserved current $j_\mu^a(x)$ (and derivatives thereof).
\item Both sides of \eqref{eq:OPEgeneral} obey C, P, T and Lorentz symmetry.%
\footnote{Here we assume that the theory has a C operation. See \cite{Hauer:1997ig} for a discussion of its properties in this context.}
\item The current commutes with itself when evaluated at different points with spacelike separation (locality).
\end{itemize}
Under these assumptions it was shown that the most general form of the above OPE is given by \cite{Luscher:1977uq,Bernard:1990jw,Hauer:1997ig}
\begin{equation}\label{eq:opeform}
f_{abc}j_\mu^b(x)j_\nu^c(0)=c_{\mu\nu}^\rho(x) j_\rho^a(0)+d_{\mu\nu}^{\sigma\rho}(x)\partial_\sigma j_\rho^a(0),
\end{equation}
with the below specifications on the OPE coefficient functions. 
In order to see that this behavior is compatible with the charge algebra \eqref{eq:chargealg}, one considers the equal time commutator:
\begin{equation}
f_{abc}\comm{j_\mu^b(x)}{j_\nu^c(0)}_\text{e.t.}=\lim_{\epsilon\to 0} f_{abc}\big[j_\mu^b(x,-i\epsilon)j_\nu^c(0)-j_\mu^b(x,i\epsilon)j_\nu^c(0)\big].
\end{equation}
Evaluating the OPE at $-x^2-\epsilon^2$ and with the normalization $\norm$ entering by \cite{Hauer:1997ig}
\begin{align}
f_{abc}\comm{j_0^b(x)}{j_\mu^c(0)}_\text{e.t.}&=-\norm \cas \delta(x) j_{a,\mu}(0),
\end{align}
one finds an expansion that sometimes goes under the name \emph{L\"uscher's theorem} \cite{Luscher:1977uq,Bernard:1990jw,Hauer:1997ig}:%
\footnote{L\"uscher derived this form of the current product for the case of the non-linear sigma model \cite{Luscher:1977uq}, while Bernard obtained similar constraints for the massive current algebras in two dimensions \cite{Bernard:1990jw}. A nice general discussion is given in \cite{Hauer:1997ig}.}
\begin{align}
c_{\mu\nu}^\rho(x)&=
a_1(x)\frac{\eta_{\mu\nu}x^\rho}{x^2}
+a_2(x) \frac{x_{(\mu}\delta_{\nu)}^\rho}{x^2}+a_3(x)\frac{x_\mu x_\nu x^\rho}{x^4},
\\
d_{\mu\nu}^{\sigma\rho}(x)&=
\frac{b_1(x)}{4}\frac{x_{[\mu} x^\rho \delta_{\nu]}^\sigma+x_{[\mu} x^\sigma\delta_{\nu]}^\rho}{x^2}
+\frac{b_2(x)}{4}\delta_{[\mu}^\sigma\delta_{\nu]}^\rho
+\frac{x^\sigma}{2} c_{\mu\nu}^\rho(x).
\end{align}
Here all coefficient functions $a_k$, $b_k$ depend on $x$ only via the Lorentz-invariant $x^2$ and are of order $\order{\abs{x}^{-0}}$ due to the asymptotic freedom of the theory.%
\footnote{The notation $\order{\abs{x}^{-0}}$ denotes possible logarithmic terms.} 
Furthermore the parameter functions depend on the normalization $\norm$, the Casimir $\cas$ and on one model-dependent function $\xi(x)$ which is a function of $\log(\mu^2 x^2)$, where $\mu$ denotes a mass scale. Using the above assumptions one can derive many constraints on the parameter functions. Current conservation translated into $\partial^\mu c_{\mu\nu}^\rho=0$ and $\partial^\mu d_{\mu\nu}^{\rho \sigma}=0$ for example implies several differential equations \cite{Luscher:1977uq,Hauer:1997ig,Abdalla:1991vua}, e.g.\ at $t=0$:
\begin{equation}\label{eq:conscoeff}
a_1(x)=-2x \frac{\dd }{\dd x}\big[b_1(x)-b_2(x)\big].
\end{equation}
Evaluating all these constraints yields the relations \cite{Luscher:1977uq,Hauer:1997ig}
\begin{align}
a_1(x)&=-2 \,\dot b_1(x)-b_1(x)+\frac{\norm\cas}{2\pi},
&
a_2(x)&=b_1(x)-\frac{\norm\cas }{2\pi},
&
a_3(x)&=2\dot b_1(x)-2b_1(x),
\end{align}
as well as
\begin{align}
b_1&=-\dot \xi(x)+\frac{\norm\cas}{2\pi},
&
b_2&=\dot \xi(x)+\xi(x)-\frac{\norm\cas}{2\pi},
\end{align}
where the dot denotes the derivative with respect to $\log(\mu^2 x^2)$.

%
\paragraph{Quantum bilocal current.}
We may now introduce the point-split version of the nonlocal current as
\begin{equation}
\widehat j_a^\mu(t,x|\delta)=Z(\delta) \epsilon^{\mu\nu}j_{a\nu}(t,x)-\half f_{abc}j_b^\mu(x) \int_{-\infty}^{x-\delta} \dd y\, j_c^0(y),
\end{equation}
where $Z(\delta)$ denotes a renormalization constant that has to be determined.
The \emph{quantized bilocal current} is defined as the limit \cite{Luscher:1977uq,Bernard:1990jw}
\begin{equation}\label{eq:limcurr}
\widehat j_{a}^\mu(t,x)=\lim_{\delta\to0} \,\widehat j_{a}^\mu(t,x|\delta).
\end{equation}
This current is finite and conserved only for a particular choice of the renormalization constant $Z(\delta)$.
To understand the divergence, we evaluate the relevant contributions to the product $j_0 j_0$ using \eqref{eq:opeform}:
\begin{align}
c_{00}^\rho(x)&=a_1(x)\frac{x^\rho}{x^2}+2a_2(x) \frac{ t \delta_0^\rho}{x^2}+a_3(x)\frac{t^2x^\rho}{x^4}
\xrightarrow{t\to 0}a_1(x)\delta^\rho_1\frac{1}{x},
\nonumber\\
d_{00}^{\sigma\rho}(x)&=\half x^\sigma c_{00}^\rho(x)
\xrightarrow{t\to 0}a_1(x)\delta^\rho_1\delta^\sigma_1.
\end{align}
This indicates that the origin of the divergence for $x\to 0$ lies in the term proportional to $a_1(x)$ in the first line.
The bilocal level-one charge is given by 
\begin{equation}
\levo(\delta)=\int_{-\infty}^\infty \dd x\,\widehat j^0(t,x|\delta),
\end{equation}
and the divergent part has the form (cf.\ \cite{Abdalla:1991vua})
\begin{equation}
\levo(\delta)\simeq Z(\delta)\int_{-\infty}^\infty \dd y\, j_1(y)-\half \int_{-\infty}^{\infty} \dd x \int_{-\infty}^{x-\delta} \dd y\,\frac{a_1(x-y)}{x-y}j_{1}(x-y).
\end{equation}
Now we may use \eqref{eq:conscoeff} to rewrite this as
\begin{align}
\levo(\delta)
&\simeq Z(\delta)\int_{-\infty}^\infty \dd y\, j_1(y)+\int_{-\infty}^{\infty} \dd x \,j_{1}(x)\int_{-\infty}^{x-\delta} \dd y\,\frac{\dd }{\dd y}\big[b_1(x-y)-b_2(x-y)\big]
\nonumber\\
&\simeq Z(\delta)\int_{-\infty}^\infty \dd y\, j_1(y)+\int_{-\infty}^{\infty} \dd x \,j_{1}(x)\big[b_1(x-y)-b_2(x-y)\big]_{y=-\infty}^{y=x-\delta}
\nonumber\\
&\simeq Z(\delta)\int_{-\infty}^\infty \dd y\, j_1(y)+\big[b_1(\delta)-b_2(\delta)\big]\int_{-\infty}^{\infty} \dd x \,j_{1}(x).
\end{align}
Note that $b_1(\delta)$ and $b_2(\delta)$ are divergent for $\delta\to 0$ and that the terms proportional to $+b_1(x+\infty)-b_2(x+\infty)$ should be finite by the conditions on the conserved local current at the boundaries of space. 
The renormalization constant $Z(\delta)$ is determined by requiring that the bilocal current is finite in the limit $\delta\to 0$ and thus
\begin{equation}
Z(\delta)\equiv b_2(\delta)-b_1(\delta)=2\dot\xi(\delta)+\xi(\delta)-\frac{\norm\cas}{\pi}.
\end{equation}
Similarly one may show that the quantum bilocal charge induced by the current \eqref{eq:limcurr} is \emph{conserved} under the above assumptions \cite{Luscher:1977uq,Hauer:1997ig}. 
\paragraph{Quantum monodromy and Lax formulation.} 
As seen in the classical case, the existence of nonlocal charges in principle allows to define a conserved generating function.
For a large class of models, in \cite{deVega:1984wk} the quantum analogue of the monodromy matrix $\mon(u)$ was constructed directly on asymptotic particle states under the following assumptions:
\begin{itemize}
\item A quantum operator $\mon(u)$ exists and is conserved.
\item $\mon(u)$ satisfies a quantum factorization principle (the RTT relations).
\item The discrete parity and time-reversal symmetries are realized in the quantum theory.
\end{itemize}
Since the monodromy matrix $\mon(u)$ provides a generating function for the nonlocal conserved charges, it furnishes an alternative way to study the symmetry constraints on observables such as the scattering matrix. However, we will not discuss this in more detail here.


\paragraph{Chiral Gross--Neveu Model.}

The chiral Gross--Neveu model represents a renormalizable and asymptotically free theory with the symmetry properties assumed above. We may thus apply the quantization procedure to the bilocal current. In order to explicitly compute the OPE expansion, one may insert the current product $f_{abc}j_\mu^b j_\nu^c$ into correlation functions that can be perturbatively evaluated and regularized by ordinary field theory methods.
The model-dependent function $\xi(x)$ is given by \cite{Hauer:1997ig}
\begin{equation}
\xi(x)=\sfrac{\norm\cas}{2\pi}\log(\mu^2 x^2),
\end{equation}
and thus the renormalization constant evaluates to
\begin{equation}
Z(\delta)=\frac{\norm\cas}{2\pi}\log(\mu^2\delta^2)+\mathcal{O}(\abs{\delta}^{1-0})
\simeq g^2\frac{N}{2\pi}\log(\mu^2 \delta^2)+\mathcal{O}(\abs{\delta}^{1-0}).
\end{equation}
Here we assume the normalization to scale as $\norm\simeq g^2$ and the generators of $\alg{su}(N)$ to be normalized such that the adjoint Casimir goes as $\cas\simeq N$.
The quantum currents for the chiral Gross--Neveu model thus take the form given in \cite{deVega:1984wk}:
\begin{align}
\levz_{a}&=\int_{-\infty}^\infty \dd x\, j^0_{a}(x) ,
&
\levo_{a}&=\lim_{\delta\to 0}\bigg[Z(\delta)\int_{-\infty}^\infty \dd x\,j^1_{a}(x)
-\sfrac{1}{2}\int_{-\infty}^\infty\int_{-\infty}^{x-\delta} \dd x \,\dd y\, 
\comm{ j^0(x)}{ j^0(y)}_a 
\bigg].
\end{align}
Above the generated mass scale $\mu=\half me^{\gamma_\text{E}}$ is related to the mass $m$ of the fundamental fermions of the chiral Gross--Neveu model and $\gamma_\text{E}$ denotes the Euler constant.


\subsection{Boost Automorphism}
\label{sec:boostauto}

Let us understand in some more detail how the abstract mathematical concepts underlying the Yangian algebra are realized in physical theories. We will see that this Hopf algebra in fact circumvents some naive expectations on the symmetry structure of physical observables:

\quoting{%
We prove a new theorem on the impossibility of combining spacetime and internal symmetries in any but a trivial way.}
{S.\ Coleman and J.\ Mandula 1967 \cite{Coleman:1967ad}%
\addtocounter{footnote}{1}\footnotemark[\thefootnote]}
\footnotetext{Apparently the assumptions of their famous theorem are not satisfied here. Nevertheless the Coleman--Mandula theorem shows that it is by no means obvious that spacetime and internal symmetries can be nontrivially related.}

Under a finite Lorentz transformation, the local current transforms according to
\begin{equation}
U(\Lambda)j^\mu(x)U(\Lambda)^{-1}=\Lambda^\mu{}_\nu \,j^\nu(\Lambda x),
\end{equation}
with
\begin{equation}
(\Lambda^\mu{}_\nu)=
\begin{pmatrix}
\cosh (u)&\sinh(u)
\\
\sinh(u)&\cosh(u)
\end{pmatrix}.
\end{equation}
From this one can read off the following vector transformation rule under the boost generator $\boo=\sfrac{\dd }{\dd u} U(\Lambda)|_{u=0}$:
\begin{equation}\label{eq:Lorentzcurrent}
\comm{\boo}{j_a^\mu(x)}= \epsilon_{\rho\sigma}x^\rho \partial^\sigma j_a^\mu(x)+\epsilon^{\mu\sigma}j_{a,\sigma}(x).
\end{equation}
For the transformation of the local charge this straightforwardly implies
\begin{equation}
\comm{\boo}{\levz_a}
=0,
\end{equation}
if $j_a(\pm \infty)=0$.

\begin{figure}
\begin{center}
\includegraphicsbox[scale=.8]{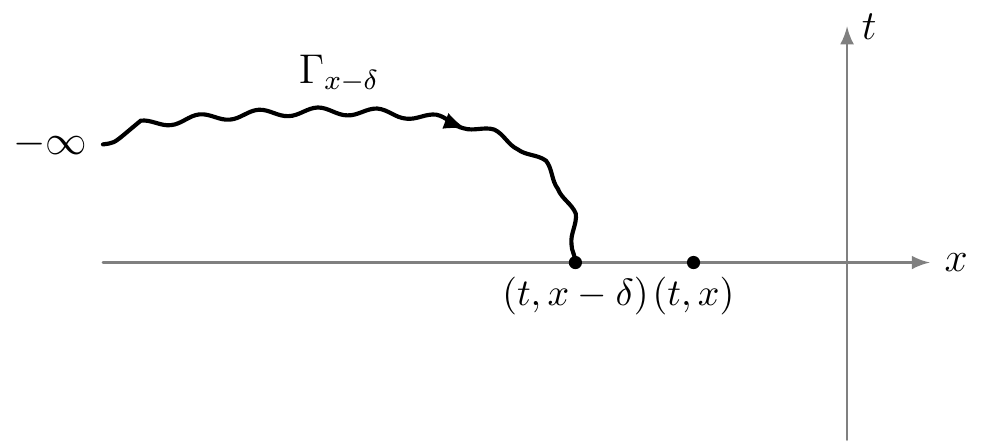}
\end{center}
\caption{Contour $\Gamma_{x-\delta}$ defining the level-one current.}
\label{fig:levonecont}
\end{figure}

\paragraph{Level-one charges and contours.}
In order to understand the transformation behavior of the bilocal level-one current we use a nice geometric argument of \cite{Bernard:1990jw}. 
Note that instead of integrating over the real axis, we may define the bilocal current via integration over a generic contour $\Gamma_{x-\delta}$ that starts at $-\infty$ and ends at $x-\delta$, see \figref{fig:levonecont}. This is possible since the current defines a flat connection and thus the integration is path-independent. 
The bilocal current then takes the form
\begin{equation}
\widehat j_a^\mu(x,t|\delta)=Z(\delta) \epsilon^{\mu\nu}j_{a\nu}(x,t)-\half f_{abc}j_b^\mu(x) \int_{\Gamma_{x-\delta}} \epsilon_{\sigma\rho}\dd y^\rho\, j_c^\sigma(y).
\end{equation}
Here $\epsilon_{\sigma\rho}\dd y^\rho\, j_c^\sigma$ represents the generalization of $\dd y j_c^0$ when going away from $t=0$.

\begin{figure}
\begin{center}
\includegraphicsbox[scale=.8]{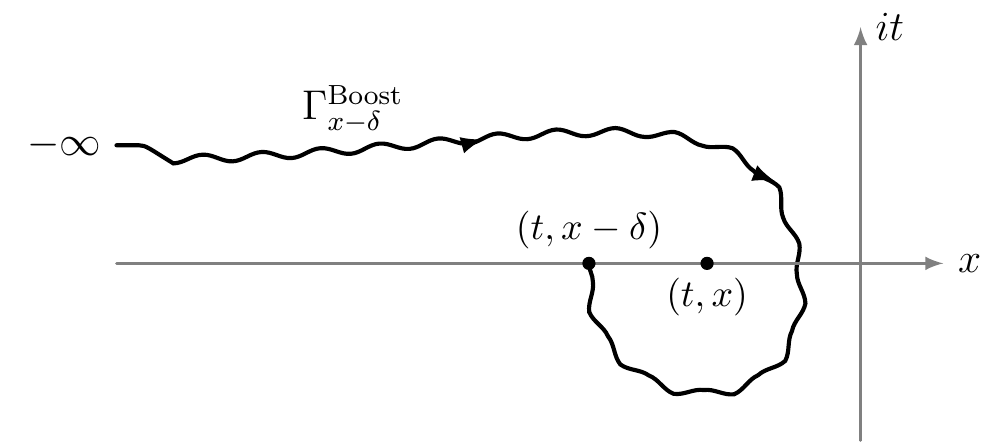}
$=$
\includegraphicsbox[scale=.8]{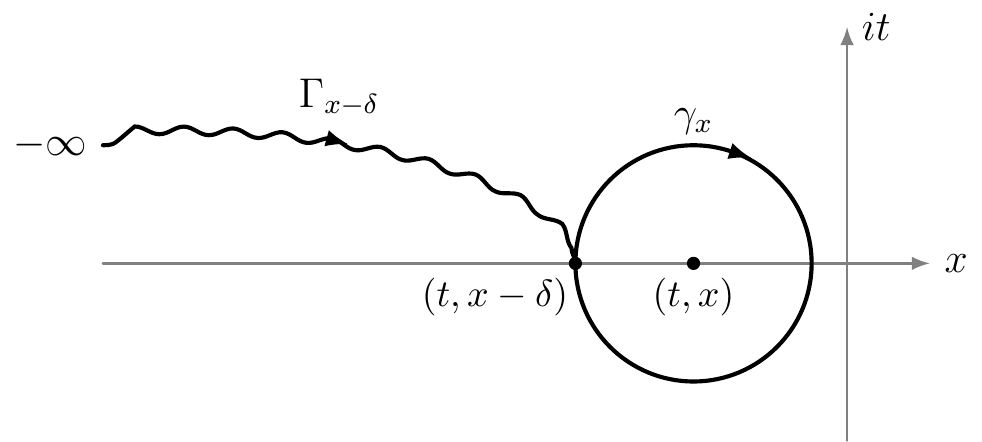}
\end{center}
\caption{Boost transformation of the contour $\Gamma_{x-\delta}$ by a rapidity $u=2\pi i$ corresponding to a Euclidean rotation.}
\label{fig:rotcontour}
\end{figure}

One may now apply a Lorentz boost $\boo_{2\pi i}$ by an imaginary rapidity $2\pi i$ to this nonlocal current. This boost corresponds to a Euclidean rotation by an angle $2\pi$.
As illustrated in \figref{fig:rotcontour}, the rotated current can be written in the original form with a contour $\Gamma_{x-\delta}$ ending at the point $x-\delta$ plus an integral around the point $x$ over a closed contour $\gamma_x$:
\begin{equation}
e^{2\pi i\boo}\,\widehat j_a^\mu(x,t|\delta)\, e^{-2\pi i \boo}=\widehat j_a^\mu(x,t|\delta)+\half f_{abc}\oint_{\gamma_x}\epsilon_{\sigma\rho}\dd y^\rho\, j_b^\mu(x) j_c^\sigma(y).
\end{equation}
We are interested in the implications of this transformation behaviour on the level-one \emph{charges}. Hence, we have to integrate the zero-component of the above expression over $x$. 
Then the last integral picks up the residue of the OPE of the product of currents and we would have to make an analysis similar to the one in \secref{sec:QuantumCharges}, where we evaluated the generic structure of the current OPE. We will not discuss this proof in more detail here but note that Bernard has shown that \cite{Bernard:1990jw}
\begin{equation}
e^{2\pi i\boo}\,\levo_a e^{-2\pi i\boo}=\levo_a -\half \cas \levz_a,
\end{equation}
with $\cas$ representing the quadratic Casimir of $\alg{g}$ in the adjoint representation, i.e.\ $-\delta_{ab}\cas=f_{acd}f_{cdb}$. Comparing with the leading orders of the expansion
\begin{equation}
\boo_{2\pi i}=\exp(2\pi i\boo)=1+2\pi i\boo+\dots,
\end{equation}
one concludes that the conserved charges transform under the boost generator as%
\footnote{In an alternative approach using form factors, the commutation relations with the boost operator were obtained by first determining the commutator of the level-one charge with the energy momentum tensor for a model with $\alg{g}=\alg{sl}(2)$ \cite{LeClair:1991cf}:
\begin{equation}
\comm{\levo^a}{T_{\mu\nu}(x)}=-\frac{\cas}{8\pi i} \big(\epsilon_{\mu \alpha}\partial_\alpha j_\nu^a (x)+\epsilon_{\nu\alpha}\partial_\alpha j_\mu^a(x)\big).
\end{equation}
Integrating the $00$-component one finds the relation \eqref{eq:boowithYang} with the boost generator $\boo=\int \dd x\, x\, T_{00}$ for $t=0$. Note also the paper \cite{Curtright:1992kp} for interesting comments on the boost commutator and the beta-function.
 }
\begin{align}\label{eq:boowithYang}
\comm{\boo}{\levz_a}&=0,
&
\comm{\boo}{\levo_a}&=-\frac{\cas}{4\pi i}\levz_a.
\end{align}
Notably, the nontriviality of the second commutator is a quantum effect (cf.\ \eqref{eq:classicalboost}). It is induced by the pole in the current OPE encircled by the contour in \figref{fig:rotcontour}.
The operator $\boo_u=\exp (u\boo)$ corresponds to a group-like finite boost transformation whereas the generator $\boo$ represents the algebra element whose primitive coproduct follows from the expansion
\begin{equation}
\boo_u\otimes \boo_u=\idop\otimes\idop+u(\boo \otimes \idop+\idop\otimes \boo)+\order{u^2}.
\end{equation}
Hence, the boost transformation couples the internal and spacetime symmetries to each other, which implies that the quantized Yangian algebra is not merely an internal symmetry. It is conceivable that this yields stronger constraints on symmetry invariants than a direct product of independent symmetry algebras.
Note that the above boost tranformation on the conserved charges exactly realizes the boost automorphism defined by Drinfel'd. 
Hence, the internal level-zero and level-one charges $\levz_a$ and $\levo_a$ together with the boost automorphism $\boo_u$ furnish the defining relations of the Yangian algebra in the first realization as given in \secref{sec:firstreal}. 

\bigskip
\begin{center}
\begin{minipage}{14cm}\small{\it
``Thus one concludes that the Yangian must actually be extended to include the Poincar\'e algebra with generators $\boo, \mathrm{P}_\mu$ in order to realize its full implications.''} A.\ LeClair and F.A.\ Smirnov 1991 \cite{LeClair:1991cf}
\end{minipage}
\end{center}
\bigskip


\subsection{Yangian Symmetry and the 2d S-matrix}
\label{sec:smatrix}

In physics, symmetries are typically used as a guiding principle to construct new models and to constrain their observables. Being an infinite-dimensional symmetry algebra, the Yangian has strong implications for the spectrum and dynamics of a theory. In \cite{Belavin:1992en}, Belavin showed for instance that the spectrum of masses of a two-dimensional quantum field theory can be computed via the Yangian symmetry. 
The prime example for the physical application of the Yangian is the scattering matrix of 
massive, relativistic, two-dimensional quantum field theories.
The S-matrix is the operator that relates asymptotic particles to each other. Since particles are defined by representations of symmetries (e.g.\ Poincar\'e symmetry), the scattering matrix is constrained by these symmetries. In a theory with Yangian symmetry, particles also transform in representations of the Yangian, which should thus be a symmetry of the S-matrix, cf.\  e.g.\ \cite{LeClair:1991cf,MacKay:1992br}. 

In order to study scattering processes in two dimensions, it is useful to consider lightcone coordinates defined by
\begin{align}
p^+&=p^0+p^1,
&
p^-&=p^0-p^1,
\end{align}
which can be expressed in terms of the often more convenient rapidities $u$ by the relations
\begin{align}
p^+&=m \exp(+u),
&
p^-&=m\exp(-u).
\end{align}
Here $m$ denotes the particle mass.
The two-dimensional momentum in the original coordinates then takes the form
\begin{equation}
(p^\mu)=
\begin{pmatrix}
m \cosh{u}\\
m\sinh{u}
\end{pmatrix},
\end{equation}
which shows that the transformation $u\to-u$ inverts the direction of the particle's movement. The transformation $u\to i\pi-u$ flips the sign of the particle's energy and thus represents a particle to antiparticle transformation. Importantly, the Lorentz boost acts additively on the variables $u$ (see below), which emphasizes their usefulness in the present context.

\paragraph{Poincar\'e symmetry and scattering states.}
We will now be interested in understanding the impact of symmetries on the scattering in this theory. We consider the Poincar\'e algebra in 1$+$1 dimensions
\begin{align}
\comm{\op{P}^+}{\op{P}^-}&=0,
&
\comm{\boo}{\op{P}^+}&=+\op{P}^+,
&
\comm{\boo}{\op{P}^-}&=-\op{P}^-,
\end{align}
where the single Lorentz transformation in two dimensions is generated by the boost $\boo$ and $\op{P}^\pm$ denotes translations into the lightcone directions.
In order to study the scattering of different particle species moving in one space dimension, we look at asymptotic scattering states. For one single particle of type $\alpha$ this state is denoted by $\ket{\alpha,u}$. In the chiral Gross--Neveu model for instance, it takes the form (with $p_1=p_1(u)$) \cite{deVega:1984wk}
\begin{equation}\label{eq:scatterosc}
\ket{\alpha,u}=(p_1^2+m^2)^{\frac{1}{4}}\,b^\dagger_\alpha(p_1)\ket{0},
\end{equation}
where $b_\alpha$, $b^\dagger_\alpha$ are oscillators that appear in the Fourier decomposition of the fundamental fermions.

Note that the Lorentz boost can be realized on one-particle states (on-shell) as
\begin{equation}
\boo\ket{\alpha,u}=\frac{\partial}{\partial u}\ket{\alpha,u},
\end{equation}
and a finite boost transformation 
$\boo_v=\exp(v\boo)$ acts on a one-particle state by a shift of the rapidity:
\begin{equation}
\boo_v\ket{\alpha,u}=\ket{\alpha,u+v}.
\end{equation}
The energy-momentum generators act on one-particle states as
\begin{align}
\op{P}^+\ket{\alpha,u}&=m\, e^{+u}\ket{\alpha,u},
&
\op{P}^-\ket{\alpha,u}&=m\, e^{-u}\ket{\alpha,u}.
\end{align}
In order to study the scattering of multiple particles, we need a notion of multi-particle scattering states. 
Importantly, the fact that the space is one-dimensional allows to order particles (i.e.\ wave-packets) with respect to their position.  It is thus natural to label the particles $1,\dots,n$ according to their space coordinate $x_1<\dots < x_n$. However, only if the fastest particle of the incoming multi-particle state is on the very left, it can cross all other particle trajectories. Thus, in order to have a nontrivial $n$-particle scattering process, the particle rapidities $u_k$ in the in-state have to have the opposite ordering as compared to the positions, i.e.\ $u_1>\dots > u_n$. After the scattering process, the situation is reversed, and the particles in the out-state with positions $x_1'<\dots < x_n'$ have rapidities ordered as $u_1'<\dots <u_n'$. This motivates to introduce the following ordered multi-particle states with rapidities ordered as $u_1>\dots>u_n$:%
\footnote{The different ordering of positions corresponds to different orders of the operators generating the individual particles from the vaccuum, cf.\ \eqref{eq:scatterosc}. In fact, the S-matrix theory can be formulated in terms of such operators spanning the so-called Zamolodchikov--Faddeev algebra. }
\begin{align}\label{eq:multistates}
&\ket{\alpha_1,u_1;\dots ;\alpha_n,u_n}_\text{in},
&
&x_1<x_2<\dots<x_n,
\\
&\ket{\beta_1,u_1;\dots ;\beta_n,u_n}_\text{out}.
&
&x_1>x_2>\dots>x_n.
\end{align}
In fact, the scattering matrix in 1$+$1 dimensions is defined as the operator that expresses an out-state in the infinite future, in the in-state basis in the infinite past, or vice versa. 
We assume that the scattering process preserves the number of particles as well as the individual rapidities, as is the case in integrable theories in two dimensions.%
\footnote{Also this can be shown using the Yangian structure of the S-matrix but we do not discuss this here.}
Then the S-matrix acts on the asymptotic states as
\begin{equation}
\ket{\alpha_1,u_1;\dots ;\alpha_n,u_n}_\text{in}=\sop^{\beta_1,\dots,\beta_n}_{\alpha_1,\dots,\alpha_n}(u_1,\dots,u_n)\ket{\beta_1,u_1;\dots ;\beta_n,u_n}_\text{out}.
\end{equation}
Note that the above definition of in- and out-states implies that the coproduct acts differently on the two bases. This serves as a motivation to introduce the notion of \emph{opposite coproduct} for the permuted coproduct \eqref{eq:oppcop}, which enters the below symmetry equation for the S-matrix.
The action of the above Poincar\'e generators on one-particle states generalizes to multi-particle states via the primitive coproduct 
\begin{align}
\Delta(\op{P}^+)&=\op{P}^+\otimes \idop+\idop\otimes \op{P}^+,
&
\Delta(\op{P}^-)&=\op{P}^-\otimes \idop+\idop\otimes \op{P}^-,
&
\Delta(\boo)&=\boo\otimes \idop+\idop\otimes \boo.
\end{align}

\paragraph{Yangian symmetry and scattering states.}
Suppose now that in addition to the above Poincar\'e symmetry, the underlying theory features a Yangian extension $Y[\alg{g}]$ of an internal symmetry algebra $\alg{g}$.
Most of the commutators between the spacetime and internal symmetries vanish, but as seen above, the boost operator has nontrivial commutation relations with the Yangian level-one generators:
\begin{align}\label{eq:booalg}
\comm{\op{P}^\pm}{\levz_a}&=0,
&
\comm{\op{P}^\pm}{\levo_a}&=0,
&
\comm{\boo}{\levz_a}&=0,
&
\comm{\boo}{\levo_a}&=-\cas\frac{\hbar}{4\pi i} \levz_a,.
\end{align}
Thus the Lorentz boost furnishes the boost automorphism of the Yangian algebra. 
The above relations imply that the one-particle states transform in an evaluation representation of the Yangian:
\begin{align}\label{eq:Qsononepart}
\rho_{\hat u}(\levz_a)\ket{u}&=\rho_0(\levz_a)\ket{u},
&
\rho_{\hat u}(\levo_a)\ket{u}&=\hat u\,\rho_0(\levz_a) \ket{u}.
\end{align}
Here $\rho_0$ again denotes a representation of the generators of our symmetry algebra acting on the particle labels $\alpha,\beta,\dots$ (cf.\ \eqref{eq:zerorep}) and $\hat u$ equals the rapidity $u$ up to a constant:
\begin{equation}
\hat u= -\frac{\hbar\cas}{4\pi i} u.
\end{equation}
The compatibility of this representation with the boost commutator in \eqref{eq:booalg} can be seen by evaluating
\begin{equation}
\rho(\comm{\boo}{\levo_a})\ket{u}=\big[\partial_u\rho_0(\levo_a)-\rho_0(\levo_a) \partial_u\big]\ket{u}=-\frac{\hbar\cas}{4\pi i}  \rho_0(\levz_a)\ket{u}.
\end{equation}
or alternatively (cf.\ \eqref{eq:boostauto})%
\footnote{Note that $\boo_v$ acts on an operator by conjugation with $\exp{(v\boo)}$.
}
\begin{equation}
\boo_v (\levo_a)\ket{u}\equiv e^{v\boo}\levo_a e^{-v\boo}\ket{u}=\Big(\levo_a-\frac{\hbar\cas}{4\pi i}  \levz_av \Big)\ket{u},
\end{equation}
where the expansion of $\exp{(v\boo)}$ yields only finitely many terms since $\comm{\boo}{\levz_a}=0$. 
In order to study the implications of the symmetry on the scattering matrix, we note that the conserved charges are time-independent and are thus the same on incoming and outgoing states, e.g.\ for the level-one charge  \cite{Luscher:1977uq}:
\begin{equation}
\levo^{a}=\lim_{t\to -\infty}\levo^a_\text{in}(t)=\lim_{t\to \infty}\levo^a_\text{out}(t).
\end{equation}
The action on multiparticle states is defined by the coproduct. Each particle transforms with a different evaluation parameter $\hat u_i$ such that we find
\begin{align}
\rho_{\hat u}(\levz_a)\ket{u_1,\dots,u_n}_{\substack{\text{in}\\\text{out}}}&=\sum_{k=1}^n \rho_0(\levz_{a,k})\ket{u_1,\dots, u_n}_{\substack{\text{in}\\\text{out}}},
\\
\rho_{\hat u}(\levo_a)\ket{u_1,\dots,u_n}_{\substack{\text{in}\\\text{out}}}
&=\Big(\sum_{k=1}^n \hat u_k\, \rho_0(\levz_{a,k})\pm\half f_{abc} \sum_{1\leq i<j\leq n} \rho_0(\levz_{b,i}) \rho_0(\levz_{c,j})\Big)\ket{u_1,\dots,u_n}_{\substack{\text{in}\\\text{out}}},\nonumber
\end{align}
where we have dropped the particle flavors for simplicity of the expression. The different signs $\pm$ arise from the application of the different coproducts $\Delta$ or $\Delta^\text{op}$ to the out- or in-state, respectively.

\paragraph{Constraints on the S-matrix.}

We would like to understand the two-particle S-matrix, which typically serves as the fundamental building block for integrable scattering matrices in two dimensions:
\begin{equation}\label{eq:twopart}
\ket{\alpha_1,u_1;\alpha_2,u_2}_\text{in}=\sop^{\beta_1,\beta_2}_{\alpha_1,\alpha_2}(u_1,u_2)\ket{\beta_1,u_1;\beta_2,u_2}_\text{out}.
\end{equation}
Lorentz symmetry is the statement $\comm{\boo\otimes \idop+\idop\otimes \,\boo}{\sop(u_1,u_2)}=0$, or explicitly
\begin{equation}
\frac{\partial}{\partial u_1}\sop(u_1,u_2)+\frac{\partial}{\partial u_2}\sop(u_1,u_2)=0,
\end{equation}
which is solved by $\sop(u_1,u_2)=\sop(u_1-u_2)$. This is equivalent to saying that the two-particle S-matrix may only depend on the Mandelstam variable $(p_1+p_2)^2=4 m^2 \cosh^2(\frac{u_1-u_2}{2})$.

Let us emphasize that the scattering matrix is the operator that relates the in- and out-representation in a scattering process:
\begin{equation}
\ket{\ldots}_\text{in}=\sop\ket{\ldots}_\text{out}.
\end{equation}
The out- and in-representations transform under different coproducts,
namely under $\Delta$ and $\Delta^\text{op}$ (cf.\ \eqref{eq:oppcop} and the definition of multi-particle states \eqref{eq:multistates}) and
hence, the internal symmetry of the theory implies that the S-matrix furnishes an intertwiner for the Yangian evaluation modules:%
\footnote{Sometimes the matrix $\check \sop=\permop \sop$ is called S-matrix, for which this condition becomes (see e.g.\ \cite{Delius:2001qh})
\begin{equation}
\rho\big(\boo_{\hat v}\otimes \boo_{\hat u}(\Delta(a))\big)\check\sop(u-v)=\check\sop(u-v)\rho\big(\boo_{\hat u}\otimes \boo_{\hat v}(\Delta(a))\big).
\end{equation}
}
\begin{equation}\label{eq:YangSmat}
\rho\big(\boo_{\hat u}\otimes \boo_{\hat v}(\Delta^\text{op}(a))\big)\sop(u-v)=\sop(u-v)\rho\big(\boo_{\hat u}\otimes \boo_{\hat v}(\Delta(a))\big),
\end{equation}
for all $a\in Y[\alg{g}]$. 
In order to remember the explicit relation of the boost to the evaluation representation, we consider again the two-particle expressions for the level-zero and level-one generators:
\begin{align}
\boo_u\otimes \boo_v (\Delta(\levz_a))
=
&\levz_a\otimes \idop+\idop\otimes \levz_a.
\\
\boo_u\otimes \boo_v (\Delta(\levo_a))
=
&\levo_a\otimes \idop+\idop\otimes \levo_a+u(\levz_a\otimes \idop)+v(\idop\otimes \levz_a)-\half  f_{abc} \levz_b\otimes \levz_c.
\end{align}
Notably, \eqref{eq:YangSmat} represents \eqref{eq:linetheo3} realized on the scattering matrix.
For the level-zero and level-one generators we therefore have via \eqref{eq:theo21} and \eqref{eq:theo22} of Theorem 2, and using the representation \eqref{eq:zerorep}:
\begin{align}\label{}
0&=\comm{\rho(\levz_a)\otimes \idop+\idop\otimes \rho(\levz_a)}{\sop(u_1-u_2)},
\\
0&=(\rho_0\otimes\rho_0)\big(\boo_{\hat u}\otimes \boo_{\hat v}(\Delta^\text{op}(\levo_a))\big)\sop(u-v)-\sop(u-v)(\rho_0\otimes\rho_0)\big(\boo_{\hat u}\otimes \boo_{\hat v}(\Delta(\levo_a))\big).
\label{eq:sconstraint}
\end{align}
We thus conclude, that the above constraints following from Yangian symmetry imply that the two-particle scattering matrix in our 1$+$1 dimensional field theory satisfies the quantum Yang--Baxter equation%
\begin{equation}
\sop_{12}(u_{12})\sop_{13}(u_{13})\sop_{23}(u_{23})=\sop_{23}(u_{23})\sop_{13}(u_{13})\sop_{12}(u_{12}),
\end{equation}
which allows to consistently factorize multi-particle scattering into two-particle S-matrices. This property represents the hallmark of an integrable theory in two dimensions. As an explicit example, we have considered the solution \eqref{eq:su2rmat} to the above constraint equations for $Y[\alg{su}(2)]$. In order to fix the scalar prefactor of the S-matrix, one has to impose further symmetry properties such as crossing and unitarity which also follow from the Yangian Hopf algebra, but which will not be discussed here in further detail.
For a more detailed discussion of the S-matrix see for instance \cite{Dorey:1996gd,Diego}. 


\section{Spin Chains and Discrete Yangian Symmetry}
\label{sec:spinchains}

As opposed to the continuous theories discussed above, Yangian symmetry also plays an important role in 1$+$1 dimensional models with a discrete space dimension. This chapter is concerned with such models on a one-dimensional lattice which are called \emph{spin chains}. Many of the features of the Yangian have a discrete nature (e.g.\ the coproduct) or take over to the discrete case by replacing continuous integrals over space by discrete sums. On the other hand there are important differences to the continuous field theories. Depending on your background, spin chains may often be the most accessible framework to discuss Yangian symmetry.


\paragraph{Spin chains and local charges.}
From quantum mechanics the spin associated to the algebra $\alg{g}=\alg{su}(2)$ is well-known. For the case of spin $\half$ for instance, the spin can be considered as a vector space that transforms under the fundamental representation of $\alg{su}(2)$. It may take the orientations up $\ket{\uparrow}$ or down $\ket{\downarrow}$.

We consider a (generalized) spin as a vector space $\spc{V}$ that transforms under some representation of a symmetry algebra $\alg{g}$. Now we may go further and form chains of spins. We define such \emph{spin chains} as physical models on a Hilbert space $\spc{H}$, which is a tensor product of the above vector spaces:
\begin{equation}
\spc{H}=\dots\,\otimes \spc{V}_k\otimes \spc{V}_{k+1}\otimes \spc{V}_{k+2}\otimes\dots\,.
\end{equation}
Here we will assume that all vector spaces are identical $\spc{V}_k=\spc{V}$. The index $k$ labels the position or site of the spin chain and the positions $k$ and $k+1$ are called \emph{nearest neighbors}. The chain may have different boundary conditions, e.g.\ periodic, open, infinite or semi-infinite boundary conditions, which we leave unspecified for the moment. We will briefly discuss different boundary
conditions and Yangian symmetry in \secref{sec:boundconds}.

\begin{figure}\centering
\includegraphicsbox{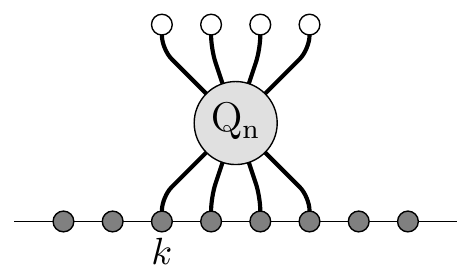}
\caption{A local charge operator $\charge_n$ acting on a spin chain (here $n=4$). Its
position $k$ on the chain is summed over, see
\protect\eqref{eq:homogeneousaction}.}
\label{fig:local}
\end{figure}

The spin chain Hilbert space $\spc{H}$ is spanned by states for which the spin at each position $k$ has a fixed orientation $v_k$, where
$v_k$ denotes a basis vector of $\spc{V}$:
\begin{equation}
\ket{\dots,v_k,v_{k+1},v_{k+2},\dots}\in \spc{H}.
\end{equation}
A physical model is typically defined by a local Hamiltonian $\ham$, whose density acts on two neighboring sites or so-called nearest neighbors.
In the case of integrable spin chains, one finds a set of integrable charges or higher Hamiltonians $\charge_n$, with the first charge given by the two-site or nearest-neighbor Hamiltonian $\ham=\charge_2$. The integrability of the model is reflected in the fact that all of the charges mutually commute:
\begin{equation}
\comm{\charge_m}{\charge_n}=0,\qquad m,n=2,3,\dots.
\end{equation}
These charge operators usually act \emph{locally} and \emph{homogeneously} 
on spin chains; this means that their density acts merely on a small number of neighboring sites and the form of the interaction encoded in the density $\charge_{n,k}\equiv\charge_n(k)$ does not depend on the position $k$, respectively. 
We will focus on such charge operators being \emph{invariant} under the symmetry $\alg{g}$ and acting on the spin chain as
\begin{align}
\charge_n &\defeq \sum_k \charge_{n,k},
&
\charge_{n,k}&:\spc{V}_k\otimes\ldots\otimes\spc{V}_{k+n-1}\to
\spc{V}_k\otimes\ldots\otimes\spc{V}_{k+n-1}.
\label{eq:homogeneousaction}
\end{align}
Here the density $\charge_{n,k}$ is a linear operator which acts on several consecutive
spins starting with site $k$, cf.\ \figref{fig:local}. 
The number $n$ of interacting sites is called the interaction range 
of the operator $\charge_n$. For most ordinary spin chains the charges are labelled such that the interaction range of $\charge_n$ is indeed $n$.

A simple example for an integrable spin chain model is the Heisenberg or XXX${}_{\frac{1}{2}}$ spin chain%
\footnote{The name XXX stems from the fact that the coefficients in the Hamiltonian \eqref{eq:XXXham} of $\sigma_k^1\otimes \sigma_{k+1}^1$, $\sigma_k^2\otimes \sigma_{k+1}^2$ and $\sigma_k^3\otimes \sigma_{k+1}^3$ are equal. If two or all three of these coefficients are chosen differently, one finds the so-called XXZ or XYZ spin chains, respectively. Choosing one coefficient to be zero yields the XX or XY model.}
with $\alg{g}=\alg{su}(2)$ symmetry. Its Hamiltonian is given by the local operator
\begin{equation}\label{eq:XXXham}
\ham_\text{XXX}=\sum_{k}\ham_{k,k+1}=\half\sum_k (\idop_k\otimes \idop_{k+1}-\sigma_k^a\otimes \sigma_{k+1}^a)
=
\sum_k\includegraphicsbox[scale=1]{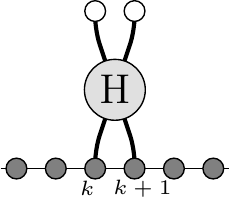}
\end{equation}
with $\sigma_k^a$ denoting the Pauli matrices acting on site $k$ of the spin chain.
Alternatively, the Hamiltonian can be expressed in terms of the permutation operator using again that the tensor Casimir of $\alg{u}(2)$ takes the form $\permop_{k,k+1}=\half(\idop_k\otimes \idop_{k+1}+\sigma_k^a\otimes \sigma_{k+1}^a)$:
\begin{equation}
\ham_\text{XXX}=\sum_k (\idop_{k,k+1}-\permop_{k,k+1}).
\end{equation}
It is convenient to introduce an even more compact notation by writing
$\ham_\text{XXX}=\perm{1,2}-\perm{2,1}$.
This square bracket notation straightforwardly generalizes to permutations of higher range:
\begin{equation}\label{eq:permnot}
 \perm{a_1,a_2,\dots,a_\ell} \ket{\field_1,X_2,\dots,\field_L}=\sum_{k}\cket{\field_1,\dots,\field_k,\field_{k+a_1},\dots,\field_{k+a_n},\field_{k+\ell+1},\dots,\field_L}.
\end{equation}
e.g.\  for the permutation operator we have
\begin{align}
 \permop \cket{\field_1,\dots,\field_L}\equiv \perm{2,1} \cket{\field_1,\dots,\field_L}&=\sum_{k}\cket{\field_1,\dots,\field_k,\field_{k+2},\field_{k+1},\field_{k+3},\dots,\field_L}.
\end{align}
In the above expression \eqref{eq:permnot}, the limits of the sum over $k$ depend on the boundary conditions. For periodic boundary conditions we have $\sum_{k=1}^L$, for open boundary conditions $\sum_{k=1}^{L+1-\ell}$, and for infinite boundary conditions we have $\sum_{k=-\infty}^\infty$.


\subsection{Lax Operator, Monodromy and Yangian Generators}
\label{sec:lax}
Consider a spin chain with spins $\ket{\spin_k}\in \spc{V}_k$, where we assume that all \emph{physical} or \emph{quantum spaces} $\spc{V}_k= \spc{V}$ are identical. The algebraic construction of spin chain models employs the concept of a so-called \emph{auxiliary space} $\spc{V}_0$. The auxiliary space typically transforms under the fundamental representation of the symmetry algebra and we label it with an index $0$ or $\bar 0$ in order to distinguish it from the physical spaces.

As a generalization of the continuous classical case \eqref{eq:Lax}, we introduce a \emph{Lax operator} on the space.
This Lax operator or Lax matrix acts on a physical space $\spc{V}_k$ and on an auxiliary space $\spc{V}_0$, i.e.\ on the product space $\spc{V}_k\otimes\spc{V}_0$.
The defining relations for the Lax operator are given by an integrability equation similar to the Yang--Baxter equation for the R-matrix, the so-called \emph{RLL-relations} defined on $\spc{V}_k\otimes\spc{V}_0\otimes\spc{V}_{\bar 0}$:
\begin{equation}\label{eq:RLL}
\rop_{0\bar 0}(u-v)\lax_{k0}(u)\lax_{k\bar 0}(v)=\lax_{k\bar 0}(v)\lax_{k0}(u)\rop_{0\bar0}(u-v).
\end{equation}
Resembling the definition of a Lie algebra via commutators, the RLL-relations relate the two products $\lax_{k0}(u)\lax_{k\bar 0}(v)$ and $\lax_{k\bar 0}(v)\lax_{k0}(u)$ to each other and can thus be understood as a generalized commutation relation defining $\lax$.
Here the R-matrix acts as an intertwiner on two auxiliary spaces $\spc{V}_{0}\otimes\spc{V}_{\bar 0}$ labeled $0$ and $\bar 0$.
Since we are interested in integrable models, we assume that the R-matrix obeys the quantum Yang--Baxter equation \eqref{eq:YBE}.
Alternatively, given the Lax operator, we may understand \eqref{eq:RLL} as a defining equation for the R-matrix, which for consistency has to obey the quantum Yang--Baxter equation.
In fact, if the auxiliary space $\spc{V}_0$ and the physical space $\spc{V}_k$ are the same, the Lax operator is often identified with the R-matrix (up to convenient shifts in the spectral parameter and overall scalar factors). In particular, for fundamental models such as the Heisenberg spin chain, where the physical and auxiliary spaces carry the same representation, one often finds
\begin{equation}\label{eq:LaxR}
\lax_{k0}(u)\simeq\rop_{k0}(u-\sfrac{i}{2}).
\end{equation}
The full power of the Lax formalism comes into play when the physical and auxiliary representations are different. Note that in principle nothing prevents us from choosing arbitrary representations on the physical and auxiliary spaces and to study possible solutions to the Yang--Baxter or RLL-equations.

One of the most important quantities in the context of integrable spin chains is the \emph{monodromy matrix} $\mon^\alpha{}_\beta(u)$ defined as a product of $L$ of the above  Lax matrices:%
\footnote{This is the discretized version of the path-ordered exponential \eqref{eq:defmon}.}
\begin{equation}
\mon_{j_1\dots j_L,\beta}^{i_1\dots i_L,\alpha}(u)=\lax_{k_2j_1}^{\alpha i_1}(u-u_1)\lax_{k_3j_2}^{k_2 i_2}(u-u_2)\dots \lax_{\beta j_L}^{k_Li_L}(u-u_L)=
\includegraphicsbox[scale=.9]{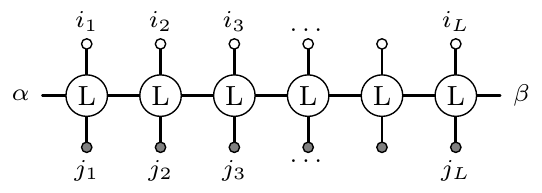}.
\label{eq:mon}
\end{equation}
Alternatively we write this monodromy in the often more user-friendly form
\begin{equation}\label{eq:monuserfriend}
\mon_{1,\dots,L,0}(u)=\lax_{10}(u-u_1)\lax_{20}(u-u_2)\dots \lax_{L0}(u-u_L).
\end{equation}
In general, one may consider inhomogeneous spin chains with non-trivial parameters $u_k$ for $k=1,\dots L$.
In the following we will restrict to homogeneous chains with inhomogeneities $u_k=0$. 

The monodromy $\mon_0\equiv\mon_{1,\dots,L,0}$ acts on $L$ physical and one auxiliary space and obeys the same equation as the underlying Lax matrix, namely the \emph{RTT-relation} that we have already encountered \eqref{eq:RTT}:%
\footnote{For a more extended pedagogical introduction to the Lax formalism and algebraic Bethe ansatz see e.g.\ \cite{Sklyanin:1991ss,Faddeev:1996iy}.}
\begin{equation}\label{eq:chainRTT}
\rop_{0\bar 0}(u-v)\mon_{0}(u)\mon_{\bar 0}(v)=\mon_{\bar 0}(v)\mon_{0}(u)\rop_{0\bar 0}(u-v).
\end{equation}
As we have seen in \secref{sec:RTT}, the Yangian algebra may be defined via this equation.
In order to identify the generators of the Yangian algebra (in the first realization) within this formalism, we will now expand the above monodromy around the point $u=\infty$. For this purpose we first require a more explicit expression for the Lax operator, which is typically written as (up to overall factors):
\begin{equation}\label{eq:Laxexplicit}
({\lax_{k0}})^\alpha{}_\beta(u)
=-i  u\idop_k\otimes (\idop_0)^\alpha{}_\beta+ i{\levz_k^a} \otimes (E_0^a)^\alpha{}_\beta 
\equiv -iu[\idop]^\alpha{}_\beta+ i[{\levz_k}]^\alpha{}_\beta
=\includegraphicsbox{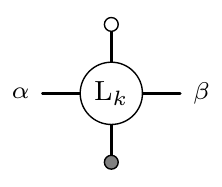}\;.
\end{equation}
Here $E^a_0$ denotes a generator in the fundamental representation of the underlying symmetry algebra $\alg{g}$ acting on the auxiliary space and $\levz^a$ may correspond to a generator in a different representation.
For convenience we suppress the fundamental indices in the physical space $\spc{V}_k$ and, as usual, we sum over the adjoint index $a$.%
\footnote{Here we have chosen our convention such that $\lax(\ihalf)=\permop$ for $\levz^a=\frac{\sigma^a}{2i}$ and $E^a=\sigma^a$, with $\sigma^a$ denoting the Pauli matrices.}
Thus, the monodromy matrix on a chain ranging from $1$ to $L$ takes the form
\begin{equation}
\mon^\alpha{}_\beta(u)
=(-iu)^L\bigbrk{[\idop_1]^\alpha{}_{\gamma_2}-\sfrac{1}{u}\, {[\levz_1]}^{\alpha}{}_{\gamma_2}}\dots\bigbrk{[\idop_L]^{\gamma_L}{}_{\beta}-\sfrac{1}{u}\, {[\levz_L]}^{\gamma_L}{}_{\beta}},
\end{equation}
such that 
\begin{equation}
\mon^\alpha{}_\beta(u)=(-iu)^L\Big([\idop]^\alpha{}_\beta-\frac{1}{u}\sum_{k=1}^L{[\levz_k]}^\alpha{}_\beta+\frac{1}{u^2}\sum_{k=1}^L\sum_{j=1}^{k-1} {[\levz_j]}^\alpha{}_\gamma\,{[\levz_k]}^\gamma{}_\beta+\mathcal{O}\bigbrk{\sfrac{1}{u^3}}\Big).
\label{eq:Yangexp}
\end{equation}

If we now consider the example of $E_a=\sigma_a$, i.e.\ the fundamental representation of $\alg{su}(2)$, we find 
\begin{equation}
 {[\levz_k]}^\alpha{}_\gamma\,{[\levz_\ell]}^\gamma{}_\beta= {\levz_k^b}\otimes (E_0^b)^\alpha{}_\gamma\,{\levz_\ell^c}\otimes (E_0^c)^\gamma{}_\beta= i\epsilon_{abc}{\levz_k^b} {\levz_\ell^c}\otimes(E_0^a)^\alpha{}_\beta+{\levz_k^b} {\levz_\ell^c}\delta_{bc}\otimes(\idop_0)^\alpha{}_\beta,
\end{equation}
where we have used that for the Pauli matrices $\sigma_b\sigma_c=i\epsilon_{bca}\sigma_a+\delta_{bc}$. 
Hence, we obtain%
\footnote{See also \cite{Dolan:2004ys} for a similar discussion.}
\begin{align}
\mon^\alpha{}_\beta(u)=(-iu)^L&\Big[-\sfrac{1}{u}\sum_{k=1}^L\levz_{k}^a+\sfrac{i}{u^2}\,\epsilon_{abc}\sum_{k=1}^L\sum_{j=1}^{k-1}  \levz_j^b\,\levz_k^c
\Big]\otimes(E_0^a)^\alpha{}_\beta
\nonumber\\
+(-iu)^L&\Big[\idop
+\sfrac{1}{u^2}\sum_{k=1}^L\sum_{j=1}^{k-1}  \levz_j^b\,\levz_k^b\Big]\otimes (\idop_0)^\alpha{}_\beta
+(-iu)^L\mathcal{O}\bigbrk{\sfrac{1}{u^3}}.
\label{eq:Yangexpansion}
\end{align}
The first bracket gives rise to the Yangian level-zero and level-one generators at orders $\frac{1}{u}$ and $\frac{1}{u^2}$, respectively (here with $f_{abc}=\epsilon_{abc}$):
\begin{equation}
\levz_a=\sum_{k=1}^L\levz_{k,a},
\qquad\quad
\levo_a=f_{abc}\sum_{k=1}^L\sum_{j=1}^{k-1} \levz_{b,j}\,\levz_{c,k}\;.
\label{eq:lev0lev1}
\end{equation}
The second line of \eqref{eq:Yangexpansion} represents a linear combination of the identity and higher powers of the level-zero charges, which are less interesting for algebraic considerations.%
\footnote{%
We use 
\begin{equation}
\sum_{k=1}^L\sum_{j=1}^{k-1}  \levz_j^a\,\levz_k^a=\half \sum_{k=1}^L\sum_{j=1}^{L} \levz_j^a\,\levz_k^a-\half \sum_k\levz_k^a\levz_k^a=\half \levz^a \levz^a-\half \sum_k\levz_k^a\levz_k^a\;,
\end{equation}
where typically $\levz_k^a\levz_k^a\simeq \idop_k$.
}
Higher orders in the $\frac{1}{u}$-expansion contain higher-level Yangian generators as well as powers of the lower-level generators.

Note that the above expansion of the spin chain monodromy is very similar to the continuous version in \eqref{eq:monexpcont}. In fact, we could have guessed these expressions for the Yangian generators by simply replacing $\int \dd x\to\sum_k$.
Here we have no local contribution to the level-one generators coming from the homogeneous monodromy matrix. We may obtain an additional local contribution $\sum_k u_k \levz_k^a$ by introducing nontrivial inhomogeneities $u_k\neq 0$ in \eqref{eq:mon}. 

In principal, checking the Yangian symmetry of a spin chain model, i.e.\ of a defining Hamiltonian, provides an integrability test. In general, however, the question of whether a Hamiltonian has exact Yangian symmetry strongly depends on the boundary conditions of the underlying model. 


\subsection{Different Boundary Conditions}
\label{sec:boundconds}

Here we briefly comment on the compatibility of the Yangian with different spin chain boundary conditions. Similar considerations apply to the case of continuous two-dimensional field theories discussed above. 
\begin{figure}
\begin{center}
$\begin{array}{ll l}
\text{Periodic:}& &\,\,\includegraphicsbox[scale=1.2]{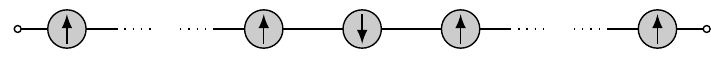}\\[3pt]
\text{Open:}& &\includegraphicsbox[scale=1.2]{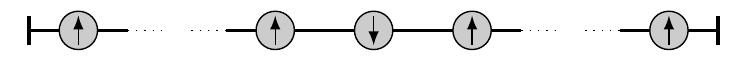}\\[3pt]
\text{Semi-Infinite:}& &\includegraphicsbox[scale=1.2]{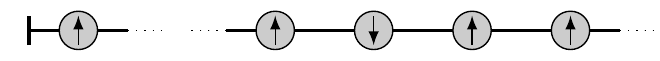}\\[3pt]
\text{Infinite:}& &\hspace{.9cm}\includegraphicsbox[scale=1.2]{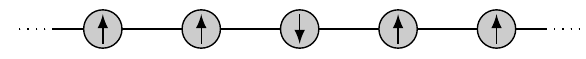}
\end{array}$
\end{center}
\caption{We distinguish between periodic ($\cyc$) and open ($\open$) as well as semi-infinite, i.e.\ half-open, and infinite spin chains.}
\label{fig:kindsofchain}
\end{figure}
We will see that an exact Yangian symmetry is generically not compatible with finite boundary conditions. This means that the  Hamiltonian defining the model does typically not commute with the Yangian level-one generators.
Nevertheless the Yangian symmetry can give nontrivial constraints in the bulk of the system, i.e.\ the symmetry equation is obeyed modulo boundary terms. While in this case the spectrum is not organized in Yangian multiplets, one may use Yangian symmetry to bootstrap a Hamiltonian, cf.\ e.g.\ \cite{Beisert:2007jv}. Suppose we make an ansatz for a Hamiltonian $\ham$, then the equation 
\begin{equation}
\comm{\levo}{\ham}=\text{boundary terms}
\end{equation}
yields non-trivial constraints on the bulk part of this Hamiltonian and in this sense represents a non-trivial symmetry of the model. In certain special cases, one may even find an exact Yangian symmetry as indicated below.

\paragraph{Periodic Boundary Conditions.}
Periodic boundary conditions are implemented by identifying the spin chain site $L+1$ with the site $1$, such that the sites $L$ and $1$ are nearest neighbors.
Generically, exact Yangian symmetry is not compatible with periodic boundary conditions. This can be seen from the ordered structure of the level-one symmetry in the first realization. Hence, to define a Yangian level-one generator for a periodic system, one has to choose an origin on the periodic chain such that the sites can be considered as being ordered according to their relative position to this origin. When applied to a periodic quantity, this typically implies boundary terms that spoil an invariance equation. 

As an example consider again the Heisenberg or XXX Hamiltonian introduced above:
\begin{equation}\label{eq:hamXXXperiod}
\ham\equiv\ham_\text{XXX}=\sum_{k=1}^L (\idop_{k,k+1}-\permop_{k,k+1}).
\end{equation}
The $\alg{su}(2)$ symmetry of the model means that
\begin{equation}
\comm{\levz_a}{\ham}=0,
\end{equation}
with $\levz_a=\sum_{k=1}^L\frac{\sigma_{a,k}}{2i}$ denoting the Lie algebra (or level-zero) generators of $\alg{su}(2)$. 
Here again $\sigma_{a=1,2,3}$ are the Pauli matrices and $\comm{\levz_a}{\levz_b}=\epsilon_{abc}\levz_c$.
The corresponding Yangian $Y[\alg{su}(2)]$ is spanned by these level-zero generators and the level-one generators
\begin{equation}\label{eq:levoXXX}
\levo_a=\epsilon_{abc}\sum_{1\leq j<k\leq L}\levz_{b,j}\levz_{c,k}.
\end{equation}
Importantly, here we have chosen a spin chain origin at site $1$ or $L$, respectively. While this point is not distinguished by the periodic Hamiltonian \eqref{eq:hamXXXperiod}, defining the level-one generator \eqref{eq:levoXXX} requires this choice.
We remember from the $\alg{su}(2)$ example discussed above that 
$\comm{ \epsilon_{abc} \levz_b\otimes \levz_c}{\permop}=\levz_a\otimes \idop-\idop\otimes \levz_a$
which implies that
\begin{equation}\label{eq:XXXbound}
\comm{\levo_a}{\ham}=\levz_{a,1}-\levz_{a,L}\equiv \levz_a\otimes \idop\otimes\dots\otimes\idop-\idop\otimes\dots\otimes\idop\otimes \levz_a.
\end{equation}
Here the dots stand for $L-3$ identity operators $\idop$.
Hence, the Heisenberg Hamiltonian commutes with the level-one symmetry up to boundary terms, i.e.\ terms that act only on the boundary of the spin chain. Similar considerations apply to the Yangian symmetry of the spin chain Hamiltonian of $\superN=4$ super Yang--Mills theory discussed in \secref{sec:dilatationop} \cite{Dolan:2003uh}.
Note that particular examples of spin chain models exist, whose periodic boundary conditions are compatible with an exact Yangian symmetry, see  \cite{Haldane:1992sj}.
\paragraph{Cyclic Boundary Conditions.}

In contrast to periodic boundary conditions which only imply that the sites $L$ and $1$ are neighbors, cyclic boundary conditions in addition require that the system is invariant under cyclic shifts $k\to k+1$ or $k\to k-1$ of the sites. This implies that the total momentum of all spin chain excitations has to be zero. In particular, this means that the Yangian generators should commute with the shift operator $\shift$, which induces cyclic permutations by one site. That this yields additional constraints can be seen from evaluating \cite{Drummond:2009fd,Beisert:2010jq}
\begin{align}\label{eq:shiftlevelone}
\comm{\shift}{\levo_a}
&=\shift\,f_{abc}\Big[\sum_{k=1}^L\sum_{j=1}^{k-1}\levz_{b,j}\levz_{c,k}-\sum_{k=2}^{L+1}\sum_{j=2}^{k-1}\levz_{b,j}\levz_{c,k}\Big]
=\shift\,\big(f_{abc}f_{dbc}\levz_{d,1}-2f_{abc}\levz_{b,1}\levz_c\big).
\end{align}
In general, the right hand side of this equation does not vanish, which emphasizes the dependence of the level-one generators on the choice of an origin of the chain. The first term is proportional to the dual coxeter number $\cox=f_{abc}f_{bcd}$, which vanishes only for some particular algebras $\alg{g}_{\cox=0}$. The second term is proportional to the level-zero generator $\levz_c$ and is generically non-zero. 

Suppose, however, we forget about Hamiltonians for the moment and instead consider Yangian invariants, i.e.\ `states' $\ket{I}$ that are annihilated by the Yangian generators. Then at least for certain algebras $\alg{g}_{\cox=0}$ one may define cyclic invariants $\ket{I}$ of the Yangian algebra:%
\footnote{See \cite{Frassek:2014bya} for a pedagogical introduction to a systematic study of Yangian invariants.}
\begin{align}\label{eq:cyclicinv}
\levo\ket{I}&=0, 
&&\text{consistent if}&
\levz\ket{I}=0,
\quad\text{and}\quad
\cox&=0.
\end{align}
These conditions imply that the right hand side of \eqref{eq:shiftlevelone} vanishes on $\ket{I}$.
This type of cyclic boundary conditions is particularly interesting since it applies to scattering amplitudes in $\superN=4$ super Yang--Mills theory discussed in \secref{sec:Neq4} with $\alg{g}_{\cox=0}=\alg{psu}(2,2|4)$.
\paragraph{Open and Semi-infinite Boundary Conditions.}
Similar to the case of periodic boundary conditions, also open boundaries are often not compatible with a full Yangian symmetry. Here the situation very much depends on the specific bulk and boundary part of the Hamiltonian:
\begin{equation}
\ham=\ham_\text{bulk}+\ham_\text{left-boundary}+\ham_\text{right-boundary}.
\end{equation}
Analogous considerations apply to semi-infinite ($=$half-open) boundaries where either the left or right boundary is absent and the chain extends to infinity on that side.

Suppose we have a system with a level-zero symmetry algebra $\alg{g}$ in the bulk and a symmetry $\alg{h}$ at the boundary (see e.g.\ \cite{Delius:2001he,MacKay:2004rz,MacKay:2010ey}) such that $(\alg{g},\alg{h})$ form a symmetric pair, which means that $\alg{g}=\alg{h}\oplus\alg{m}$, with
\begin{align}
\comm{\alg{h}}{\alg{h}}&\subset \alg{h},
&
\comm{\alg{h}}{\alg{m}}&\subset \alg{m},
&
\comm{\alg{m}}{\alg{m}}&\subset \alg{m}.
\end{align}
Furthermore we assume to have a Yangian symmetry $Y[\alg{g}]$ in the bulk. Then the whole system, including the boundary, often still has a \emph{twisted Yangian} symmetry $Y[\alg{g},\alg{h}]$ whose level-zero generators are $\levz_i$ (note the index $i$), while the level-one generators have the modified form (note the index $p$)
\begin{equation}\label{eq:twistlev1}
 {\mathrm{\tilde J}}_p=\widehat \levz_p+\half f_{pqi}\levz_i \levz_q,
 = \levo_p+\quarter\comm{\cas_\alg{h}}{\levz_p},
\end{equation}
with $\cas_\alg{h}$ representing the quadratic Casimir of $\alg{g}$ restricted to $\alg{h}$.
Importantly, 
the indices $i,j,\dots$ correspond to generators of $\alg{h}$ and the indices $p,q,\dots$ to generators of $\alg{m}$. The index sums thus only run over subsets of all index values. 
For further details and references see for instance \cite{MacKay:2004tc}.

\paragraph{Infinite Chain: No Boundaries.}
As seen above, the Yangian generators are most naturally defined with boundaries at $\pm \infty$, i.e.\ with no boundaries at all. 
On such an infinite chain, the right hand side of \eqref{eq:XXXbound} vanishes and the Hamiltonian has exact Yangian symmetry. However, typically such infinite spin chain systems are rather of formal interest and do not directly represent physical models. In particular, their spectrum is not quantized due to the underlying noncompact space.


\subsection{Periodic Chains, Transfer Matrix and Local Charges}
\label{sec:trans}
\begin{figure}\centering
$\transfer(u+\ihalf)=$\,
\includegraphicsbox[scale=.9]{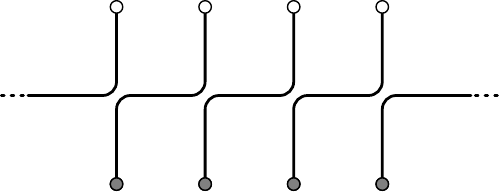}
\,\,$+\,iu\,\,{\sum\limits_k}\,\,$
\raisebox{-.2cm}{\includegraphicsbox[scale=.9]{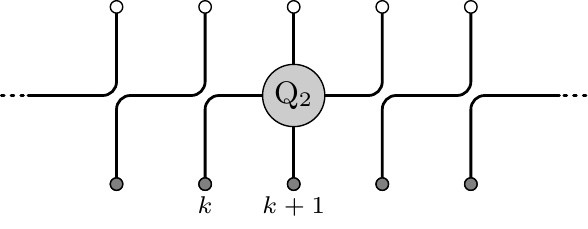}}
\,$+\,\,\order{u^2}$
\caption{Expanding the transfer matrix around $u=\ihalf$ gives rise to the shift operator $\shift$ at zeroth order, to the shift operator times the Hamiltonian $\shift\charge_2$ at order $u$, and an infinite set of additional commuting charges at higher orders of the spectral parameter $u$.}%
\label{fig:transfer}%
\end{figure}

In the case of periodic spin chain boundary conditions, the so-called \emph{transfer matrix} $\transfer(u)\equiv\transfer_{1\dots L}(u)$ 
is defined as the trace over the monodromy matrix (the trace implements the periodicity)
\begin{equation}
{\transfer}_{j_1\dots j_L}^{i_1\dots i_L}(u)=\Tr \mon_{j_1\dots j_L,\beta}^{i_1\dots i_L,\alpha}(u)\equiv\transfer_{j_1\dots j_L,\alpha}^{i_1\dots i_L,\alpha}(u),
\label{eq:cyctransfer}
\end{equation}
and acts on the whole spin chain:
\begin{equation}
\transfer(u):\spc{V}_1\otimes\dots\otimes\spc{V}_L\to\spc{V}_1\otimes\dots\otimes\spc{V}_L.
\end{equation}
In our alternative notation we may write this trace over \eqref{eq:mon} as
\begin{equation}\label{eq:transferuserfriend}
{\transfer}(u)=\Tr\nolimits_0\lax_{10}(u)\lax_{20}(u)\dots \lax_{L0}(u),
\end{equation}
where we suppress the free indices and only indicate the vector spaces $\spc{V}_k$ on which the Lax-matrices act. The index $0$ denotes the auxiliary space which is traced, e.g.\ $\Tr_0(\lax_{10})_{j_1}^{i_1}(\lax_{20})_{j_2}^{i_2}=\lax_{kj_1}^{\ell i_1}\lax_{\ell j_2}^{k i_2}$.
It follows from the RTT-relations \eqref{eq:chainRTT} that this transfer matrix commutes with itself when evaluated at different spectral parameters (see e.g.\ \cite{Faddeev:1996iy,Nepomechie:1998jf}):
\begin{equation}
\comm{\transfer(u)}{\transfer(v)}=0.
\label{eq:transcomm}
\end{equation}
Let us furthermore assume that the considered system forms part of the so-called \emph{fundamental models}, for which the auxiliary and quantum spaces are the same (or isomorphic) and for which a special point $u=\ihalf$ exists such that $\lax(\ihalf)= \permop$, cf.\ \cite{Tarasov:1983cj}. This is the case for simple examples of integrable spin chains such as the Heisenberg model. Hence the point $u=\ihalf$ is dinstinguished and we expand the transfer matrix around this point in order to obtain $u$-independent conserved charge operators (cf. \figref{fig:transfer}) 
\begin{equation}
\transfer(u+\ihalf)=\shift \exp{i \sum_{r=2}^{L}u^{r-1}\charge_r}
=\shift+i u\, \shift \charge_2+u^2\, \brk{i \,\shift \charge_3-\half\, \shift \charge_2 \charge_2}+\order{u^3},
\label{eq:transexp}
\end{equation}
Here $\shift=\transfer(\ihalf)$ denotes the shift operator that is given by a product of iterative permutations $\permop_{k,k+1}$ and induces a cyclic permutation of all spin chain sites, e.g.\
\begin{equation}
\shift \ket{X_1,X_2\dots,X_L}=\ket{X_2,\dots,X_L,X_1}.
\end{equation}
Since we may pull out the overall factor $\shift$ in \eqref{eq:transexp}, the $\charge_r$ denote local operators on the spin chain.
It follows from \eqref{eq:transcomm} that all operators $\charge_r$ mutually commute such that the transfer matrix furnishes a generating functional for local integrable charge operators \cite{Baxter:1972pc,Tarasov:1983cj}:
\begin{equation}
\charge_r=-\frac{i}{(r-1)!}\frac{\dd}{\dd u^{r-1}}\left.\log{\transfer(u)}\right |_{u=\ihalf},\qquad r\geq2.
\label{eq:gencharges}
\end{equation}
Remember that the first of these charges $\charge_2=\ham$ is typically the Hamiltonian that is chosen to define the model's dynamics.
The logarithmic derivative of the transfer matrix generates $L-1$ mutually commuting charges and thereby naturally associates the interaction range of the longest operator to the dimension of the transfer matrix. Note that taking the logarithmic derivative ensures that the charges are local \cite{Luscher:1976pb} and that we have divided the definition of the charges by the common shift operator
\begin{equation}
\shift=\transfer(\ihalf)=e^{i \charge_1}.
\end{equation}
Hence, the first two powers in the expansion of the transfer matrix define the momentum operator and the Hamiltonian of the model (cf.\ \figref{fig:transfer}):
\begin{align}
\charge_1&=-i \log  \transfer(\ihalf),
&
\charge_2&=-i \frac{\dd}{\dd u}\log{\transfer(u)} \big |_{u=\ihalf}.
\label{eq:firsttwo}
\end{align} 
Expressing the momentum operator $\charge_1$ as the logarithm of an operator only formally illustrates the analogy to the other charges. In particular, \eqref{eq:firsttwo} implies that the density of the Hamiltonian is proportional to the logarithmic derivative of the Lax operator:
\begin{equation}
\ham_{k,k+1}=\charge_{2,k,k+1}\simeq \frac{\dd}{\dd u}{\log \lax_{k,k+1}(u)} \big |_{u=\ihalf}.
\end{equation}
Note that this expression makes only sense for fundamental models with $\spc{V}_j\simeq\spc{V}_0$ where the Lax operator $\lax_{k0}$ acts on the same space as the Hamiltonian density $\ham_{k,k+1}$, namely on $\spc{V}_k\otimes\spc{V}_0\simeq\spc{V}_k\otimes\spc{V}_{k+1}$.

In general, i.e.\ for \emph{non-fundamental models}, the Yangian generators and the local Hamiltonians do not have to originate from the same monodromy. If the auxiliary and physical spaces are not identical, one may impose the following version of the RLL-relation \eqref{eq:RLL} (now exchanging the roles of auxiliary and quantum spaces) defined on $\spc{V}_k\otimes\spc{V}_{k+1}\otimes \spc{V}_0$:
\begin{equation}\label{eq:RLL2}
\rop_{k,k+1}(v)\lax_{k0}(u+v)\lax_{k+1,0}(u)=\lax_{k+1,0}(u)\lax_{k,0}(u+v)\rop_{k,k+1}(v).
\end{equation}
This equation follows from the quantum Yang--Baxter equation \eqref{eq:YBE} by the identification $\rop_{12}\to\rop_{k,k+1}$, $\rop_{13}\to\lax_{k0}$ and $\rop_{23}\to\lax_{k+1,0}$.
Given a Lax operator $\lax_{k0}$, this RLL-relation defines an R-matrix on $\spc{V}_k\otimes \spc{V}_{k+1}$ and allows to define a Hamiltonian via its logarithmic derivative (cf.\ \cite{Tarasov:1983cj}):
\begin{equation}
\ham_{k,k+1}=\charge_{2k,k+1}=i\frac{\dd}{\dd u} \log \rop_{k,k+1}(u)\big |_{u=0}.
\end{equation}
Higher local charges can be obtained from the expansion of a transfer matrix of the form \eqref{eq:transferuserfriend} with the replacement $\lax\to\rop$ and the trace taken over a quantum space.
\begin{figure}
\begin{center}
$ \charge_2=\,\,$\includegraphicsbox[scale=1]{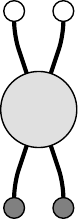}\,,\qquad\qquad
$\charge_3=\,\,$\includegraphicsbox[scale=1]{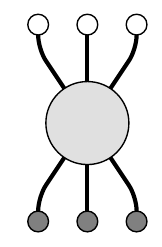}\,,\qquad\qquad
$\charge_4=\,\,$\includegraphicsbox[scale=1]{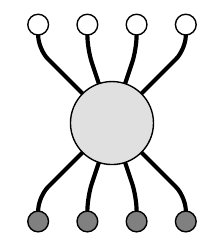}\,,\,
\quad$\dots$
\caption{The interaction range of the charge $\charge_n$ is $n$, i.e.\ the charge operator acts on $n$ neighboring sites at the same time.}\label{fig:chargerange}
\end{center}
\end{figure}%
\paragraph{Example.}

In the case of the Heisenberg spin chain
the first few local charges take the form,%
\footnote{In order to obtain exactly these expressions for the charges (which annihilate the ferromagnetic vaccuum state) from the above formalism, one should modify the definition of the Lax matrix \eqref{eq:Laxexplicit} by an overall function of $u$. This modification does not change the physics of the model. We also note that the $\alg{gl}(N)$ symmetric spin chain has the same charges.} cf.\ \figref{fig:chargerange}
\begin{align}
	\charge_2&=[1]-[2,1],\nln
	\charge_3&=\ihalf ([3,1,2]-[2,3,1]),\nln
	\charge_4&=\sfrac{1}{3} (-[1]+2[2,1]-[3,2,1]+[2,3,4,1]-[2,4,1,3]-[3,1,4,2]+[4,1,2,3]),
\label{eq:nncharges}
\end{align}
where we have used the notation of \eqref{eq:permnot} to display the permutations which furnish the charges' building blocks. We note that on a periodic spin chain we have $[1,2]\equiv [1]$ since the difference of these two operators only acts nontrivially on boundaries, which are absent on a periodic chain.

\subsection{Master Symmetry and Boost Automorphism}
\label{sec:srboostop}

In this subsection we briefly demonstrate the role of the discrete version of the Lorentz boost.

\paragraph{Boost automorphism and monodromy matrix.}
Let us derive an interesting relation for the monodromy matrix of an ordinary integrable short-range spin chain along the lines of the original  paper by Tetel'man \cite{Tetelman:1981xx}.  
First of all we introduce the Hamiltonian density using the logarithmic derivative of an R-matrix with $\rop(0)=\permop$:
\begin{equation}
\ham_{k,k+1}=i\permop_{k,k+1}\dot \rop_{k,k+1}(0) , \qquad \dot\rop_{k,k+1}(0)=\left.\frac{\dd }{\dd u}\rop_{k,k+1}(u)\right|_{u=0}.
\end{equation}
This density acts on the spin chain sites $k$ and $k+1$, i.e.\ on $\spc{V}_k\otimes \spc{V}_{k+1}$. 
We consider the RLL-relation in the form of \eqref{eq:RLL2}:
\begin{equation}
\rop_{12}(v)\lax_{10}(u+v)\lax_{20}(u)=\lax_{20}(u)\lax_{10}(u+v)\rop_{12}(v).
\end{equation}
Differentiating with respect to $v$ and setting $v$ to zero afterwards yields the following differential equation for the Lax-matrix:
\begin{equation}
\dot\lax_{10}(u)\lax_{20}(u)-\lax_{10}(u)\dot\lax_{20}(u)=i\comm{\ham_{12}}{\lax_{10}(u)\lax_{20}(u)}.
\end{equation}
Here we used that the R-matrix obeys the initial condition $\rop_{12}(0)=\permop_{12}$. 
Now we rename $1\to k$ and $2\to k+1$ and multiply this equation from the left by $\prod_{1=j}^{k-1} \lax_{0j}$ and from the right by $\prod_{\ell=k+1}^L \lax_{0\ell}$. This yields
\begin{equation}
\Big(\prod_{j=1}^{k-1} \lax_{0j}\Big) \dot\lax_{0k}\Big(\prod_{\ell=k+1}^{L} \lax_{0\ell}\Big)
-
\Big(\prod_{j=1}^{k} \lax_{0j}\Big) \dot\lax_{0,k+1}\Big(\prod_{\ell=k+2}^{L} \lax_{0\ell}\Big)
=
i\Big[\ham_{k,k+1},\prod_{j=1}^L \lax_{0j}\Big].
\end{equation}
If we now furthermore multiply this equation by $k$ and sum over $k$ from $1$ to $L$ , this yields 
\begin{equation}\label{eq:transboostbound}
\frac{\dd \mon(u)}{\dd u}+0\times \dot\lax_{01}\Big(\prod_{j=2}^L\lax_{0j}\Big)-L\times \Big(\prod_{j=1}^L\lax_{0j}\Big)\dot\lax_{0,L+1}
=
i\Big[\sum_{k=1}^{L} k\,\ham_{k,k+1},\mon(u)\Big],
\end{equation}
where $\mon(u)$ denotes the monodromy matrix \eqref{eq:mon} with suppressed indices. Note that the appearence of only right boundary terms originates in our choice of labeling the first and last spin chain leg $1$ and $L$.

If we take the limit of an \emph{infinite spin chain}, i.e.\ $-\infty \leftarrow 1$ and $L\to \infty$, the boundary terms on the left hand side of \eqref{eq:transboostbound} drop out and we find the equation
\begin{equation}
\frac{\dd \mon(u)}{\dd u}=i\comm{\boo}{\mon(u)},
\label{eq:monboost}
\end{equation}
where we have defined the \emph{spin chain boost operator}
\begin{equation}\label{eq:spinchainboost}
\boo=\sum_{k=-\infty}^{\infty} k\, \ham_{k,k+1}.
\end{equation}
Note that this is the discrete version of the field theory boost \eqref{eq:boostmoment}, obtained by replacing $\int dx\,x\to \sum_k k$.
Using the expansion \eqref{eq:Yangexpansion} of the monodromy matrix,  we see that this equation implies the relation 
\begin{equation}
\comm{\boo}{\levo_a}\simeq \levz_a.
\end{equation}
Hence, the spin chain boost operator defined above corresponds to Drinfel'd's boost automorphism of the Yangian algebra \eqref{eq:boostauto}.
\paragraph{Master symmetry and transfer matrix.}
The concept of \emph{master symmetry} of integrable models dates back to 1981 \cite{Fokas:1981cd,Fokas:1983xx}. It denotes a symmetry whose iterative application to the constants of motion leaves their commutator invariant. Consequently, the master symmetry maps integrable charges to integrable charges and thereby generates a set of infinitely many commuting operators.

In order to identify such a symmetry we now take the trace $\Tr\nolimits_0$ over the monodromy matrix in \eqref{eq:monboost} to obtain the following equation for the transfer matrix
\begin{equation}
\frac{\dd \transfer(u)}{\dd u}=i\comm{\boo}{\transfer(u)}.
\label{eq:transboost}
\end{equation}
We assume that the expansion of the transfer matrix $\transfer(u)$ yields the local charges via \eqref{eq:gencharges}.
Hence, \eqref{eq:transboost} is a remarkable statement since it implies
\begin{align}
\charge_2&=\,\shift^{-1}\comm{\boo}{\shift}= \shift^{-1}\boo\,\shift-\boo,\label{eq:recQ2}\\[1ex]
\charge_3&=+\ihalf \shift^{-1}\comm{\boo}{\shift\charge_2}-\ihalf \charge_2\charge_2=\ihalf \comm{\boo}{\charge_2},\\[1ex]
\charge_4&=\sfrac{i}{3} \comm{\boo}{\charge_3},\\
&\hspace{-.2cm}\dots\nonumber
\end{align}
The spin chain boost operator $\boo=\boost{\charge_2}$, the first moment of the Hamiltonian, therefore recursively generates the algebra of local integrable charges
\begin{equation}\label{eq:nngen}
 \charge_{r+1}=\frac{i}{r}\,\comm{\boo}{\charge_r}.
\end{equation}
It represents a master symmetry of the short-range integrable model. Note that the charges $\charge_3^\SR$ and $\charge_4^\SR$ given in \eqref{eq:nncharges} may be obtained using the boost operator in this way.%
\footnote{Notably, this method produces additional boundary terms which vanish on infinite or periodic spin chains and can therefore be dropped.}

\paragraph{Poincar\'e algebra.}
Let us compare the algebra spanned by the local charges and the boost operator to the ordinary two-dimensional Poincar\'e algebra, cf.\ \cite{Thacker:1985gz}:
\begin{equation}\label{eq:contpoin}
\comm{\gen{P}}{\gen{H}}=0,\qquad
\comm{\gen{B}}{\gen{P}}=\gen{H},\qquad
\comm{\gen{B}}{\gen{H}}=\gen{P}.         
\end{equation}
Here space and time translations $\gen{P}$ and $\gen{H}$ are `rotated' into each other by the Lorentz boost $\gen{B}$.
For our integrable spin chain the Poincar\'e algebra is enhanced according to
\begin{align}\label{eq:Poinalg}
&\gen{H}\to \charge_2,\charge_3,\dots,
&
&\gen{P}\to \charge_1
&
\gen{B}\to \boo,
\end{align}
and we have the commutation relations
\begin{align}
\comm{\charge_1}{\charge_2}&=0& \comm{\boo}{\charge_1}&\simeq\charge_2& \comm{\boo}{\charge_2}&\simeq\charge_3&\comm{\boo}{\charge_3}&\simeq\charge_4&&\dots\nonumber\\
\comm{\charge_2}{\charge_3}&=0&&&&&&&&\nonumber\\
\comm{\charge_3}{\charge_4}&=0&&&&&&&&\nonumber\\
\dots&&&&&&&&&
\end{align}
\begin{figure}
 \begin{center}
  \includegraphicsbox{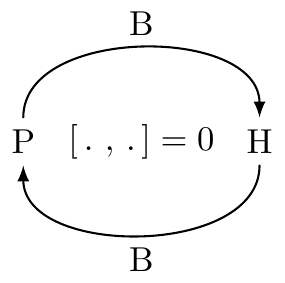}
\qquad\qquad
  \includegraphicsbox{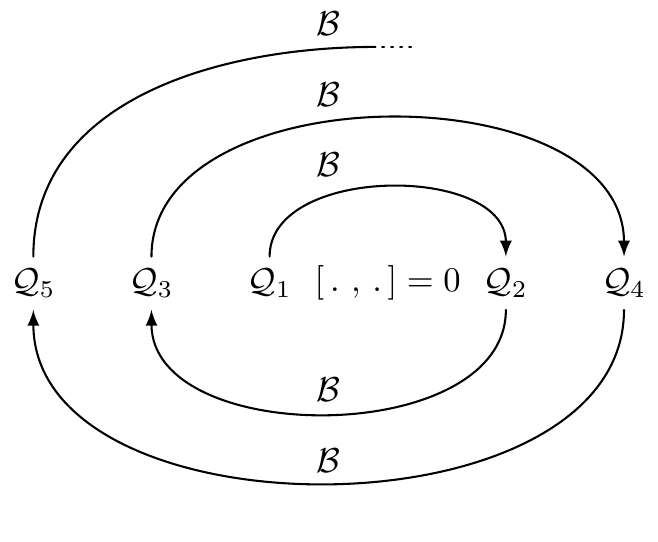}
 \end{center}%
\caption{Analogy between the Poincar\'e algebra with generators $\gen{B}$, $\gen{P}$ and $\gen{H}$ and the algebra of the boost and the integrable charges.}
\label{fig:algenhanced}
\end{figure}%
Space and time translation are supplemented by an infinite set of symmetries. The Lorentz boost translating between the two symmetries $\gen{P}$ and $\gen{H}$ of the ordinary Poincar\'e algebra \eqref{eq:Poinalg}, here takes the role of a ladder operator. For the integrable spin chain the sequence of symmetries does not close and we find a tower of conserved charge operators, cf.\ \figref{fig:algenhanced}.%
\footnote{Also in continuous field theories one may construct a tower of local charges, which are mapped onto themselves by the Lorentz boost, see e.g.\ \cite{Evans:1999mj}}
Mapping conserved charges to conserved charges, the boost thus represents a master symmetry in the above sense.

\paragraph{Periodic Chain.}
To get rid of the boundary terms in \eqref{eq:transboostbound}, we have taken the limit of an infinite spin chain. On a periodic spin chain, the definition of the boost operator is obsolete since it crucially depends on the choice of a spin chain origin. Also in this respect, the boost operator resembles the Yangian level-one generators. Nevertheless, the short-range charge operators of the infinite spin chain are the same as those of the periodic chain since in both cases boundary operators vanish. Hence, while only properly defined on an infinite chain, the boost operator can be formally used to generate the set of periodic integrable charges. In fact, if the boost operator were well-defined on periodic chains, the finite transformation corresponding to \eqref{eq:transboost} would merely constitute a similarity transformation.

\subsection{More Boosts and Long-Range Spin Chains}
\label{sec:moreboosts}

We have seen above that integrable spin chains feature a tower of commuting charges $\charge_n$, $n=1,2,\dots$. Furthermore, we have seen that the boost operator
\begin{equation}
\boo=\sum_k k \,\ham_k=\sum_k k\, \charge_{2,k},
\end{equation}
plays the role of a master symmetry. Certainly, we may also define higher `boost operators' associated with the higher local charges:
\begin{equation}
\boo[\charge_n]=\sum_k k\, \charge_{n,k}.
\label{eq:genboost}
\end{equation}
Here $\boo=\boo[\charge_2]$ denotes the boost operator encountered in the previous subsections. A natural question is, which role the higher boosts play for the considered spin chain models. We will see that they allow to define an integrable (so-called \emph{long-range}) deformation of the above \emph{short-range} spin chain models, which plays an important role in the context of the gauge/gravity duality further discussed below (cf.\ \secref{sec:dilatationop}).

In fact, one may define deformed charge operators $\charge_n(\lambda)$ via the evolution equation
\begin{equation}
\frac{\dd}{\dd\lambda}\,\charge_n(\lambda)
=i\alpha_k(\lambda)\bigcomm{\boost{\charge_k(\lambda)}}{\charge_n(\lambda)},
\qquad
k>2.
\label{eq:deformationboost}
\end{equation}
Here we sum over $k$ and $\alpha_k(\lambda)$ denotes some function of $\lambda$ whose form specifies the precise deformation.
If the charges $\charge_n(0)$ commute among each other, also the solutions $\charge_n(\lambda)$ to the defining equation \eqref{eq:deformationboost} commute. Hence, the deformed charges $\charge_n(\lambda)$ define an integrable system. The deformation of the local charges typically increases their interaction range on the spin chain. Suppose we denote the local short-range charges of the previous subsections by $\charge_n^\SR\equiv \charge_n(0)\equiv\charge_n^{(0)}$. Then the charges $\charge_n(\lambda)$ define a long-range spin chain model, i.e.\ a model whose defining Hamiltonian and higher integrable charges have longer and longer range of interaction when going to higher powers of the parameter $\lambda$.
Again, the long-range Hamiltonian $\ham(\lambda)=\charge_2(\lambda)$ defines the dynamics of the model and the parameter $\lambda$ may be understood as a coupling constant.

In general, perturbatively long-ranged spin chains may be defined as
deformations of the above short-range chains, e.g.\ the Heisenberg chain. 
The short-range charges $\charge_n$ 
are taken to be the leading order $\charge_n^{(0)}$
in a power series
\begin{equation}
\charge_n(\lambda)
=\charge_n^{(0)}
+\lambda\charge_n^{(1)}
+\lambda^2\charge_n^{(2)}
+\order{\lambda^3},
\qquad
\charge_n^{(0)}=\charge_n,
\label{eq:expansionlambda}
\end{equation}
such that the interaction range of the charges grows 
with the perturbative order in $\lambda$.
The long-range charges can still be written as linear combinations 
of local and homogeneous operators~$\mathrm{O}_k$
\begin{equation}
	\charge_r(\lambda)=\sum_k c_{r,k}(\lambda) \mathrm{O}_k,
\end{equation}
but now with coefficients $c_{r,k}(\lambda)$
which are formal power series in $\lambda$
starting at a certain order.
The charges have to obey the integrability condition $\comm{\charge_r(\lambda)}{\charge_s(\lambda)}=0$
order by order in $\lambda$, which is guaranteed by the deformation equation \eqref{eq:deformationboost}.
\begin{figure}
\begin{center}
\includegraphicsbox[scale=1.05]{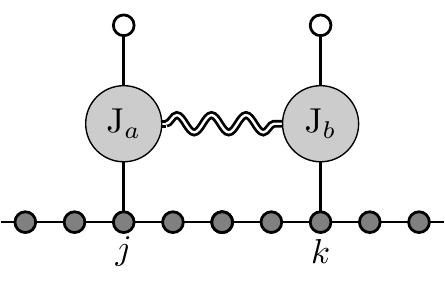}
\quad
\includegraphicsbox[scale=1.05]{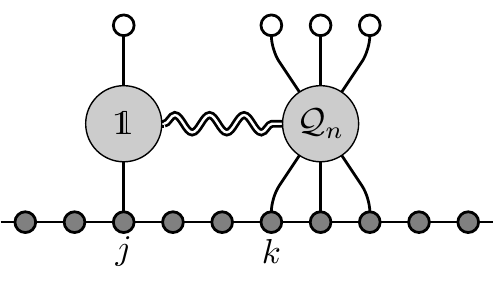}
\quad
\includegraphicsbox[scale=1.05]{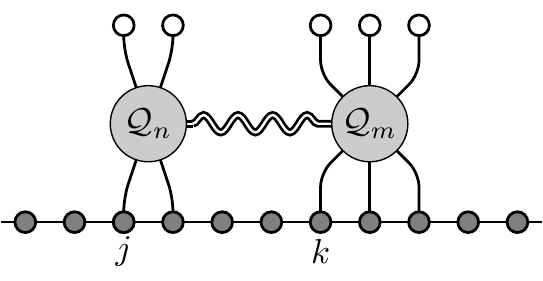}
\end{center}
\caption{Overview over bilocal operators: Yangian level-one generators $\levo$, boost operators $\boost{\charge_n}\equiv \biloc{\idop}{\charge_n}$ and bilocal charges $\biloc{\charge_n}{\charge_m}$.}
\label{fig:overviewbiloc}
\end{figure}

To make connection to the Yangian algebra, we note that also the Yangian generators should be deformed in order to preserve their commutation relations with the local (long-range) Hamiltonians. By assumption, the level-zero generators $\levz_a$ commute with the charges $\charge_n$ and thus also with the boost operators $\boost{\charge_n}$. For the level-one generators one uses a deformation equation analogous to \eqref{eq:deformationboost}:
\begin{equation}
\frac{\dd }{\dd \lambda}\,\levo_a(\lambda)
=i\alpha_k(\lambda) \bigcomm{\boost{\charge_k(\lambda)}}{\levo_a(\lambda)}.
\label{eq:deformationyang}
\end{equation}
This equation defines the long-range level-one Yangian generators.
In fact, we may remember that in analogy to the level-one symmetry, also the boost can be understood as a formally bilocal expression, i.e.\ $\boost{\charge_n}=\biloc{\idop}{\charge_n}$ (see \eqref{eq:boostbi} for the continuous case). Taking this consideration further one can define even more general bilocal charge operators, cf.\ \figref{fig:overviewbiloc}:
\begin{equation}
\biloc{\charge_n}{\charge_m}=\sum_{j<k}\charge_{n,j}\charge_{m,k}.
\end{equation}
These may also be employed to generate long-range deformations of spin chain Hamiltonians via similar differential equations as \eqref{eq:deformationboost} and \eqref{eq:deformationyang}. We note that these bilocal charge generators induce so-called dressing phase contributions to the dilatation operator of $\superN=4$ super Yang--Mills theory, cf.\ \secref{sec:Neq4}. However, discussing these in more detail is beyond the scope of these lectures; see \cite{Bargheer:2008jt,Bargheer:2009xy} for further elaboration.%
\footnote{Note that this construction of long-range spin chains using boost operators may also be generalized to chains of trigonometric type \cite{Beisert:2013voa} or to chains with open boundary conditions \cite{Loebbert:2012yd}. Furthermore interesting relations to (inhomogeneous versions of) Baxter's corner transfer matrix exist \cite{Itoyama:1986ad,Jiang:2014mja}.}
\paragraph{Example.}
To illustrate the deformation of level-one generators via \eqref{eq:deformationyang}
let us once more consider $\alg{g}=\alg{su}(2)$ with generators $\levz_a=\frac{\sigma_a}{2i}$ in the fundamental representation. We choose $\alpha_3=1$ and  $\alpha_{k>3}=0$, i.e.\ we deform the level-one generators only using the higher boost $\boost{\charge_3}$. The first higher charge $\charge_3$ of the $\alg{su}(2)$ Heisenberg chain is given by \eqref{eq:nncharges} and we can plug it into the definition of the generalized boost operator \eqref{eq:genboost}.
Then we find the following leading-order deformation of the level-one generator:
\begin{equation}\label{eq:defYang}
\levo_a(\lambda)
=\levo_a(0)+\lambda\, i\comm{\boost{\charge_3}}{\levo_a}+\order{\lambda^2}
=\epsilon_{abc}\sum_{j<k}\levz_{b,j}\levz_{c,k}+\lambda\, \epsilon_{abc}\sum_{k}\levz_{b,k}\levz_{c,k+1}+\order{\lambda^2}.
\end{equation}
Here the level-zero generators $\levz_a(\lambda)\equiv\levz_a$ remain undeformed. Notably, this deformation of the level-one generators of the Heisenberg model corresponds to the two-loop deformation of the Yangian symmetry of the $\alg{su}(2)$ dilatation operator of $\superN=4$ super Yang--Mills theory, cf.\ \cite{Haldane:1992sj,Serban:2004jf,Beisert:2007jv}. In this context the parameter $\lambda$ represents the 't~Hooft coupling constant, cf.\ \secref{sec:Neq4}. 

\section{Yangian Symmetry in 4d Field Theory}
\label{sec:Neq4}

\quoting{The models analyzed in this paper, formulated in two space-time dimensions, are clearly unrealistic. However, we believe that the phenomenon exhibited by these models is indicative of what one would expect in more realistic models. In fact
the restriction to two dimensions is only in order
to have an asymptotically free theory in which one
has an explicit expansion parameter ($N$). The only
asymptotically free theory in four dimensions
necessarily involves gauge fields and does not
lend itself to any simple approximation.}{D.\ Gross and A.\ Neveu 1974 \cite{Gross:1974jv}}

Let us see that in fact Yangian symmetry can also be found in dimensions greater than two. In order to get a glimpse on how this happens, we first have to introduce a very interesting model, the maximally supersymmetric Yang--Mills theory in four dimensions. For a selection of helpful reviews see for instance
\cite{Aharony:1999ti,'tHooft:2002yn,Beisert:2004ry,Minahan:2006sk}.


\subsection{$\superN=4$ Super Yang--Mills Theory}

Four-dimensional $\superN=4$ super Yang--Mills (SYM) theory was originally introduced in the $1980$s as the dimensional reduction of a ten dimensional super Yang--Mills theory with fermions \cite{Gliozzi:1976qd,Brink:1976bc}. Compactification on a six dimensional torus gives rise to a four-dimensional field content comprised of a gauge field $\gf{A}_\mu$, four Dirac spinors $\Psi_a$, $\bar\Psi^a$ as well as six scalars $\Phi_k$, $k=1,\dots,6$. Furthermore we have an adjoint covariant derivative $\gf{D}_\mu=\partial_\mu-ig \comm{\gf{A}_\mu}{\,\cdot\,\,}$.
The Lagrangian reads 
\begin{align}
\mathcal{L}_{\mathrm{YM}}=\Tr \Big[&
\quarter  \gf{F}^{\mu\nu}\gf{F}_{\mu\nu}
+\half  \gf{D}^\mu \Phi^n \gf{D}_\mu \Phi_n 
+ \bar\Psi_{\dot \alpha}^a\sigma_\mu^{\dot \alpha \beta}\gf{D}^\mu\Psi_{\beta a}
- \quarter g^2 \comm{\Phi^m}{\Phi^n}\comm{\Phi_m}{\Phi_n}
\nonumber\\
&- \half i g   \Psi_{a\alpha}\Sigma^{ab}_m\eps^{\alpha\beta}\comm{\Phi^m}{\Psi_{\beta b}}
- \half i g \bar\Psi_{\dot \alpha}^a\Sigma^m_{ab}\eps^{\dot \alpha \dot \beta}\comm{\Phi_m}{\bar \Psi_{\dot \beta}^b}
\Big],
\label{eq:Lagrange}
\end{align}
with a field strength defined by $\gf{F}_{\mu\nu}=i g^{-1}\comm{\gf{D}_\mu}{\gf{D}_\nu}$.%
\footnote{Here, spacetime indices are denoted by $\mu, \nu,.\,.=1,.\,.\,,4$ while spinor indices of $\alg{su}(2)$ given by $\alpha,\beta,.\,.$ or $\dot\alpha,\dot\beta,.\,.$ take two values. Vector and spinor indices of $ \alg{so}(6)\simeq\alg{su}(4)$ are denoted by $m,n,.\,.$ and $a,b,.\,.$ and range from $1$ to $6$ or $1$ to $4$, respectively.}  The four- and six- dimensional sigma matrices obey the relations
\begin{equation}
 \{\sigma^\mu,\sigma^\nu\}=\eta^{\mu\nu},
\qquad\quad
\{\Sigma^m,\Sigma^n\}=\eta^{mn}.
\end{equation}
All fields transform in the adjoint representation of a semisimple Lie group. In what follows we take this gauge group to be $\grp{SU}(N)$. The \sym\ action is uniquely determined up to two free paramters; the dimensionless coupling constant $g_{\mathrm{YM}}$ and the rank of the gauge group $N$.%
\footnote{Here we ignore a topological term $\sim\theta \,\gf{F}\gf{\tilde F}$.}
\paragraph{Symmetry.} 
The different indices appearing in the Lagrangian represent different symmetries. Contraction of all indices shows the symmetry invariance of the respective terms:
\begin{itemize}
\item Spacetime symmetry: The indices $\mu,\nu$ correspond to the vector representation of the Lorentz group. Indices $\alpha, \beta$ and $\dot\alpha,\dot \beta$ represent the left and right handed spinor representations of the Lorentz group.
\item Global internal symmetry (R-symmetry): The Lagrangian has a global internal $\grp{SO}(6)\simeq \grp{SU}(4)$ symmetry acting on the $\grp{SO}(6)$ vector indices $m,n=1,\dots,6$.
\item Local internal symmetry (Gauge Symmetry): The Lagrangian has a local $\grp{SU}(N)$ gauge symmetry. Above the respective indices are hidden in the trace $\Tr$.
\end{itemize}
In fact, the Poincar\'e and R-symmetry of the Lagrangian are enhanced to \emph{superconformal symmetry}. To be precise, the action of this four-dimensional quantum field theory is invariant under the set of generators%
\footnote{Note that the $\gen{Q}$'s here denote supercharges and should not be confused with the local charges $\mathcal{Q}_n$ discussed in the previous section.}
\begin{equation}
\{\gen{L},\gen{\bar L},\gen{P},\gen{K},\gen{D},\gen{R}|\gen{Q},\gen{\bar Q},\gen{S},\gen{\bar S}\}\quad\in\quad\alg{psu}(2,2|4), 
\end{equation}
which span the $\superN=4$ superconformal algebra divided by its center, cf.\ e.g.\ \cite{Beisert:2010kp}.
 The above generators correspond to the set of Lie algebra (level-zero) generators $\levz$ that we encountered in the previous sections. 
They satisfy the graded (due to the fermionic supersymmetry generators) commutation relations%
\footnote{In this section we distinguish between upper and lower adjoint indices.} 
\begin{equation}
\scomm{\levz_a}{\levz_b}=\struc_{ab}{}^c\,\levz_c.
\end{equation}
This Lie algebra contains the Poincar\'e Lorentz- and momentum generators $\gen{L}$ ,$\gen{\bar L}$ and $\gen{P}$ as well as the momentum supercharges $\gen{Q}$ and $\gen{\bar Q}$. Being conformal, the symmetry algebra also encloses the conformal boost $\gen{K}$, the dilatation generator $\gen{D}$ and the conformal supercharges $\gen{S}$ and $\gen{\bar S}$; all fields of the theory are massless. The action is invariant under an $\alg{su}(4)$ internal symmetry contained in $\alg{psu}(2,2|4)$ and identified with the generators $\gen{R}$. This R-symmetry rotates the supercharges into each other.%
\footnote{This form of supersymmetry with generators transforming non-trivially under the internal R-symmetry was referred to as \emph{hypersymmetry} in the original work on \sym\ \cite{Fayet:1975yi,Gliozzi:1976qd}.}

One of the most remarkable features of \sym\ is the fact that its coupling constant is constant, i.e.\ independent of the renormalization scale $\mu$, cf.\ \cite{Sohnius:1981sn,Mandelstam:1982cb,Howe:1983sr,Brink:1982pd}:
\begin{equation}
 \beta =\mu\frac{\partial g}{\partial \mu}=0.
\end{equation}
  This scale independence implies that (super)conformal symmetry is preserved at the quantum level making this model the paradigm of four-dimensional quantum field theories.
In fact, this large amount of symmetry is further enhanced in the so-called planar limit, which turns this gauge theory into an integrable model.

\paragraph{Planar Limit.}

We consider the action of \sym\ in terms of the standard Yang--Mills coupling $g_\mathrm{YM}$:
\begin{equation}
 S_{\mathrm{YM}}=\frac{2}{g_{\mathrm{YM}}^2} \int \dd ^4x \,\mathcal{L}_\mathrm{YM}(g=1).
\end{equation}
For the large-$N$ limit it proves useful to redefine the free parameters of the theory by introducing the so-called 't Hooft coupling  \cite{'tHooft:1973jz}
\begin{equation}
 \lambda=g_{\mathrm{YM}}^2 N.
 \label{eq:thooft}
\end{equation}
The limit $N\to\infty$, $\lambda=\text{fixed}$ is called the \emph{'t~Hooft}, \emph{large-$N$} or \emph{planar limit}. The latter name stems from the fact that in this limit only Feynman diagrams contribute that can be drawn on a plane (as opposed to different topologies).
Taking $N$ to infinity and restricting to the leading perturbative order, the 't Hooft coupling $\lambda$ is the essential expansion parameter in the planar limit. It is related to the coupling constant $g$ in \eqref{eq:Lagrange} by $\lambda=8 \pi^2 g^2$.

The above large-$N$ limit was originally introduced in 1973 by 't Hooft who investigated $\grp{U}(N)$ gauge theories with regard to inseparable quark bound states as found in QCD \cite{'tHooft:1973jz}. The 't Hooft limit was his approach to simplifying the strong coupling behavior of QCD. He also noticed that in this limit the expansion of correlators very much resembles the genus expansion in a string theory with coupling $g_\mathrm{s}=1/N$. Later this became manifest in form of the AdS/CFT correspondence, c.f.\ \secref{sec:adscft}.

In the planar limit \sym\ acquires an additional symmetry, namely \emph{integrability} which leads to many simplifications in explicit calculations and which is realized on several types of observables in the form of Yangian symmetry (see below).

\paragraph{Gauge invariant operators and spin chain picture.}
The prime observables of a conformal field theory are correlators of gauge invariant local operators. In fact, the knowledge of all two point-functions is equivalent to knowing the spectrum of the theory, while the three-point functions encode the conformal structure constants and thus the dynamics. Hence, it is important to understand how these local operators look like.
 
All fields of \sym\ transform in the adjoint representation of the gauge group $\grp{SU}(N)$. Thus, we can associate a fundamental and an anti-fundamental color index to each of the fields $X^\gi{a}{}_\gi{b}\in\{\gf{D}_\mu,\Psi_{\alpha a},\bar\Psi_{\dot \alpha}^a,\Phi_k,\gf{F}_{\mu\nu}\}^\gi{a}{}_\gi{b}$. A gauge transformation acts as $X\mapsto UXU^{-1}$ .
Taking color traces of products of fields 
transforming homogeneously under gauge transformations, we can thus construct gauge invariant local operators as: 
\begin{equation}
\mathcal{O}(x)= \Tr [X_1 X_2 \dots X_L](x)
=(X_1)^{\gi{a}_1}{}_{\gi{a}_2} (X_2)^{\gi{a}_2}{}_{\gi{a}_3}\dots (X_L)^{\gi{a}_L}{}_{\gi{a}_1}(x).
\label{eq:traces}
\end{equation}
Here all fields $X_i$ are understood to be evaluated at the same spacetime point $x^\mu$.%
\footnote{Note that the gauge field $\gf{A}_\mu$ cannot be used for the construction of such states because it transforms inhomogeneously under gauge transformations.}
Since all gauge indices are contracted, it is clear that operators of this form are gauge singlets. 
In the following we will refer to the trace operators \eqref{eq:traces} also as local gauge invariant \emph{states} and make use of the identification
\begin{equation}
 \Tr[{X_1\dots X_L}]\equiv \ket{X_1\dots X_L}.
\end{equation}

Obviously, also products of traces will be gauge invariant operators. However, in the strict%
\footnote{If we speak of the \emph{strict} large~$N$ limit here, we mean that only the leading order in $1/N$ is kept.} 
large-$N$ limit these are not relevant for the correlators we are interested in.
The set of states of the form \eqref{eq:traces} thus forms a (cyclic) basis for local gauge invariant states in \sym\ at large~$N$.

The above trace operators can be considered as a tensor product of fields transforming under a representation of the theory's symmetry algebra plus additional cyclic boundary conditions. In other words we may call the above basis states \emph{spin chains} with cyclic boundary conditions, e.g.\
\begin{align}
&\Tr{\Phi_1\,\dots\,\Phi_1\,\Phi_2\,\Phi_1\,\dots\,\Phi_1}.\nonumber\\
&\hspace{-.3cm}\includegraphicsbox[scale=.7]{FigCyclicChainOneMagnon}
\end{align}
In particular, the representation of the superconformal symmetry takes the tensor product form
\begin{equation}\label{eq:symtraces}
\levz_a=\sum_{k=1}^L\levz_{a,k},\qquad \levz_a\in\alg{psu}(2,2|4),
\end{equation}
where $\levz_{a,k}$ denotes the representation on one of the fields.
Except for the Lorentz- and internal rotations $\gen{L}$, $\gen{\bar L}$ and $\gen{R}$, the representations of all symmetry generators of \sym\ acquire radiative corrections in the coupling constant when promoted to higher loop orders
\begin{equation}
\levz_a(g)=\levz_a^{(0)}+g\,\levz_a^{(1)}+g^2\,\levz_a^{(2)}+\dots,\qquad \levz_a\in\alg{psu}(2,2|4).
\end{equation}
The graded commutation relations of $\alg{psu}(2,2|4)$ with structure constants $\struc_{ab}{}^c$ are not affected by these deformations
\begin{equation}
\scomm{\levz_a(g)}{\levz_b(g)}=\struc_{ab}{}^c\,\levz_c(g).
\end{equation}
In what follows we will refer to the perturbative order $g^{2\ell}$ as the $\ell$-loop order.

\subsection{Dilatation Operator}
\label{sec:dilatationop}
In this section we briefly indicate, how Yangian symmetry acts on the dilatation operator alias the Hamiltonian of $\superN=4$ super Yang--Mills theory.
In fact, studying eigenvalues of the dilatation operator is equivalent to studying the energy spectrum of this theory. 
This can be seen in a particular radial quantization scheme (cf.\ e.g.\ \cite{Fubini:1972mf,Minwalla:1997ka}), where the dilatation operator generates time shifts 
\begin{equation}
 \gen{ D}=-i r\frac{\partial}{\partial r}=-i \frac{\partial}{\partial t}.
\end{equation}
Hence, studying conformal dimensions $\Delta$ of gauge invariant states $\mathcal{O}$ with $\gen{D}\mathcal{O}=\Delta \mathcal{O}$
is equivalent to the study of the energy spectrum of these states. Thus, one often does not distinguish between the terms energy spectrum and anomalous dimensions, and the dilatation operator is referred to as the Hamiltonian of the theory.

\paragraph{Integrable structures in the \texorpdfstring{$\alg{su}(2)$}{su(2)} subsector at one loop.}
As indicated above, solving the spectral problem of local operators in \sym\ reduces to the problem of finding the spectrum of the dilatation operator $\gen{D}$. Furthermore we have seen that a basis for gauge invariant local operators is given by traces of the form \eqref{eq:traces}:
\begin{equation}
\mathcal{O}=\Tr[\field_1 \field_2\dots \field_L].
\end{equation}
Therefore it suggests itself to think about diagonalizing $\gen{D}=\gen{D}_0+\delta \gen{D}(g)$ on this basis set of states. 

Diagonalizing the anomalous part of the dilatation operator $\Da(g)$ is a very challenging task even at one-loop order. Thus, it appears reasonable to think of dividing the problem into smaller pieces, i.e.\ to diagonalize the dilatation generator on a subset of local gauge invariant states. All such subsectors closed under the action of the dilatation generator (i.e.\ closed under renormalization) were identified in \cite{Beisert:2003jj}. As a result one finds several closed sectors, each characterized by its field content and the residual symmetry. For this classification of subsectors, it proves useful to combine the six scalar fields $\scalar_i$ of \sym\ into complex fields 
\begin{equation}
\mathcal{X}=\scalar_1+i\scalar_2,\quad
\mathcal{W}=\scalar_3+i\scalar_4,\quad
\mathcal{Z}=\scalar_5+i\scalar_6,
\end{equation}
and their conjugates. The simplest subsector is then given by the (half-BPS) states of the form
\begin{align}
\Tr{\mathcal{Z}^L}=&\Tr{\mathcal{Z}\mathcal{Z}\mathcal{Z}\dots\mathcal{Z}},\nonumber\\
&\hspace{-1.2cm}\includegraphicsbox[scale=.7]{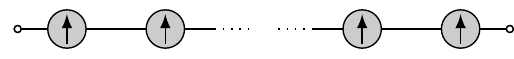}
\end{align}
where the picture again emphasizes the analogy between trace operators and spin chains. We will refer to these states as the cyclic vacuum states of length $L$.
The name vacuum already indicates the relation to the ferromagnetic vacuum of a spin chain. 
The vacuum has the residual symmetry%
\footnote{Note that there is no known vacuum state with residual symmetry $\alg{psu}(2,2|4)$ which would appear natural in \sym.} 
\begin{equation}
\alg{psu}(2|2)\times\alg{psu}(2|2) \times\alg{u}(1)^3,
\end{equation}
where all generators except for the length measuring operator act trivially. That is, the length $L$ is the only non-vanishing quantum number characterizing the cyclic vacuum state. Hence, diagonalizing the anomalous dilatation generator on these states is trivial.

In order to proceed to the simplest \emph{non-trivial} subsector, we excite also the field $\mathcal{W}$ to consider states of the form
\begin{align}
&\Tr{\mathcal{Z}\dots\mathcal{Z}\mathcal{W}\mathcal{Z}\dots\mathcal{Z}}\nonumber\\
&\hspace{-.45cm}\includegraphicsbox[scale=.7]{FigCyclicChainOneMagnon}
\end{align}
The residual symmetry of this sector is
\begin{equation}
 \alg{su}(2)\times\alg{u}(1)^2,
\end{equation}
and it is therefore referred to as the $\alg{su}(2)$ sector of \sym.
The fields $\mathcal{Z}$ and $\mathcal{W}$ transform under the fundamental representation of $\alg{su}(2)$ and the $\alg{u}(1)$ charges are given by the length operator and the anomalous dimension $\Da$. The resulting states can be identified with spin chains built out of fundamental modules of $\alg{su}(2)$, i.e.\ with chains of the Heisenberg type encountered above.

The one-loop dilatation operator in the $\alg{su}(2)$ subsector can be found by explicitly renormalizing the involved Feynman diagrams and reading off the renormalization constants. At one loop order Minahan and Zarembo found the following expression in their famous paper \cite{Minahan:2002ve}:%
\footnote{Note that including all scalar fields of \sym\ leads to the $\alg{so}(6)$ subsector which is closed only at one loop order in perturbation theory. In the original work \cite{Minahan:2002ve}, however, one-loop integrability was shown in the whole scalar $\alg{so}(6)$ sector including the $\alg{su}(2)$ sector.}
\begin{equation}
 \Da_{\alg{su}(2)}(g)=\frac{g^2}{2}\sum_{k=1}^L (\idop_k-\permop_{k,k+1})+\order{g^3}=\frac{g^2}{2}(\perm{1}-\perm{2,1}).
\label{eq:Dsu2}
\end{equation}
Here $\idop_k$ and $\permop_{k,k+1}$ denote the identity and permutation operator acting on sites $k$ and $k+1$, respectively. Due to the periodicity of the trace states on which this operator acts, we identify the sites $L+1$ and $1$. Remarkably, the one-loop anomalous dilatation generator in \eqref{eq:Dsu2} equals the Hamiltonian of the Heisenberg $\text{XXX}_{\frac{1}{2}}$ spin chain (with cyclic boundary conditions), the prime example of an integrable model. Its discovery was a huge breakthrough in the study of \sym.

\paragraph{Higher Loop Integrability: Long-Range Integrable Spin Chains.}
\label{sec:higherloop}
It was shown in \cite{Beisert:2003tq} that the integrable structures of the spectral problem in \sym\ extend to higher loop orders. Noting the existence of degeneracies in the spectrum at two-loop order, it was possible to construct perturbative corrections to the first two integrable charges in the $\alg{su}(2)$ subsector
\begin{equation}
\charge_2(g)=\charge_2^{(0)}+g^2\charge_2^{(1)}+\order{g^3},\qquad
\charge_3(g)=\charge_3^{(0)}+g^2\charge_3^{(1)}+\order{g^3},
\end{equation}
such that these operators commute
\begin{equation}
\comm{\charge_2^{(0)}}{\charge_3^{(0)}}
+g^2\bigbrk{\comm{\charge_2^{(1)}}{\charge_3^{(0)}}
+\comm{\charge_2^{(0)}}{\charge_3^{(1)}}}+\order{g^3}=0.
\end{equation}
Importantly, as opposed to the one-loop level, these deformed charges are only \emph{perturbatively integrable}, i.e.\ they commute up to higher powers in the coupling constant.

Also the interaction range of the higher charge orders increases with the power of the coupling constant, e.g.\ the two loop correction to the dilatation generator acts on three sites at the same time, see also \figref{fig:longcharge}:
\begin{equation}
\charge_{2,\alg{su}(2)}^{(1)}=-3\perm{1}+4\perm{2,1}-\perm{3,2,1}.
\end{equation}
In fact, this property is expected with regard to Feynman diagram calculations. The more powers of the coupling contribute to a given perturbative order, the more fields can be involved into the interactions. 

This is exactly the type of long-range spin chain that we considered in \secref{sec:moreboosts}. We note that in fact the higher perturbative orders of the dilatation operator can be generated by the generalized boost operators discussed above \cite{Bargheer:2008jt,Bargheer:2009xy}. The corresponding two-loop deformation of the Yangian level-one generators was given in \eqref{eq:defYang}.

\begin{figure}
\begin{center}
$\charge_2(g)=$
\,\,
\includegraphicsbox[scale=1]{FigTwoLeg}
$+\,\,g^2$\hspace{-.4cm}
\includegraphicsbox[scale=1]{FigThreeLeg}
$+\,\,g^4$
\includegraphicsbox[scale=1]{FigFourLeg}
$+\quad\order{g^5}$
\caption{The interaction range of the integrable long-range charges grows with increasing order of the coupling constant.}
\label{fig:longcharge}
\end{center}
\end{figure}

\paragraph{Yangian symmetry of the dilatation operator.}

Having identified the one-loop dilatation operator with the Heisenberg spin chain Hamiltonian, the above example \eqref{eq:XXXbound} immediately shows that the Yangian algebra $Y[\alg{su}(2)]$ commutes with this operator up to boundary terms. As demonstrated by Dolan, Nappi and Witten \cite{Dolan:2003uh}, this property extends to the complete one-loop dilatation operator of \cite{Beisert:2003jj} which commutes with the level-one generators of $Y[\alg{psu}(2,2|4)]$ into boundary terms:
\begin{equation}
\comm{\levo_a}{\delta \gen{D}^{(1)}}\simeq \levz_{a,1}-\levz_{a,L}.
\end{equation}
In order to extend this relation to higher loops one major difficulty is that not even the two-loop dilatation operator is known for the complete theory, i.e.\ on the full $\alg{psu}(2,2|4)$ spin chain. However, the (asymptotic) higher loop dilatation operator is known in certain subsectors and some statements on the Yangian symmetry can be made, see e.g.\ \cite{Zwiebel:2006cb,Bargheer:2008jt,Bargheer:2009xy}. In the $\alg{su}(2)$ sector for instance, the representation of the Yangian generators may be deformed using the boost and bilocal charges of \secref{sec:moreboosts}.
\subsection{Scattering Amplitudes}

In this section we briefly indicate how Yangian symmetry is realized on the scattering matrix of \sym.

\paragraph{Four Dimensional Kinematics.}
We are interested in scattering of $n$ massless fields in \sym. Therefore it is useful to express the $n$ external four-momenta $p_k^\mu$ as bi-spinors $p_k^{\alpha\dot \alpha}=(\sigma_\mu)^{\alpha\dot \alpha} p_k^\mu$ and explicitly solve the on-shell condition $p_k^2=0$ for all external particles in terms of commuting spinors \cite{Nair:1988bq} 
\begin{equation}
p_k^{\alpha\dot \alpha}=\lambda_k^\alpha\bar\lambda_k^{\dot \alpha},\quad \mbox{for}\quad k=1,\dots,n.
\label{eq:spindec}
\end{equation}
Here $\lambda^\alpha_k$ and $\bar\lambda^{\dot \alpha}_k$ are complex conjugate bosonic Lorentz-spinors with indices $\alpha,\beta,\ldots=1,2$ and $\dot \alpha,\dot \beta,\ldots=1,2$.
The spinor decomposition \eqref{eq:spindec} of massless momenta in four dimensions is unique only up to a complex rescaling
\begin{equation}
\lambda^\alpha\to c\lambda^\alpha,\qquad
\bar\lambda^{\dot \alpha}\to c^{-1} \bar\lambda^{\dot \alpha}.
\end{equation}
All physical quantities should therefore be independent of this transformation.%
\footnote{Physical scattering amplitudes require at least two negative energy particles. This is due to the two constraining equations $\sum_{k=1}^np_k=0$ and $p_n^2=\Big(\sum_{k=1}^{n-1}p_k\Big)^2=0$. Here we will focus on positive energy solutions and consider all particles as incoming in what follows. The arguments in this chapter generalize to the inclusion of negative energy particles which, however, results in less clear expressions, cf.\ \cite{Bargheer:2009qu}.}

It is straightforward to construct invariants under Lorentz symmetry out of the momentum spinors according to
\begin{equation}\label{eq:spinbrac}
\spaa{ij}:=\spaa{\lambda_i \lambda_j}=\eps_{\alpha \beta}\lambda_i^\alpha\lambda_j^\beta,\qquad
\spbb{ij}:=\spbb{\bar\lambda_i \bar\lambda_j}=\eps_{\dot \alpha\dot \beta}\bar \lambda_i^{\dot \alpha}\bar\lambda_j^{\dot \beta}.
\end{equation}
These spinor brackets furnish fundamental building blocks for the construction of scattering amplitudes as we will see below.
\paragraph{Color Ordering.}
Tree-level scattering amplitudes in $\grp{SU}(N)$ \sym\ can be expanded according to
\begin{equation}\label{eq:colorord}
 \hat A_n(\{\lambda_i,h_i,a_i\})=
 \sum_{\sigma\in S_n/\mathbb{Z}_n} A_n(\lambda_{\sigma(1)},h_{\sigma(1)},\dots,\lambda_{\sigma(n)},h_{\sigma(n)}) \Tr\gen{T}^{a_{\sigma(1)}}\gen{T}^{a_{\sigma(2)}}\dots \gen{T}^{a_{\sigma(n)}},
\end{equation}
such that the amplitude's color structure is encoded in traces over gauge group generators $\gen{T}^a$ of $\alg{su}(N)$.%
\footnote{At higher loop orders and at the same time going beyond the planar limit also multi-trace contributions have to be added to this expansion. Here we will be interested in the planar tree level where only single traces appear.} Here the symbol $h_i$ denotes the helicity of the $i$th particle. This straightforward separation of color and kinematical structure allows to reduce the non-trivial scattering problem to the kinematical part of the amplitude $A_n$. The cyclicity of the trace implies that this kinematical scattering amplitude $A_n$ is a function invariant under cyclic permutations of its arguments.

\paragraph{Superfield.}
In order to compute scattering amplitudes in \sym\ it is most convenient to make use of the fact that fields with different helicity transform in different representations of the internal R-symmetry. We may thus introduce fermionic spinors $\eta^A$, $A,B,\ldots=1,\dots,4$, of $\alg{su}(4)$ and collect all fields in a chiral on-shell superfield
 \cite{Mandelstam:1982cb,Brink:1983pd}
\begin{align}
\superfield(\lambda,\bar\lambda,\eta)
=
&G^+(\lambda,\bar\lambda)
+\eta^A \Gamma_A(\lambda,\bar\lambda) 
+\sfrac{1}{2}\eta^A\eta^B S_{AB}(\lambda,\bar\lambda) 
\nonumber\\
&+\sfrac{1}{3!} \varepsilon_{ABCD} \eta^A\eta^B \eta^C {\bar \Gamma}^D(\lambda,\bar\lambda) 
 +\sfrac{1}{4!}\varepsilon_{ABCD}\eta^A\eta^B \eta^C\eta^D  G^-(\lambda,\bar\lambda).
\end{align}
Here each power of the Gra{\ss}mann parameters $\eta$ corresponds to a different representation of the R-symmetry.  The on-shell gluons $G^\pm$, fermions $\Gamma$/$\bar\Gamma$, and scalars $S$ have helicity $\pm 1$, $\pm \half$ and $0$.

\begin{figure}
\begin{center}
\includegraphics[scale=1]{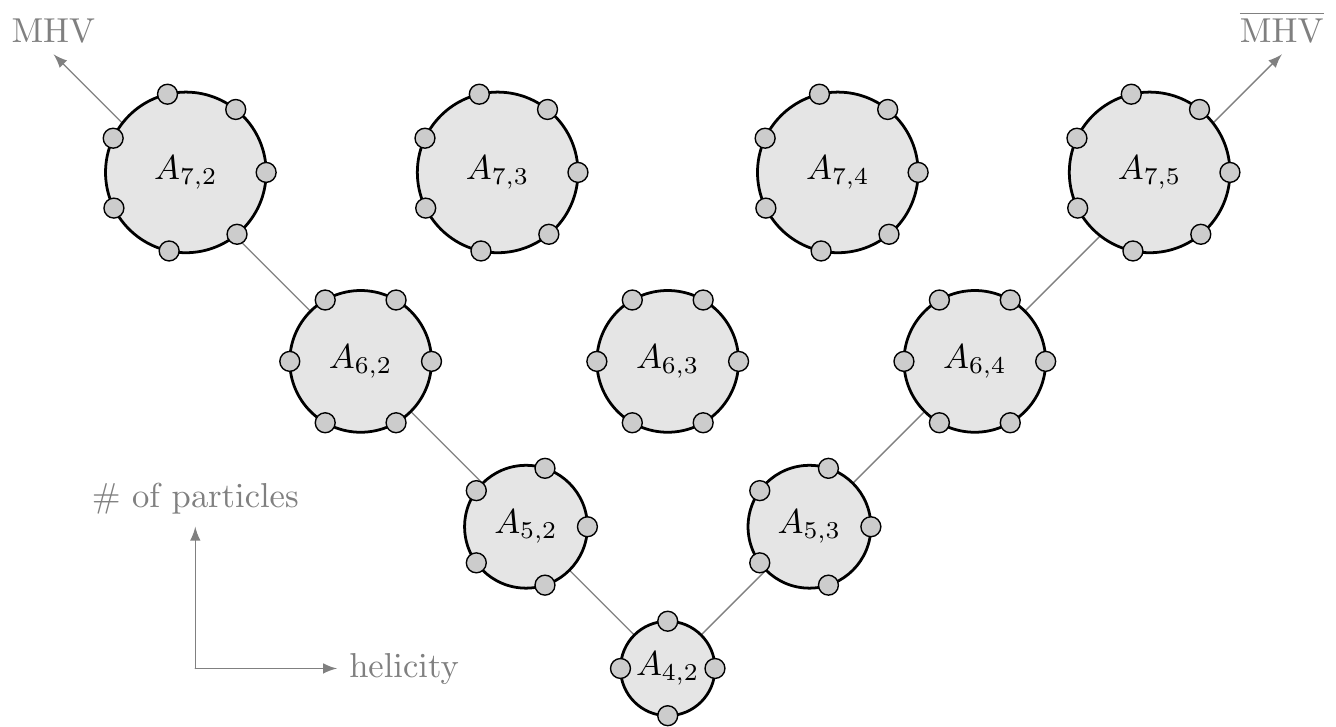}
\caption{Scattering amplitudes $A_{n,k}$ in \sym\ are nontrivial for $n>3$ external legs (and real momenta). They can 
be classified according to their helicity configuration measured by the respective power of the fermionic spinors $\eta^{4k}$. The simplest so-called $\text{MHV}$ and $\overline{\text{MHV}}$ amplitudes are those on the left and right boundary of the above triangle.}
\label{fig:ampetaexpansion}
\end{center}
\end{figure}

Every analytic function of the superfield $\superfield$ can be expanded in terms of the Gra{\ss}mann superspace coordinates $\eta$. The fields contributing to a certain order in this expansion are determined by the respective power in $\eta$. In particular, singlets of $\alg{su}(4)$, i.e.\ symmetry invariant functions of the superfield, are proportional to $\eta^{4}=\sfrac{1}{4!}\eps_{ABCD}\eta^A\eta^B\eta^C\eta^D$. Considering the $n$-particle scattering amplitude as a superspace function 
\begin{equation}
 A_n(\superfield_1,\ldots,\superfield_n)=A_n(\superfield(\Lambda_1),\ldots,\superfield(\Lambda_n)),
\qquad 
\Lambda_k=(\lambda_k,\bar\lambda_k,\eta_k)
\end{equation}
we may thus expand it in terms of component amplitudes given by the coefficients of powers of $\eta^4$
\begin{equation}
A_n=\sum_{k=2}^{n-2}A_{n,k}, \qquad\qquad \hel A_{n,k}=4 k A_{n,k}.
\label{eq:ampexpan}
\end{equation}
Here we have introduced the $\eta$ counting generator $\hel=\eta^A\partial/\partial \eta^A$ and used that supersymmetry implies $A_{n,1}=A_{n,n-1}=0$ \cite{Grisaru:1976vm,Grisaru:1977px}. In Minkowski signature the three-particle scattering amplitude $A_3$ of massless particles vanishes by kinematical arguments. Hence, the lowerst non-trivial amplitude is $A_4$, see \figref{fig:ampetaexpansion}.

As seen in \eqref{eq:ampexpan}, scattering amplitudes in \sym\ are categorized according to their helicity configuration $A_n=A_{n,2}+A_{n,3}+\dots$. Remarkably, the so-called maximally helicity violating (MHV) amplitudes $A\supup{MHV}_n=A_{n,2}$ can be written in a very compact fashion \cite{Parke:1986gb,Berends:1987me,Nair:1988bq}
\begin{equation}
A\supup{MHV}_n
=
\frac{\deltad{4}(P)\,\deltad{8}(Q)}{\tprods{1}{2}\tprods{2}{3}\ldots \tprods{n}{1}}\,,
\qquad\quad
P^{\alpha\dot \beta}=\sum_{k=1}^n \lambda^\alpha_{k}\bar\lambda^{\dot \beta}_{k}\,,
\quad
Q^{\alpha B}=\sum_{k=1}^n \lambda^\alpha_{k}\eta^B_{k}\,,
\label{eq:MHV}
\end{equation}
with the Lorentz-invariant spinor brackets defined in \eqref{eq:spinbrac}. This definition of the amplitude ensures conservation of the overall momentum $P$ and super-momentum $Q$.

\paragraph{Level-Zero Symmetry.}

Using spinor helicity superspace coordinates
the one-particle representation of the superconformal algebra $\alg{psu}(2,2|4)$
was written down by Witten
\cite{Witten:2003nn}:
\begin{equation}\label{eq:freerep}
\begin{array}[b]{@{}rclcrcl@{}}
\gen{L}^a{}_b\eq \lambda^a\partial_{b}-\half\delta^a_b \lambda^c\partial_c,
&&
\gen{\bar L}^{\dot \alpha}{}_{\dot \beta}\eq 
\bar\lambda^{\dot \alpha}\bar\partial_{\dot \beta}
-\half\delta^{\dot \alpha}_{\dot \beta} \bar\lambda^{\dot \gamma}\bar \partial_{\dot \gamma},
\\[1ex]
\gen{D}\eq \half\partial_\gamma\lambda^\gamma+\half\bar\lambda^{\dot \gamma}\bar \partial_{\dot \gamma},
&&
\gen{R}^A{}_B\eq \eta^A\partial_B-\quarter \delta^A_B\eta^C\partial_C,
\\[1ex]
\gen{Q}^{\alpha B}\eq \lambda^\alpha\eta^B ,
&&
\gen{S}_{\alpha B}\eq \partial_\alpha\partial_B  ,
\\[1ex]
\bar{\gen{Q}}^{\dot \alpha}_{B}\eq \bar\lambda^{\dot \alpha} \partial_B ,
&&
\bar{\gen{S}}_{\dot \alpha}^B\eq \eta^B \bar\partial_{\dot \alpha} ,
\\[1ex]
\gen{P}^{\alpha\dot \beta}\eq \lambda^\alpha\bar\lambda^{\dot \beta} ,
&&
\gen{K}_{\alpha\dot \beta}\eq \partial_\alpha\bar\partial_{\dot \beta},
\end{array}
\end{equation}
where we use the short-hand notation $\partial_\alpha=\partial/\partial\lambda^\alpha$, 
$\bar\partial_{\dot \alpha}=\partial/\partial\bar\lambda_{\dot \alpha}$ 
and $\partial_A=\partial/\partial\eta^A$. 
The above one-particle representation $\eqref{eq:freerep}$ is promoted to a representation on tree-level scattering amplitudes in \sym\ by taking the tensor product, i.e.\ the primitive coproduct:
\begin{equation}\label{eq:RepLO}
\levz_a=\sum_{k=1}^n \levz_{a,k}.
\end{equation}
Here $\levz_{a,k}$ is the representation of the 
conformal symmetry generator $\levz_a$ 
on the $k$-th leg $(\lambda_k,\bar\lambda_k,\eta_k)$ 
of $A_n$ as specified in \eqref{eq:freerep}. Hence, the representation of the symmetry algebra on scattering amplitudes very much resembles the spin chain symmetry acting on local gauge invariant states \eqref{eq:symtraces}. In fact, one may check that the scattering amplitude $A_n$ is invariant under the action of the above generators \cite{Witten:2003nn}:%
\footnote{Importantly, there are further corrections to the above expressions for the conformal level-zero generators in the limit where two external momenta of the amplitude become collinear. These limits require careful treatment and can also be tackled by algebraic considerations, see e.g.\ \cite{Bargheer:2009qu,Beisert:2010gn,Bargheer:2011mm}.}
\begin{equation}\label{eq:levzA}
\levz_a A_n=0.
\end{equation}
\paragraph{Level-One Symmetry.}

We may define the level-one generators in the expected form
\begin{equation}
\levo_a=f^{a}{}_{bc} \sum_{1\leq j<k\leq n}\levz_{j}^b \,\levz_{k}^c.
\end{equation}
with a vanishing one-site representation as in \eqref{eq:zerorep}.
Due to the commutation relation \eqref{eq:Yangcomms} of the Yangian algebra given by
\begin{align}\label{eq:gradedYang}
[\levz_a,\levz_b\}&=f_{abc}\levz^c,
&
[\levz_a,\levo_b\}&=f_{abc}\levo^c,
\end{align}
it suffices to show the invariance of $A_n$ under the level-zero symmetry and \emph{one} level-one generator. This will imply invariance under the whole Yangian algebra via \eqref{eq:gradedYang}. For the explicit calculation it makes sense to choose the simplest level-one generator which is the level-one momentum operator $\gen{\widehat P}$ being linear in derivatives. This generator takes the explicit form
\begin{equation}
\gen{\widehat P}^{\alpha\dot\alpha}=\sum_{1\leq j<k\leq n}\Big[\gen{P}_j^{\gamma\dot \gamma}\big(\gen{L}^\alpha_{k,\gamma}\delta_{\dot \gamma}^{\dot \alpha}+ \gen{\bar L}_{k,\dot \gamma}^{\dot\alpha}\delta_\gamma^\alpha+\gen{D}_k \delta_\gamma^\alpha\delta_{\dot \gamma}^{\dot \alpha}\big)+\gen{Q}_j^{\alpha C} \gen{\bar Q}_{k,C}^{\dot \alpha}-(j\leftrightarrow k)\Big].
\end{equation}
Here the $(j\leftrightarrow k)$ stands for all the previous terms under the sum but with $j$ and $k$ interchanged.
Acting with the level-one momentum operator on the MHV amplitude, one finds (see \cite{Beisert:2010jq} for more details)
\begin{equation}
\gen{\widehat P}^{\alpha \dot\alpha} A_{n}^\text{MHV}=\frac{\lambda_1^\gamma\lambda_n^\beta+\lambda_1^\beta\lambda_n^\gamma}{\spaa{1n}}\epsilon_{\gamma\delta}\gen{P}^{\delta\dot\alpha}A_{n}^\text{MHV}=0.
\end{equation}
Remember that due to \eqref{eq:colorord}
scattering amplitudes are invariant under cyclic shifts. Hence, we have a case of cyclic boundary conditions discussed in \secref{sec:boundconds}. Fortunately, the dual coxeter number $\cox$ of the symmetry algebra $\alg{psu}(2,2|4)$ of $\superN=4$ SYM theory is zero and scattering amplitudes are invariant under the level-zero symmetry \eqref{eq:levzA} \cite{Drummond:2009fd}. Thus, the case of \eqref{eq:cyclicinv} applies and we have a consistent realization of cyclic Yangian invariants.%
\footnote{The particular Yangian $Y[\alg{psu}(2,2|4)]$ allows for further special features such as the occurence of so-called bonus or secret symmetries, see e.g.\ \cite{Matsumoto:2007rh,Beisert:2011pn,deLeeuw:2012jf}.}

In fact, Yangian symmetry of the tree-level S-matrix of $\superN=4$ SYM theory was first understood in the language of the so-called \emph{dual conformal symmetry} \cite{Drummond:2008vq,Drummond:2009fd}. Furthermore, there is a map between all tree-level scattering amplitudes and certain contributions to the dilatation operator \cite{Zwiebel:2011bx}. In particular, the four-point superamplitude furnishes the integral kernel for the one-loop dilatation operator \cite{Zwiebel:2011bx,Wilhelm:2014qua}. Hence, the Yangian symmetry of the four-point amplitude and the dilatation operator (both discussed above) can be shown to be consistent with each other \cite{Brandhuber:2015dta}.

\paragraph{4d versus 2d S-matrix.}
Let us finally compare the tree-level S-matrix of  $\superN=4$ SYM theory to the scattering matrix of the two-dimensional field theories considered in \secref{sec:smatrix}. In particular, we are interested in two-to-two particle scattering processes. As indicated, the four-point amplitude of $\superN=4$ SYM theory obeys
\begin{equation}\label{eq:4ptYang}
\levo^a A_4=f^a{}_{bc}\sum_{k=1}^4\sum_{j=1}^{k-1}\levz_j^b\levz_k^c\,A_4=0.
\end{equation}
Here the generators of the Poincar\'e algebra enter the definition of the level-one generators $\levo^a$, i.e.\ the Yangian symmetry is not merely a Yangian of an internal symmetry algebra (cf.\ \eqref{eq:freerep}).

On the other hand, the two-particle S-matrix $\sop(u)$  of the above two-dimensional theories is subject to \eqref{eq:sconstraint}, which for $u=0$ becomes
\begin{equation}\label{eq:2dSYang}
f^{a}{}_{bc}\levz_1^b\levz_2^c\,\sop(0)+\sop(0)\,f^{a}{}_{bc}\levz_3^b\levz_4^c=0.
\end{equation}
Here we think of the S-matrix as an operator that maps the ingcoming particles $3,4$ to the outgoing particles $1,2$.
We set $u=0$ since the 2d rapidities are quantum numbers of the Poincar\'e algebra, which a priori does not form part of the 2d Yangian. In fact, one may rewrite \eqref{eq:4ptYang} in the form (see \cite{Brandhuber:2015dta} for more details)
\begin{equation}\label{eq:A4likeS}
\levo^a A_4 =f^{a}{}_{bc}\levz_1^b\levz_2^c \, A_4+f^{a}{}_{bc}\levz_3^b\levz_4^c\, A_4=0,
\end{equation}
where it was used that the expression
\begin{equation}
f^a{}_{bc}\sum_{k=1}^4\sum_{j=1}^{k-1}\levz_j^b\levz_k^c-f^{a}{}_{bc}(\levz_1^b\levz_2^c+\levz_3^b\levz_4^c)=f^{a}{}_{bc}(\levz_1^b\levz_3^c+\levz_1^b\levz_4^c+\levz_2^b\levz_3^c+\levz_2^b\levz_4^c)
\end{equation}
annihilates the four-point amplitude. Notably, \eqref{eq:A4likeS} now looks very close to \eqref{eq:2dSYang}. This illustrates the similarity between the four-point amplitude $A_4$ of $\superN=4$ SYM theory and the 2d S-matrix $\sop(0)$ evaluated at $0$. In fact, one may define a deformation $A_4(u)$ of $A_4(0)\equiv A_4$ \cite{Ferro:2012xw,Ferro:2013dga}, such that $A_4(u)$ transforms under an evaluation representation of the Yangian algebra with non-vanishing rapidity-parameter $u$ in analogy to the 2d S-matrix $\sop(u)$, cf.\ \eqref{eq:sconstraint}. While algebraically consistent, the physical interpretation of the 2d rapidity-like parameter $u$ remains to be understood in the 4d theory.

\subsection{Two Dimensions in Disguise: The AdS/CFT Correspondence}
\label{sec:adscft}

\begin{figure}
\begin{center}
\includegraphicsbox[width=8cm]{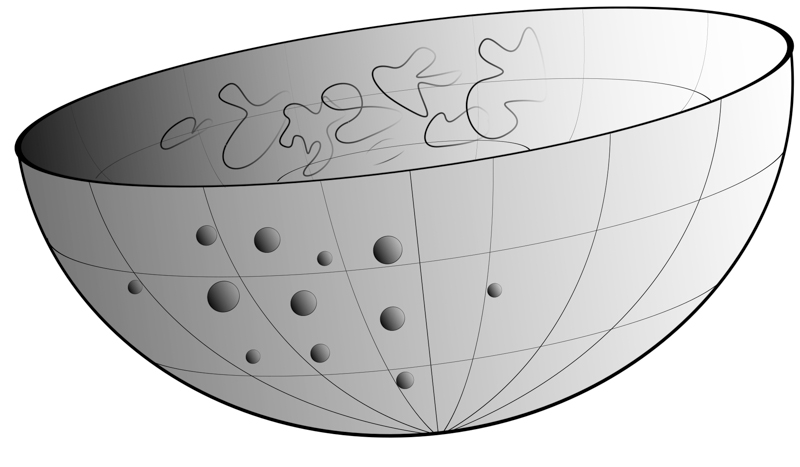}
\caption{Illustration of the AdS/CFT correspondence: Gauge theory particles on the boundary of space are dual to gravity described by strings in the bulk. While the gauge theory lives in four-dimensional spacetime, the string theory is effectively described by a two-dimensional worldsheet theory.}
\label{fig:adscft}
\end{center}
\end{figure}

What makes \sym\ an outstanding example in the class of quantum gauge theories is its relation to gravity via string theory. The four-dimensional quantum field theory discussed in this section is conjectured to be the dual description of type IIB string theory on $\mathrm{AdS_5}\times\mathrm{S}^5$ \cite{Maldacena:1998re}. As such, it describes gravitational excitations via gauge degrees of freedom.
The fact that the flat Minkowski background of the gauge theory represents the conformal boundary of the string geometry, makes this correspondence even more appealing, cf.\ \figref{fig:adscft}. The explicit map between the corresponding coupling parameters is given by
\begin{align}
\lambda&= g^2_\mathrm{YM}N=\frac{R^4}{\alpha'^2},
&
\frac{1}{N}&=\frac{4\pi g_\mathrm{s}}{\lambda}.
\end{align}
Here $R$ denotes the common radius of $\mathrm{AdS_5}$ and $\mathrm{S}^5$ while $\alpha'$ represents the string tension. The string coupling constant is given by $g_\mathrm{s}$. 

In particular, the conjectured correspondence maps the strong coupling regime of the gauge theory to weak string coupling and vice versa. While explicit calculations in \sym\ require a small coupling expansion $\lambda\ll 1$, its string dual is only accessible for small curvature $R^4/\alpha'^2=\lambda\gg 1$.
This, on the one hand, represents an obstacle for proving the duality. On the other hand, weak coupling results in either gauge or string theory provide information on the strong coupling limit of its counterpart and thus open a new door to largely unexplored areas of research. This, however, requires a verification of the AdS/CFT correspondence which, for the moment, is most promising in the limit of large $N$, where the dual theories are believed to be integrable.

Notably, the super-string theory on $\mathrm{AdS_5}\times\mathrm{S}^5$ is described by a two-dimensional worldsheet theory that vaguely resembles the principal chiral model briefly discussed in \secref{sec:nonlocandnoether} (but is more complicated). Indeed, for this classical super-string theory nonlocal charges were constructed, which are similar to the ones of \secref{sec:classint}. This shows the classical integrability of the string theory \cite{Bena:2003wd,Hatsuda:2004it}. 
Also on the gauge theory side classical integrability has been discussed, formulating the equations of motion in the language of Lax pairs or the inverse scattering method \cite{Volovich:1984kr,Volovich:1984ra,Arefeva:1985pr,Abdalla:1986xq} (see also \cite{Witten:1978xx}).
However, \sym\ represents the first four-dimensional gauge theory that was found to be integrable and many of its features remain to be understood. Certainly, the Yangian plays an important role for this ongoing journey.
In fact, Yangian symmetry was also observed for other AdS/CFT observables lying beyond the scope of this review, e.g.\ for 2d scattering matrices \cite{Beisert:2007ds,deLeeuw:2008dp,Arutyunov:2009mi,Torrielli:2011gg}, Wilson loops \cite{Muller:2013rta,Munkler:2015gja,Munkler:2015xqa,Beisert:2015uda} or tree-level three-point functions \cite{Kazama:2014sxa,Jiang:2014cya}.

\section{Summary and Outlook}

Hidden symmetries have great appeal. They explain mysterious simplifications and their identification poses exciting riddles. In these lectures we have discussed the Yangian, a particular class of hidden symmetry which appears in various physical contexts.
\medskip

In \secref{sec:classint}, we first looked at classical field theories in two dimensions. We have seen that thinking outside the box of ordinary Noether symmetries, one may find nonlocal charges that were a priori hidden. After this classical  discourse, the obvious question for the corresponding quantum symmetry arose. 
In order to understand this point, we made a step into a more mathematical direction. We followed Drinfel'd who defined the Yangian algebra to tackle an a priori unrelated problem, namely to solve the quantum Yang--Baxter equation. We have seen that addressing this problem leads to the rich mathematical framework of quantum groups and we have discussed the place of the Yangian in this context.
\medskip

We then went back to 2d field theory in order to apply our supplemented mathematical background. With L\"uscher, we understood how the quantum version of the above classical nonlocal symmetries can be defined by renormalizing their bilocal generators. A consistent definition of these charges at hand, we followed Bernard and identified them as the generators of the Yangian algebra and the field theory Lorentz boost as a realization of Drinfel'd's boost automorphism. We also realized that the scattering matrix of a 2d field theory with Yangian symmetry furnishes a solution to the quantum Yang--Baxter equation, i.e.\ the quantity that Drinfel'd was after when introducing the Yangian.
\medskip

Having studied Yangian symmetry in the context of continuous two-dimensional field theories, we thought about a discretized version of the Yangian on spin chains. We understood the role played by local charges or Hamiltonians defining the spin chain dynamics and how these may co-exist with the nonlocal Yangian symmetry. Different boundary conditions were discussed and we indicated the existence of certain long-range spin chains and their connection to a generalization of Drinfel'd's boost automorphism.
\medskip

Finally, we tried to better understand whether Yangian symmetry is tied to two dimensions. In \secref{sec:Neq4} we briefly introduced the four-dimensional $\superN=4$ super Yang--Mills theory. We found that the color structure of this gauge theory in the planar limit allows to introduce a two-dimensional discrete space, namely the space of color traces, on which Yangian generators may be defined. As a consequence, we have seen the bulk Yangian symmetry of the theory's dilatation operator as well as the Yangian symmetry of tree-level scattering amplitudes. Lastly, we briefly sketched the duality of $\superN=4$ SYM theory to string theory described by a 2d worldsheet theory.

\medskip
The Yangian provides in many respects a special and interesting realization of integrability. 
It represents one of three members within the family of integrable quantum group symmetries.
In fact, the Yangian may be deformed and one obtains more general quantum groups, which typically do not allow to scale away the quantum deformation parameter $\hbar$ (as it was possible for the Yangian discussed above). To be more precise, the solutions to the classical Yang--Baxter equation (and hence to its quantum deformation) fall into three categories via the Belavin--Drinfeld theorem \cite{Belavin:1982} (cf.\ \appref{app:BelavinDrinfeld}): 
1.\ Rational solutions,
2.\ Trigonometric solutions and
3.\ Elliptic solutions. 
Describing \emph{rational} quantum R-matrices, the Yangian corresponds to the simplest of these categories and thereby to the most accessible mathematical structure. Hence, a lot remains to be discovered when going beyond this class.
\medskip

Due to the limited scope of this review, certainly many interesting mathematical facts about the Yangian as well as physical applications of this algebra were not discussed or even touched in these lectures. For further reading on the Yangian and related topics, let us mention the very helpful and at many places complementary reviews by Bernard \cite{Bernard:1992ya} and MacKay \cite{MacKay:2004tc}. Also in the special context of the AdS/CFT correspondence, several useful reviews on Yangian symmetry exist, see e.g.\ \cite{
Beisert:2010jq,
Bargheer:2011mm,
Torrielli:2011gg,
Spill:2012qe}.
Note also the more general collection of reviews on integrability in AdS/CFT \cite{Beisert:2010jr}.

\paragraph{Acknowledgments.}

This review arose from lectures given at the \emph{Young Researchers Integrability School} at Durham University (UK). First of all I would like to acknowledge the Department of Mathematical Sciences at Durham University for hosting this school and for providing a great atmosphere.
I am happy to thank the organizers and the scientific committee of this school for their organizational efforts and for the invitation to lecture on this interesting topic. I am thankful to the participants for curious questions, helpful remarks and stimulating discussions. I would also like to thank the other lecturers in Durham for inspiring exchange and the fruitful coordination of the different courses. 
Special thanks to Till Bargheer, Niklas Beisert, Thomas Klose, Dennis M\"uller, Hagen M\"unkler, Anne Spiering and Alessandro Torrielli for helpful discussions and comments on the manuscript.
Let me finally thank all my collaborators on various related projects for the opportunity to study Yangian symmetry from different perspectives and to learn about this subject by working on very interesting problems.


\appendix
\addtocontents{toc}{\protect\setcounter{tocdepth}{-1}}
\section{Belavin--Drinfeld Theorem}
\label{app:BelavinDrinfeld}

According to a theorem by Belavin and Drinfel'd, the (nondengenerate) solutions $r(u)$ to the classical Yang--Baxter equation can be classified via the discrete subgroup $\Gamma\subset \mathbb{C}$ of their poles in the complex plane \cite{Belavin:1982} (see also \cite{Jimbo:1989qm}). One finds three different categories:
\begin{enumerate}
\item Rational functions with $\text{rank}(\Gamma)=0$.
\item Trigonormetric functions with $\text{rank}(\Gamma)=1$, i.e.\ functions of the form $f(e^{ku})$ with $f$ being a rational function.
\item Elliptic functions with $\text{rank}(\Gamma)=2$.
\end{enumerate}
The Yangian corresponds to quantum deformations of the first class of solutions. The last category leads to elliptic quantum groups while the second class corresponds to quantum affine algebras related to trigonometric R-matrices.

\pdfbookmark[1]{\refname}{references}
\bibliographystyle{nb}
\bibliography{LectureYangian}

\end{document}